\documentclass[aip, amsmath,amssymb, reprint]{revtex4-1}

\usepackage{graphicx}
\usepackage{dcolumn}
\usepackage{bm}

\usepackage[utf8]{inputenc}
\usepackage[T1]{fontenc}
\usepackage{mathptmx}
\usepackage{etoolbox}

\makeatletter
\def\@email#1#2{%
 \endgroup
 \patchcmd{\titleblock@produce}
  {\frontmatter@RRAPformat}
  {\frontmatter@RRAPformat{\produce@RRAP{*#1\href{mailto:#2}{#2}}}\frontmatter@RRAPformat}
  {}{}
}%
\makeatother
\begin{document}

\preprint{AIP/123-QED}

\title{Multiscale equilibration of highly entangled isotropic model polymer melts}

\author{Carsten Svaneborg}
\affiliation{University of Southern Denmark, Campusvej 55, DK-5230 Odense M, Denmark}
\email{zqex@sdu.dk}

\author{Ralf Everaers}
\affiliation{ENSL, CNRS, Laboratoire de Physique and Centre Blaise Pascal de l'École Normale Supérieure de Lyon, F-69342 Lyon, France}
\email{ralf.everaers@ens-lyon.fr}

\date{\today}

\begin{abstract}
We present a computationally efficient multiscale method for preparing equilibrated, isotropic long chain model polymer melts. 
As an application we generate Kremer-Grest melts of $1000$ chains with $200$ entanglements and $25000$-$2000$ beads per chain, which cover the experimentally relevant bending rigidities up to and beyond the limit of the isotropic-nematic transition.
In the first step, we employ Monte Carlo simulations of a lattice model to equilibrate the large-scale chain structure above the tube scale
 while ensuring a spatially homogeneous density distribution.
We then use theoretical insight from a constrained mode tube model to introduce the bead degrees of freedom together with random walk conformational statistics all the way down to the Kuhn scale of the chains.
This is followed by a sequence of simulations with carefully parameterized force-capped bead-spring models, which slowly introduce the local bead packing while reproducing the larger scale chain statistics of the target Kremer-Grest system at all levels of force-capping.
Finally we can switch to the full Kremer-Grest model without perturbing the structure.
The resulting chain statistics is in excellent agreement with literature results on all length scales accessible in brute-force simulations of shorter chains.
\end{abstract}

\maketitle


Polymeric systems exhibit a wide range of characteristic time and length scales.
In the case of natural rubber or {\it cis-}PI~\citep{Everaers2020Mapping},
important static scales comprise (i) the Kuhn length, $l_K \approx 1\mbox{nm}$, characterizing the crossover to the large-scale random-walk behavior, (ii) the tube diameter, $d_T \approx 5\mbox{nm}$ with $N_{eK}=24$ Kuhn segments between entanglements, for the onset of entanglement effects, (iii) the coil diameter,  $\langle R^{2}\rangle = l_K L \approx 100 \mbox{nm}$, and (iv) the contour length, $L \approx 10 \mu\mbox{m}$ for chains with a typical length of  $N_K=10^4$ Kuhn segments with $Z = N_K/N_{eK}=390$ entanglements per chain. 
The spread is even larger between the characteristic time scales (here estimated at $T=298K$).
There are already almost three orders of magnitude between the Kuhn time, $\tau_K\sim 1\times 10^{-7}$s, and the entanglement time, $\tau_e\sim 7\times 10^{-5}$s when monomers diffuse over distances comparable to $l_K$ and $d_T$ respectively. The Rouse time of $\tau_{R} = N_K^2 \tau_K \sim 10^8 \tau_K \sim 10$s governs {\em fast} processes such as the tension equilibration inside the tube~\cite{DoiEdwards86}, while the estimated maximal relaxation time is $\tau_{max} =3 \tau_e Z^3 \sim 3$h.

Different aspects of polymeric behavior can be studied with models representing different levels of coarse-graining.\citep{faller2007coarse,MultiscalePeterKremerFaradayDiss2010,gooneie2017review}
For microscopic models, it remains a challenge to prepare well-equilibrated initial states in the long chain limit, since standard simulation methods following the physical dynamics are not able to bridge the gap between the time step and the longest relaxation in the system.
Alternative approaches fall into two classes. Variants of the double-bridging algorithm~\citep{karayiannis2002novel,karayiannis2002atomistic,karayiannis2003advanced, auhl2003equilibration, dietz2022facile}
introduce connectivity-altering moves allowing pivot-like~\citep{madras1988pivot} moves in dense systems for the equilibration of the large-scale chain structure. Schemes building on the Flory ideality hypothesis~\citep{flory49} superpose pre-equilibrated chain configurations with the proper large-scale random walk statistics in the simulation box and gradually remove the local overlap between monomers~\citep{kremer90,auhl2003equilibration,carbone2010fine,moreira2015direct,SvaneborgEquilibration2016}. This approach can be generalized to a systematic multi-scale approach, which equilibrates density fluctuations and chain conformations from the largest scales down to the monomer~\citep{zhang2014equilibration,zhang2015communication,SvaneborgEquilibration2016} or even atomic scale~\citep{zhang2019hierarchical}.

Here we follow the latter route in building on our previous work~\citep{SvaneborgEquilibration2016}, where we have already used a simple multi-chain lattice model to equilibrate the chain and melt structure beyond the tube scale.
The novelty of the present work lies firstly in our use of the constrained mode tube model formalism \cite{WarnerEdwards,read1997lozenge,everaers1998constrained,mergell2001tube} for the fine-graining step from the lattice to the bead scale, which allows us to minimize lattice artifacts and to extend the method to polymers with higher intrinsic stiffness.
Secondly, we have parameterized the intrinsic bending stiffness for an entire family of force-capped bead-spring models such that they can be employed for an iso-configurational introduction of excluded volume interactions.
As a result, we are now able to prepare long-chain KG melts for bending rigidities, which cover the entire range from intrinsically flexible chains corresponding to commodity polymers~\citep{Everaers2020Mapping} up to and beyond the limit~\citep{faller1999local,dietz2022facile} of the isotropic-nematic transition.

The paper is structured as follows:
In Section~\ref{sec:Introduction} we provide a more detailed introduction into the problem, review the state of the art and sketch the content of the present work.
Section~\ref{sec:Background} provides detailed background information on the employed models, methods and the notation. 
Section~\ref{sec:CMM finegraining} deals with fine-graining from the tube to the Kuhn scale, 
Section~\ref{sec:Push-off} with the iso-configurational push-off procedure.
In the Results Section~\ref{sec:results-and-discussion} we (i) analyze the conformational properties of the generated KG melts, (ii) compare them to target values and (iii) test if they are stable during subsequent simulations with the full force-field.
We finish with a brief conclusion in Section~\ref{sec:Conclusions}.

\section{Introduction and state of the art \label{sec:Introduction}}

\subsection{Modelling polymers on different time and length scales}

The choice of the polymer model in a theoretical or computational investigation is largely dictated by the character of the target properties of interest.
Bulk densities, the temperatures below which materials become glassy\cite{ngai2007temperature}, or the ability of polymers to form semi-crystalline phases\cite{patlazhan2012structural} depend on specific chemical details at the monomer scale and require atom-scale modeling,
see e.g. Refs. \cite{theodorou1986atomistic,kotelyanskii1996building,doherty1998polymerization,faller2001local,abrams2003combined,milano2005mapping,sun2005systematic,tsolou2005detailed,tzoumanekas2006atomistic,harmandaris2006hierarchical,neyertz2008molecular},
or at least the inclusion of selected molecular details in models for specific polymer chemistries see e.g. Refs.\cite{padding2002time,liu2013coarse,salerno2016resolving,faller2002modeling,faller2003properties,li2011primitive,maurel2015prediction,harmandaris2006hierarchical,chen2007viscosity,fritz2011multiscale,karimi2008fast,eslami2011coarse,chen2008comparison,maurel2015prediction,tschop1998simulation,abrams2003combined,hess2006long,strauch2009coarse,milano2005mapping,maurel2015prediction}
to preserve a certain degree of transferability~\cite{baschnagel2000bridging,carbone2008transferability,MultiscalePeterKremerSOftMatter2009} and representability~\cite{johnson2007representability}.

In contrast, characteristic properties of polymeric systems like the variation of the melt viscosity with the molecular weight of the chains, are controlled by the large-scale conformational statistics and dynamics of long entangled chains~\cite{DoiEdwards86}. 
Such properties are universal~\cite{deGennes79,DoiEdwards86} in the sense that a large number of chemically different systems exhibit identical behavior, if measurements are reported in suitable units such as those related to the Kuhn scale.~\citep{Everaers2020Mapping}
Universal properties can be studied with generic polymer models like the Kremer-Grest (KG) bead-spring polymer model\citep{grest1986molecular}. Tuning a single  parameter controlling the chain stiffness~\cite{faller1999local,faller2000local,faller2001chain} the KG model is able to represent the conformational properties of all commodity polymers at the Kuhn scale.\citep{Everaers2020Mapping} Predicting experimental results using computer simulations necessitates the availability of well equilibrated simulation models that accurately capture the conformational properties of real polymer materials.

\subsection{Building melt configurations for bead-spring models}

Consider the application of bead-spring models to investigate
the dynamics of polymer melts, e.g. Refs. \citep{kremer1990dynamics,hou2010stress,Svaneborg2020Characterization},
of the effect of cross-linking on the structure-property relations of rubber-like materials, e.g. Refs. \citep{grest90a,duering94,SGE_prl_04,svaneborg2008connectivity,Gula2020Entanglement},
the viscoelastic response to strain, e.g. Refs. \citep{everaers1996elastic,everaers1999entanglement,cao2015simulating,oconnor2018relating,xu2018molecular},
or the glass transition e.g. Refs. \citep{hoy2006strain,hoy2011understanding}.
Clearly their success depends on the ready availability of properly equilibrated melt configurations characterized by 
1) local chain conformations in agreement with those measured in shorter chain melts amenable to brute-force equilibration
2) large-scale chain conformations matching the expected and/or extrapolated random walk statistics, 
3) the expected and/or extrapolated density of topological entanglements, and 
4) the near absence of density fluctuations typical for nearly incompressible bulk polymers.
These requirements couple the melt structure at micro-, meso- and macroscopic scales, making it highly non-trivial to generate well equilibrated melts. 

As already pointed out in the Introduction, brute force equilibration using a physically realistic dynamics is only computationally feasible for melts of
short chains, since polymer dynamics is slow in general and entanglements cause the conformational relaxation time to increase even more drastically. 
Double bridging Monte Carlo moves\citep{karayiannis2002novel,karayiannis2002atomistic,karayiannis2003advanced} offer an attractive alternative.
They introduce large-scale conformational changes into MC or hybrid MC/MD simulations\citep{sides2004effect} by swapping the tails of a pair of chains at a common contact point.
For atomistic and (stiffer) bead-spring models, double-bridging moves require a certain amount of softening of the local interactions to have a finite success rate
~\citep{karayiannis2002novel, bobbili2020simulation, dietz2022facile}. Some care needs to be taken to ensure that such a modification do not alter the chain statistics.
Unfortunately, the computational efficiency of the approach decreases with chain length, since the density of suitable contact points drops as the move alters in general the lengths of the involved chains.

For (by present standards) modest system sizes, it is actually quite feasible to generate viable melt conformations from the random superposition of pre-equilibrated chain configurations in the simulation box~\citep{kremer90}. 
As the random overlap of Lennard-Jones type beads leads to numerical instabilities, this requires a push-off process where the liquid-like bead packing is gradually introduced in simulations with non-diverging pair interactions.  With the bead motion during the push-off limited to the monomer scale, the large scale chain statistics remains unaffected. 
Unfortunately, for longer chains the push-off  leads to swelling of the chain conformations at intermediate scales. 
Auhl {\it et al.}\citep{auhl2003equilibration}
addressed this problem by 1) reducing the initial density fluctuations by pre-packing the chains,
2) using force-capped pair interactions to ensure realistic local
bead packing during the push-off, and 3) retaining the full interaction between bonded
neighbors to produce the correct local bond statistics. This approach
was used to equilibrate melts comprising $80$ chains of lengths up
to $7000$ beads. Moreira et al.\citep{moreira2015direct} improved
this method by using a freely-rotating chain model to reproduce the
desired chain statistics combined with a push-off process using a
feedback control loop and were able to generate melts with
$1000$ chains of $2000$ beads. 

\subsection{Multi-scale approaches to building melt configurations}

The wide range of relevant time and length scales in polymeric systems makes them natural targets for multi-scale modelling.\cite{faller2007coarse,gooneie2017review,MultiscalePeterKremerFaradayDiss2010} In particle-based models, the resolution ranges from the atom scale to DPD-like descriptions, where entire chains are represented by one or two soft spheres or ellipsoids\cite{hahn2001simulation,muller2013speeding}. 
While systematic coarse-graining is an important aspect of the multi-scale view of polymeric systems, the framework also lends itself to fine-graining, i.e. the insertion of additional molecular details into a more coarse-grain model \cite{faller2007coarse,gooneie2017review,MultiscalePeterKremerFaradayDiss2010,carbone2010fine}.

In the present context, pre-packing and push-off schemes can be generalized to a systematic multi-scale approach, which equilibrates density
fluctuations and chain conformations from the largest scales down.
Zhang et al.\citep{zhang2014equilibration,zhang2015communication,zhang2019hierarchical}
represented the chains in a polymer melt via
a hierarchy of soft blob models with matching invariant degrees
of polymerization. Each blob model can then be fine-grained to an
equivalent lower resolution model until a scale comparable to KG models
is reached. The melts generated with this approach were comparable
to those of Moreira et al.\citep{moreira2015direct}

\subsection{Improvements introduced in the present work}

\begin{figure*}
\includegraphics[width=1.0\textwidth]{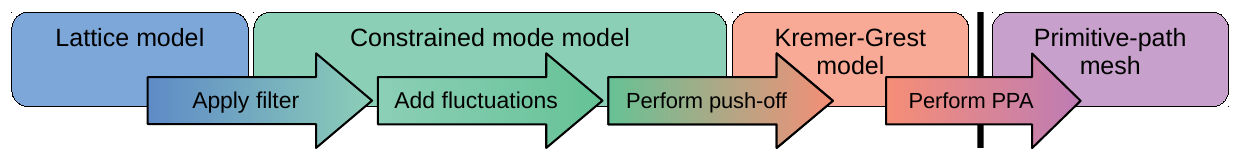}
\includegraphics[width=0.2\textwidth]{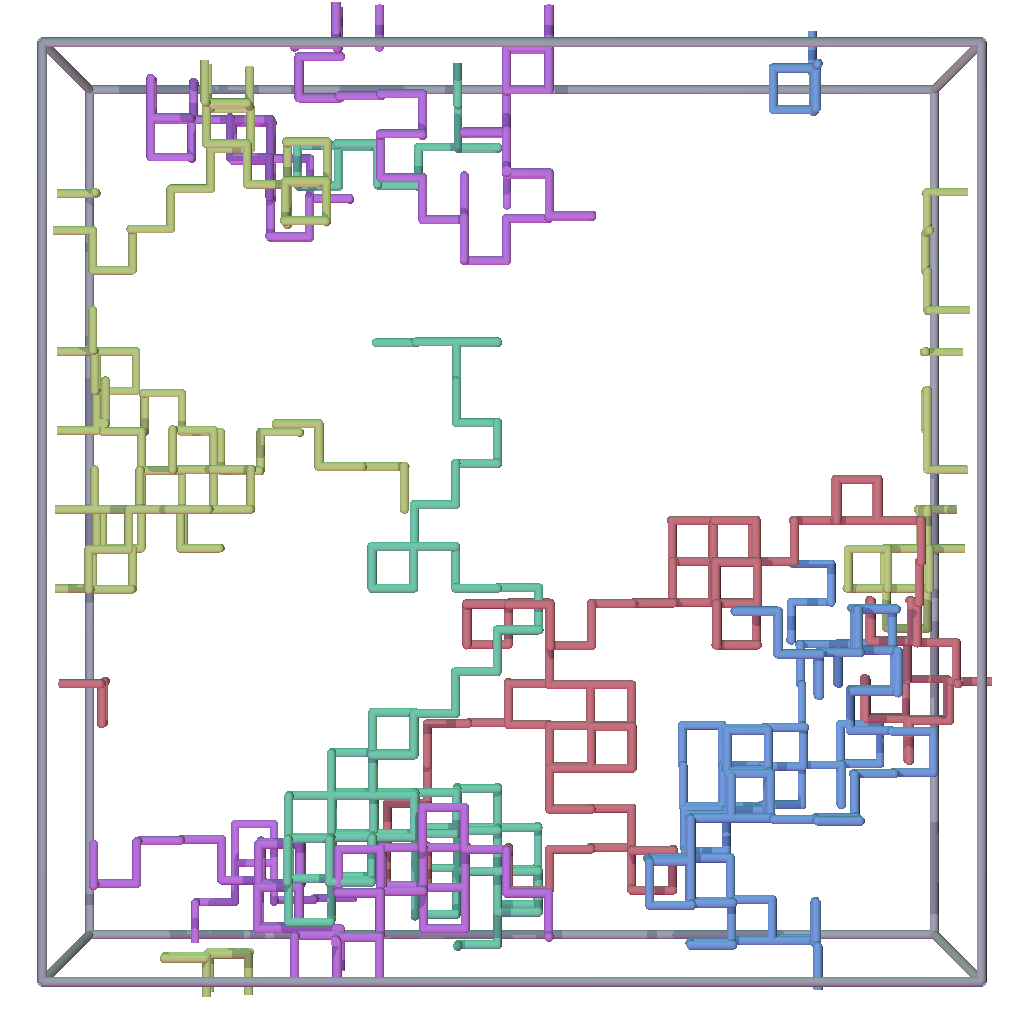}%
\includegraphics[width=0.2\textwidth]{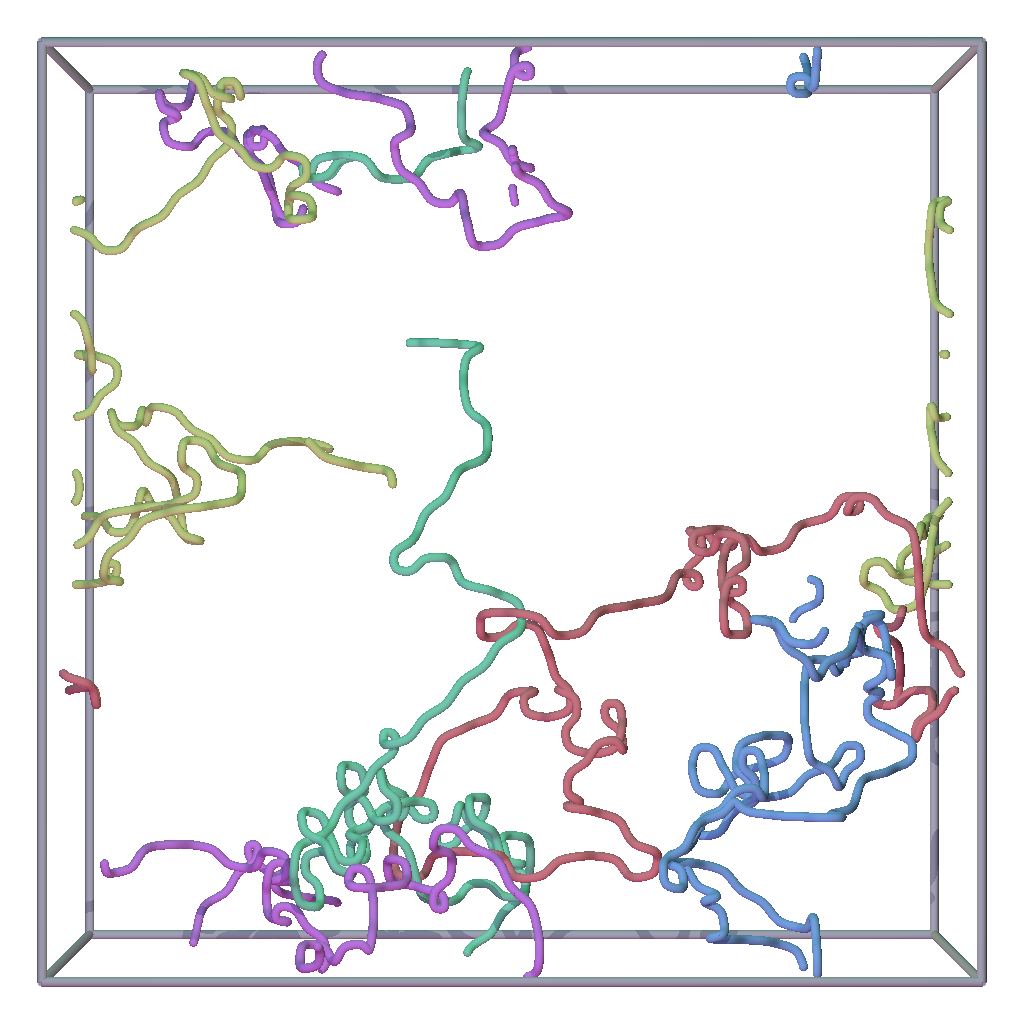}%
\includegraphics[width=0.2\textwidth]{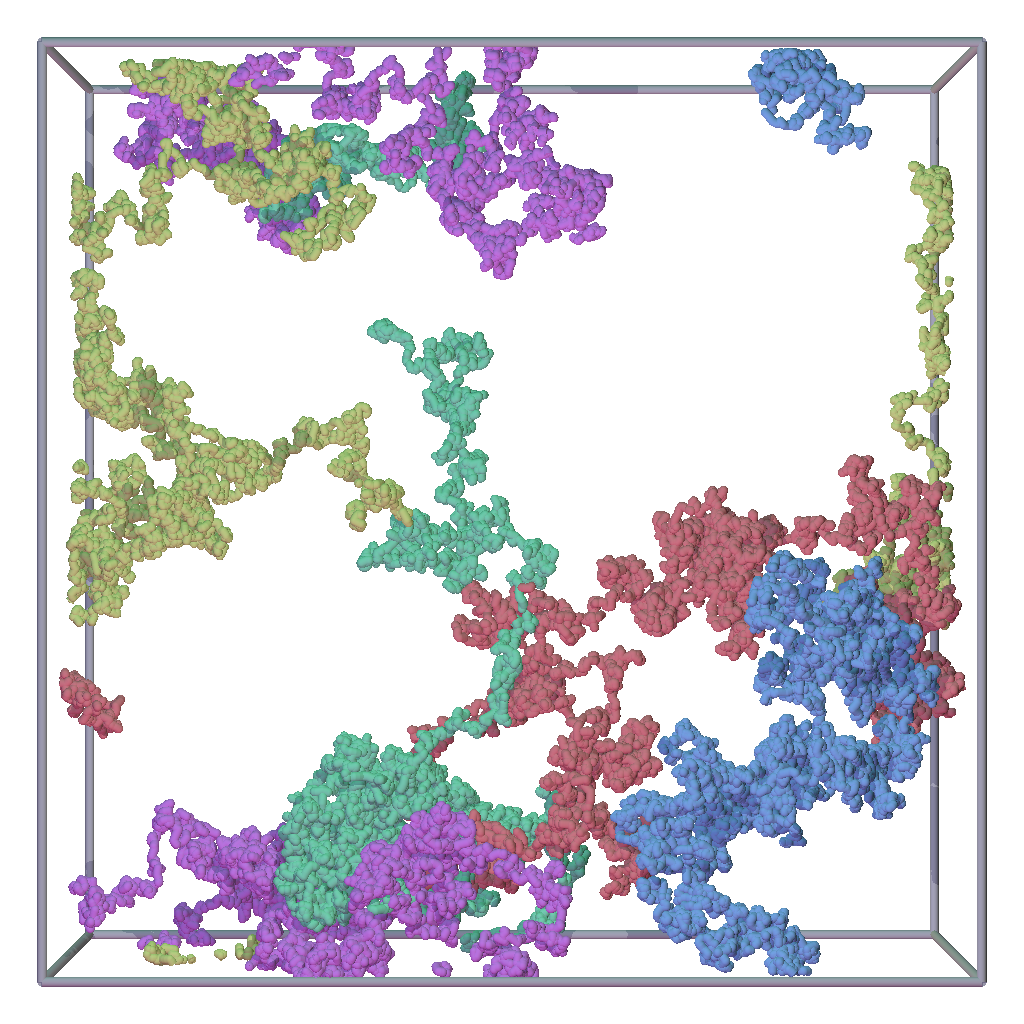}%
\includegraphics[width=0.2\textwidth]{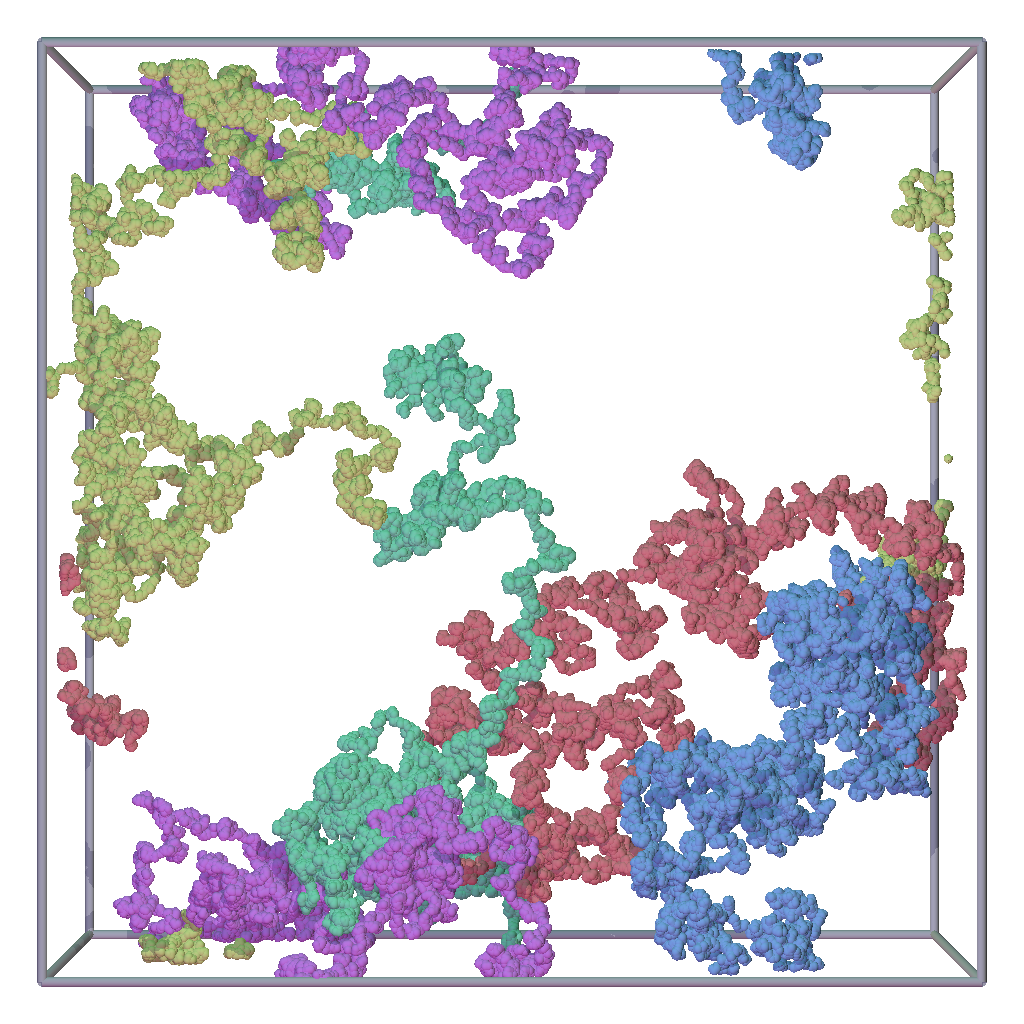}%
\includegraphics[width=0.2\textwidth]{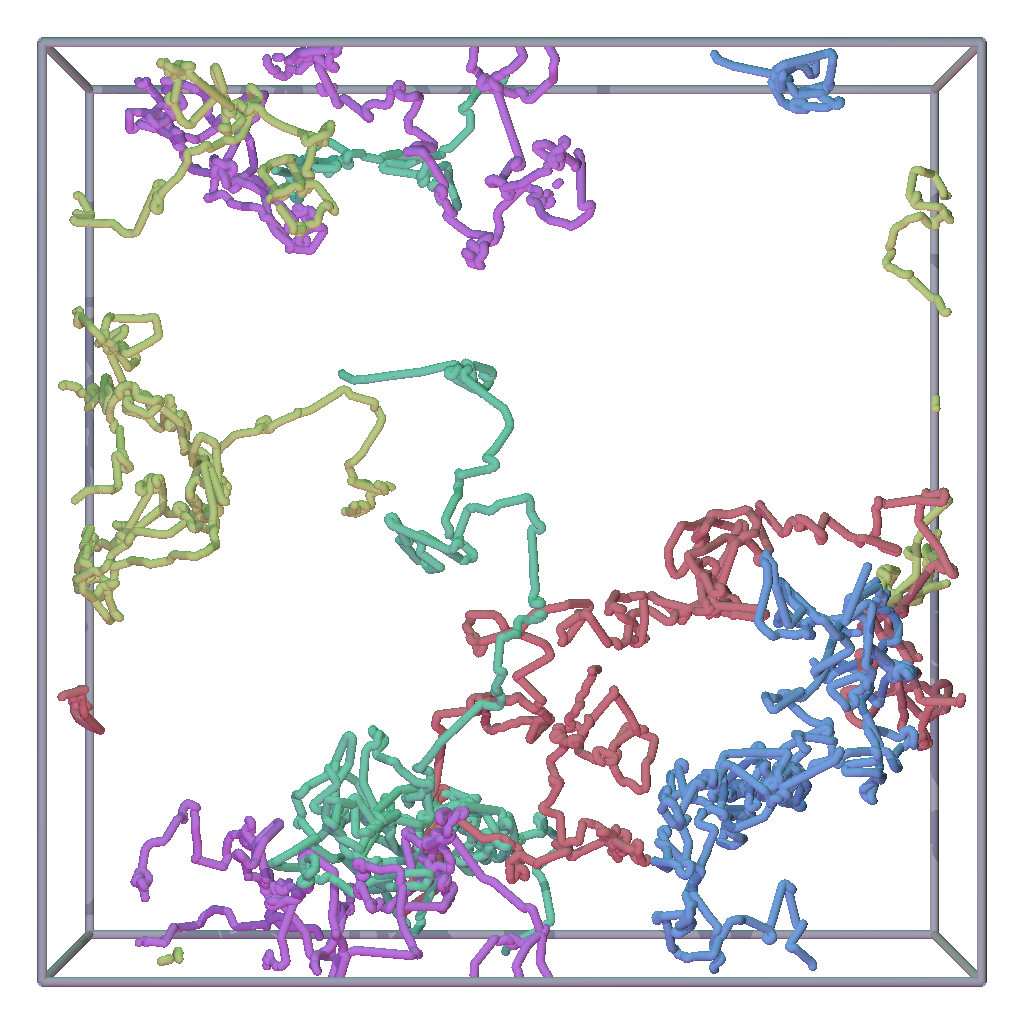}%

\includegraphics[width=0.2\textwidth]{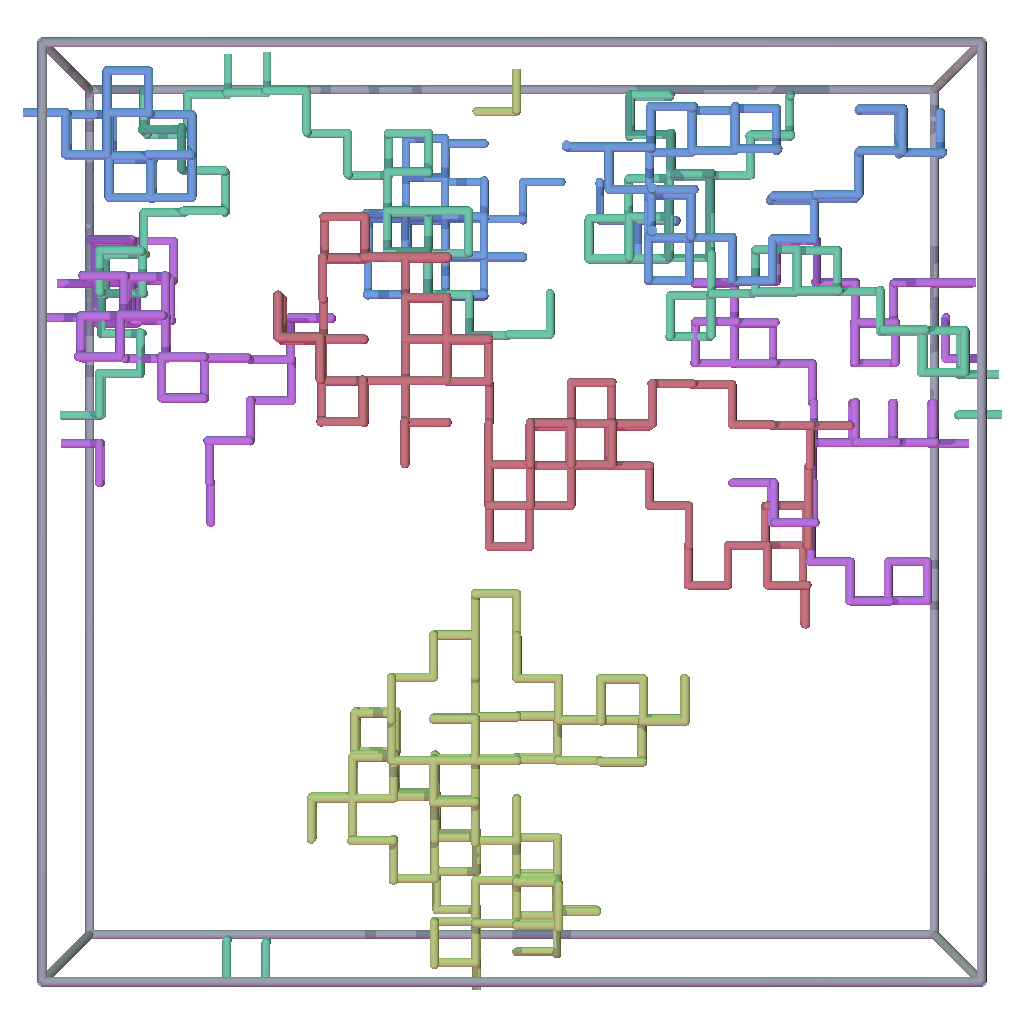}%
\includegraphics[width=0.2\textwidth]{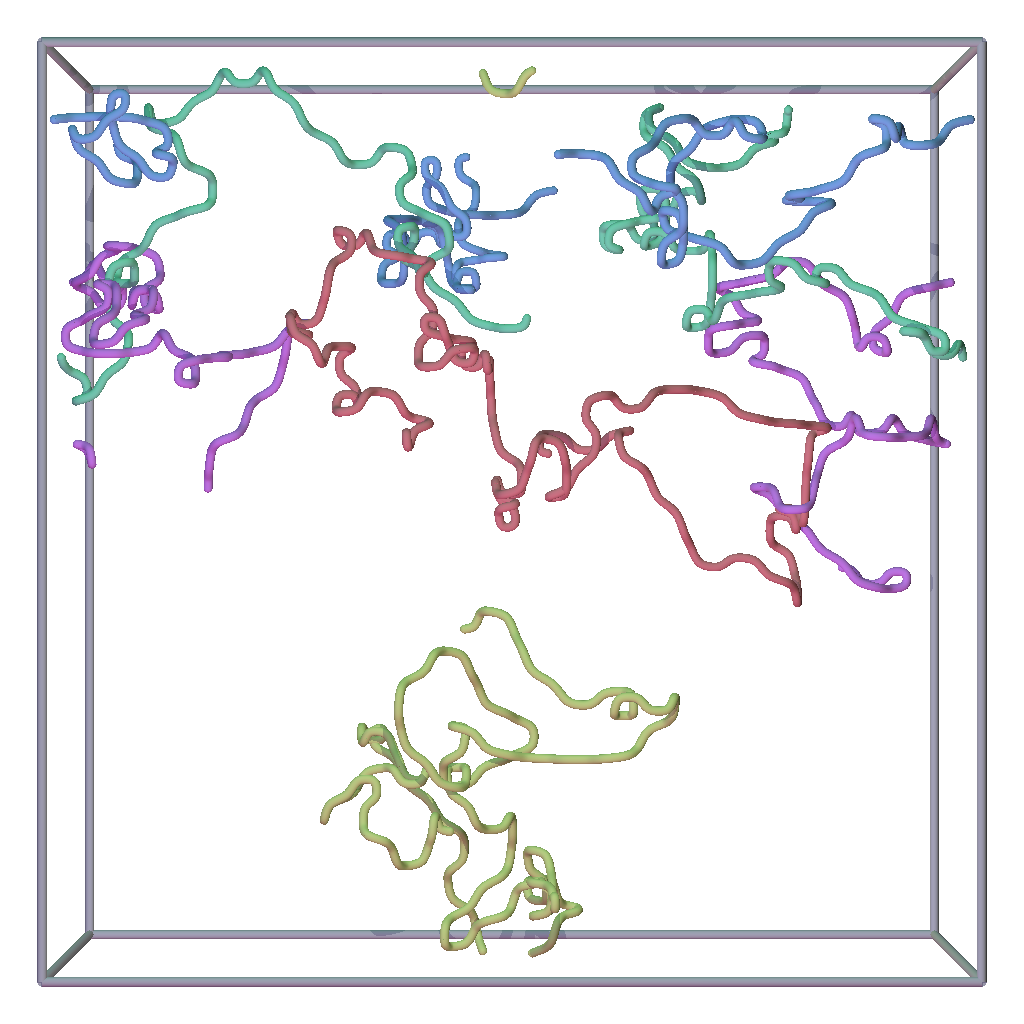}%
\includegraphics[width=0.2\textwidth]{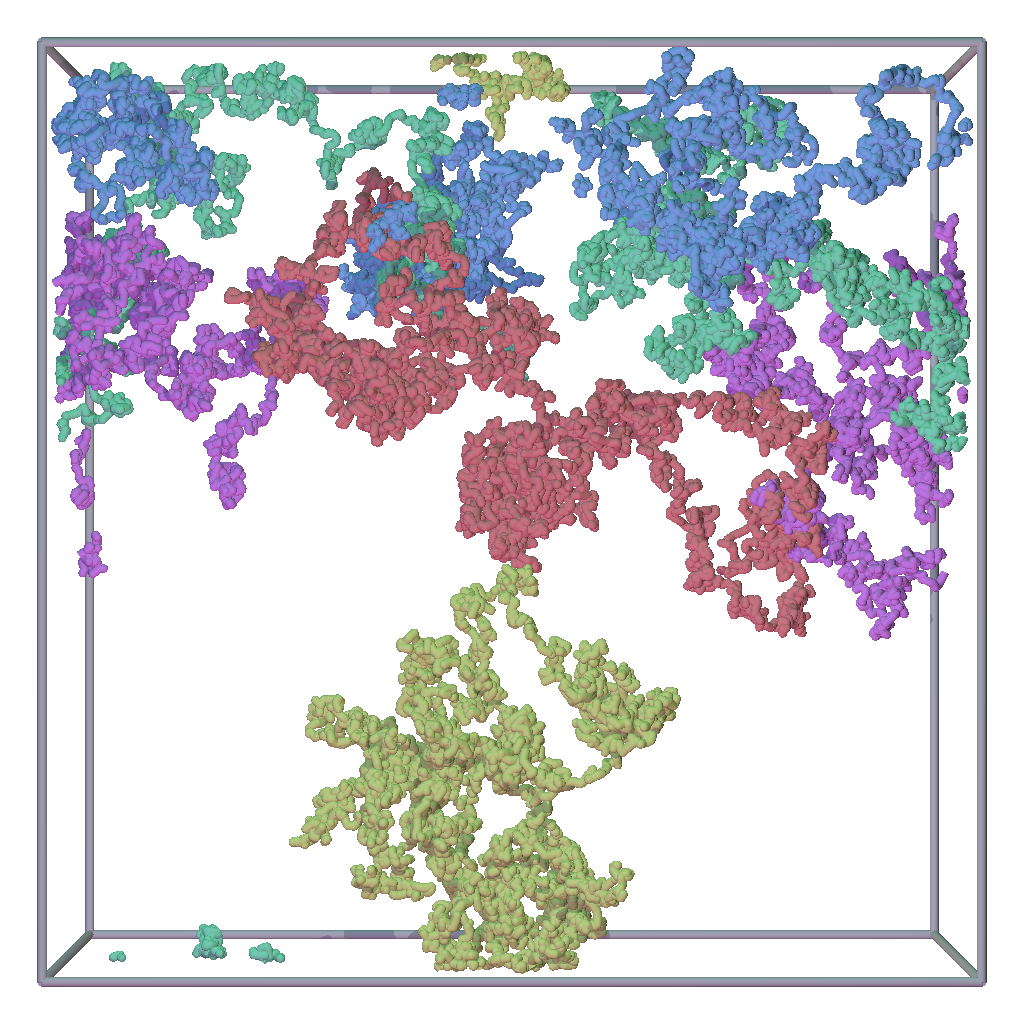}%
\includegraphics[width=0.2\textwidth]{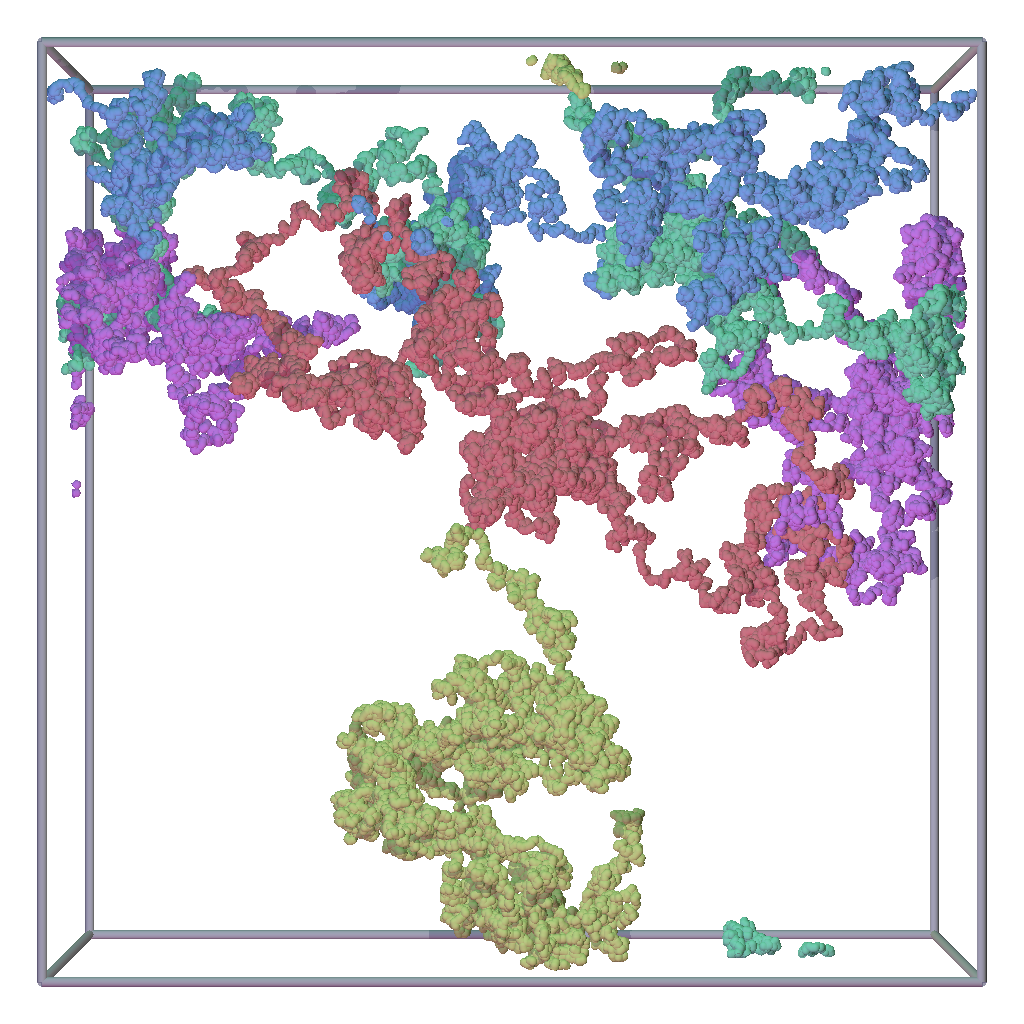}%
\includegraphics[width=0.2\textwidth]{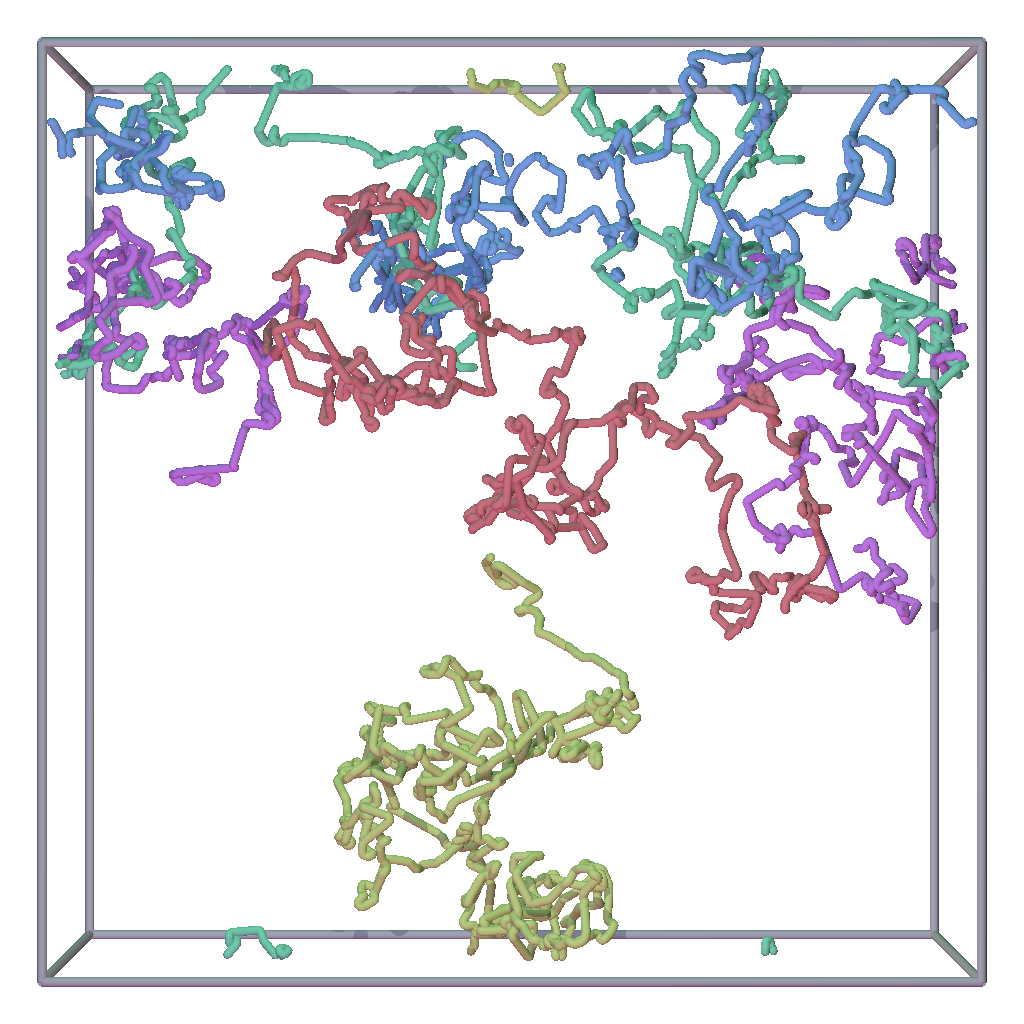}%

\includegraphics[width=0.2\textwidth]{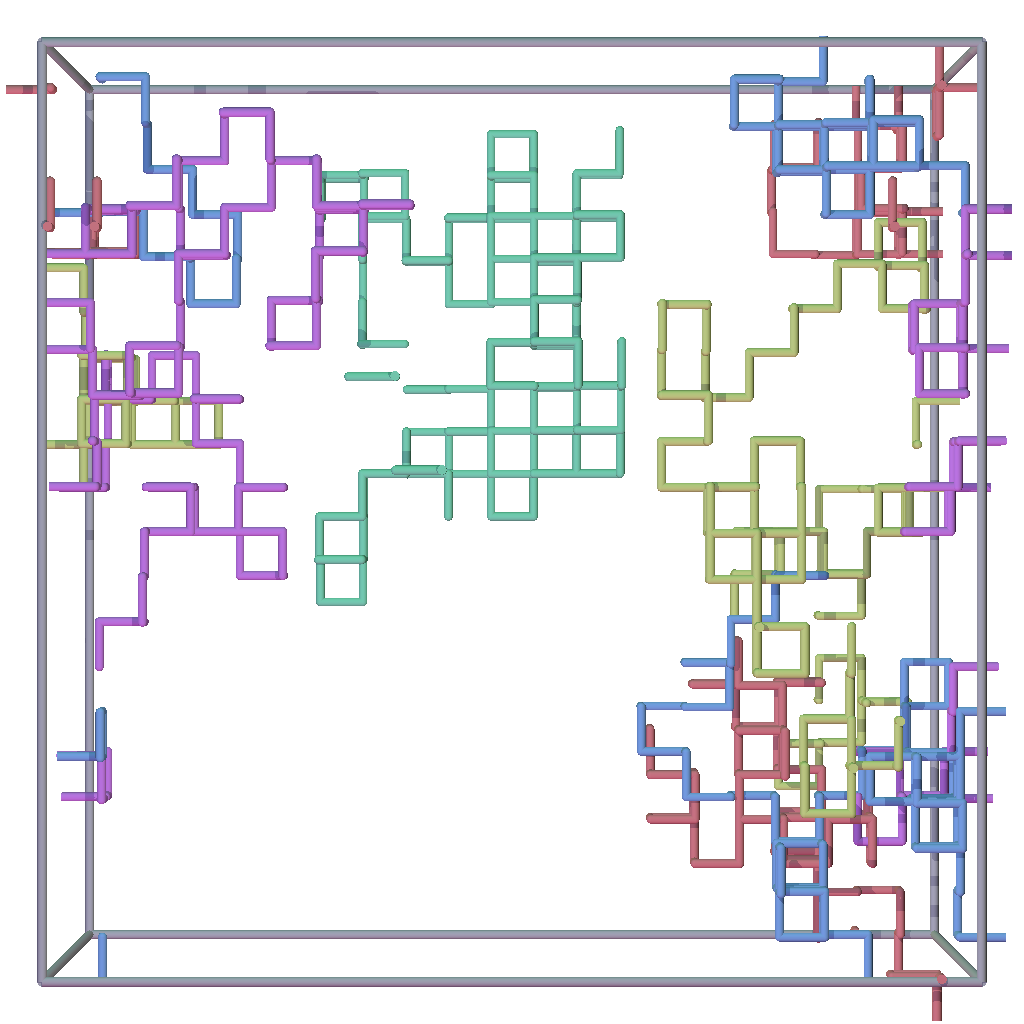}%
\includegraphics[width=0.2\textwidth]{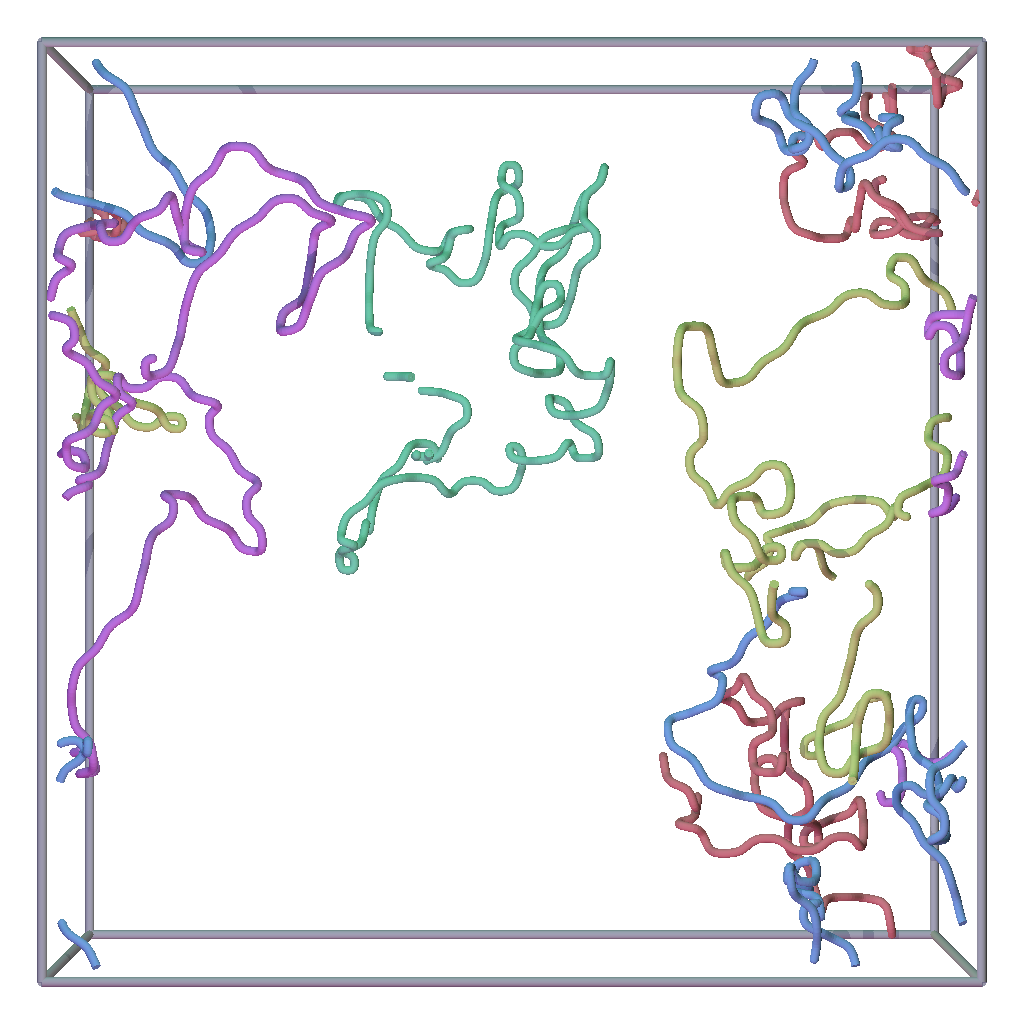}%
\includegraphics[width=0.2\textwidth]{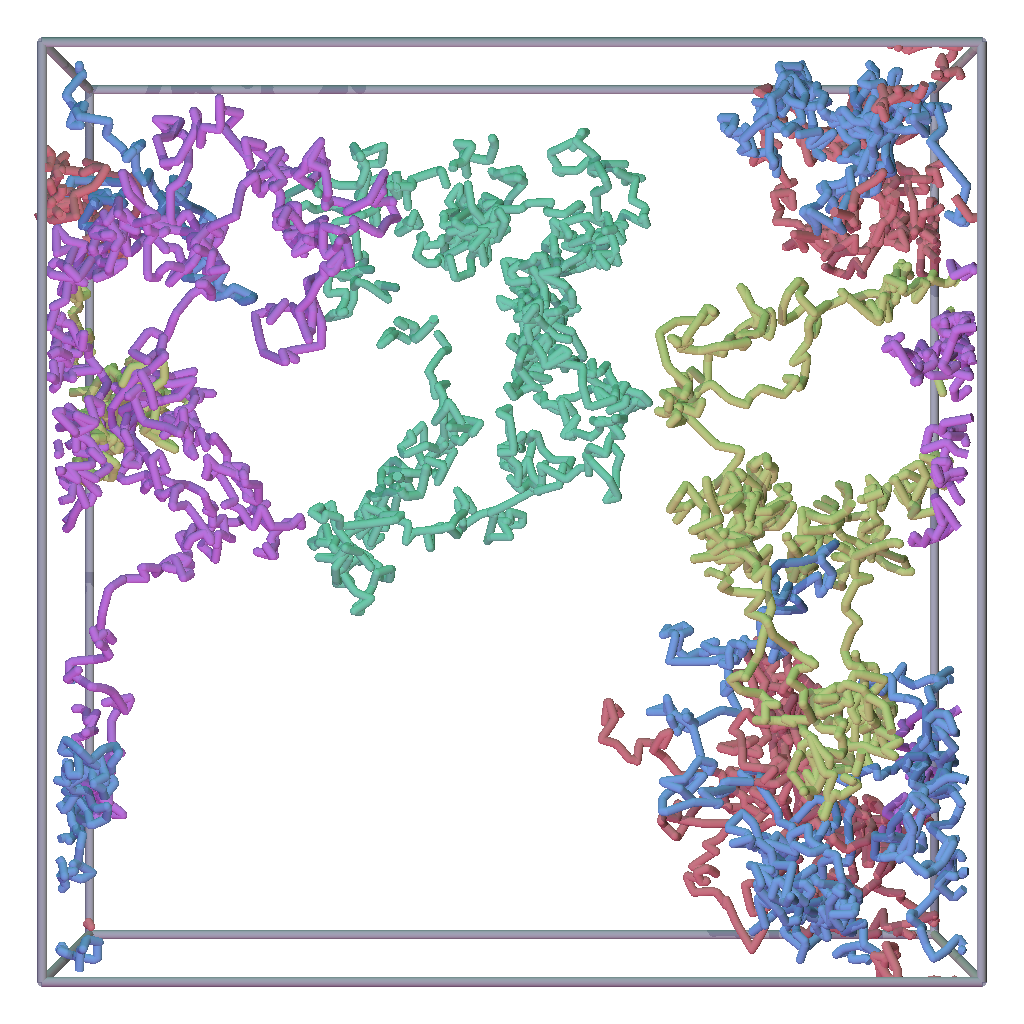}%
\includegraphics[width=0.2\textwidth]{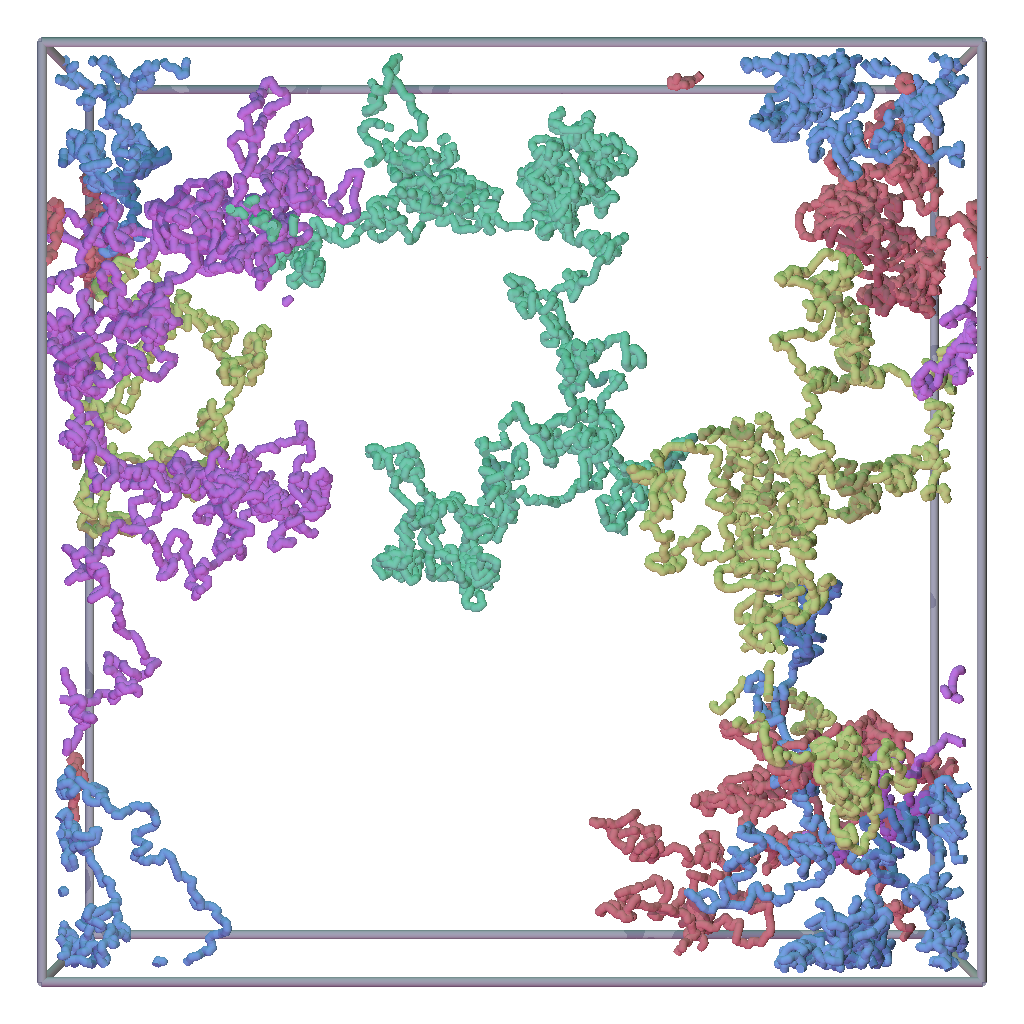}%
\includegraphics[width=0.2\textwidth]{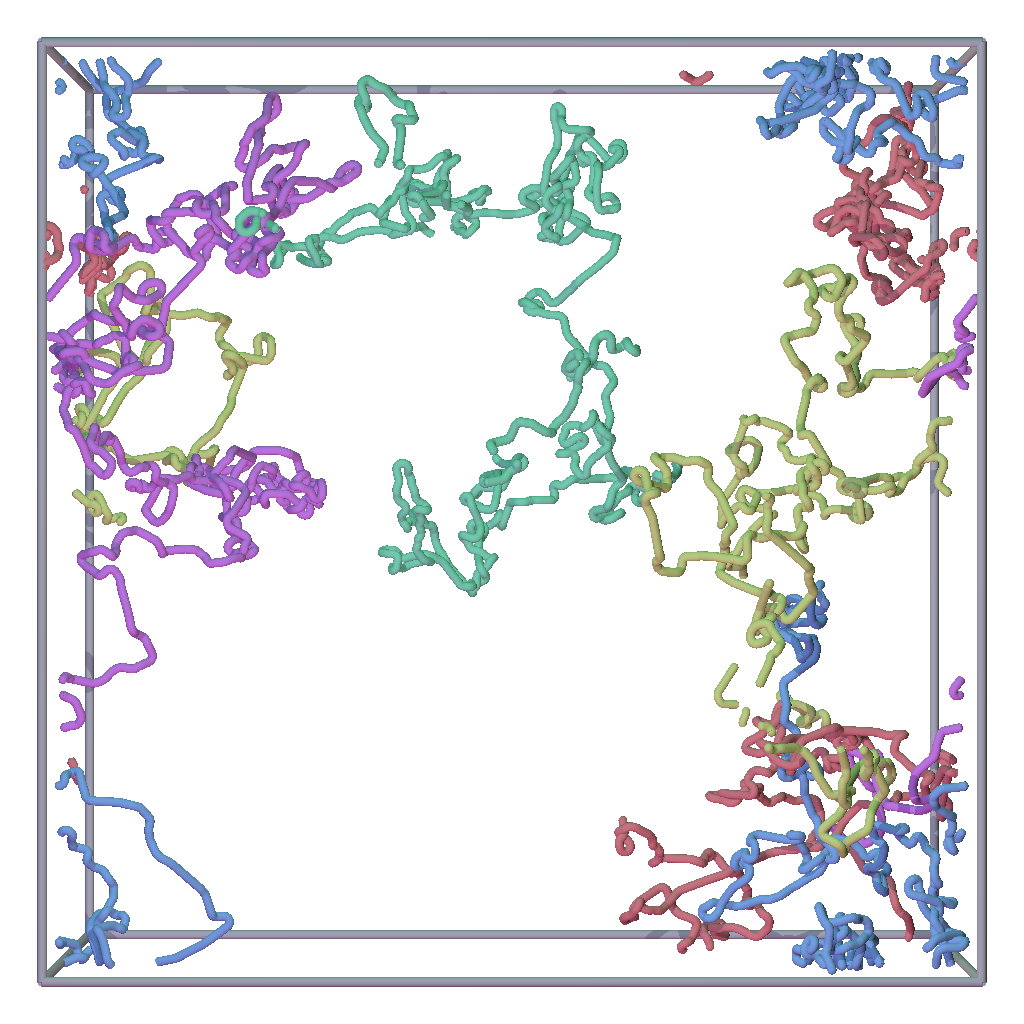}%

\includegraphics[width=0.2\textwidth]{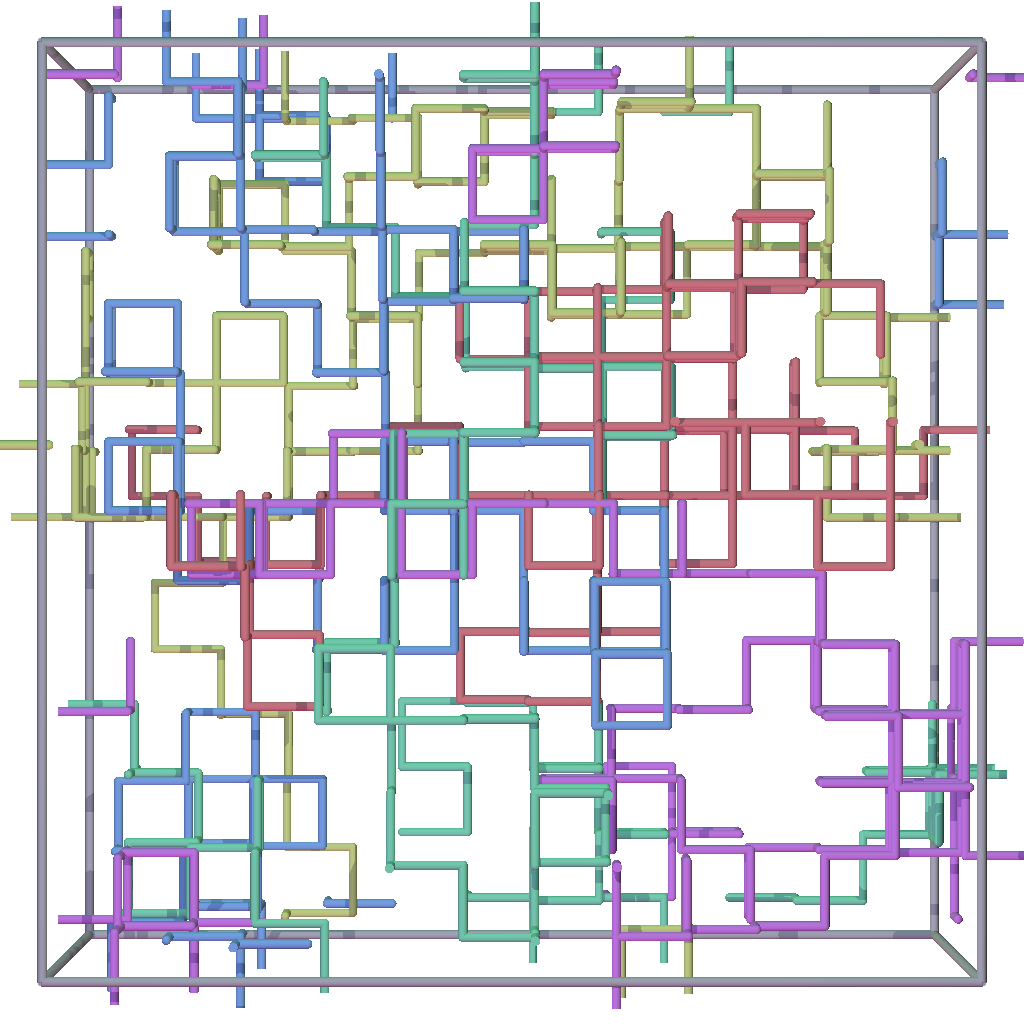}%
\includegraphics[width=0.2\textwidth]{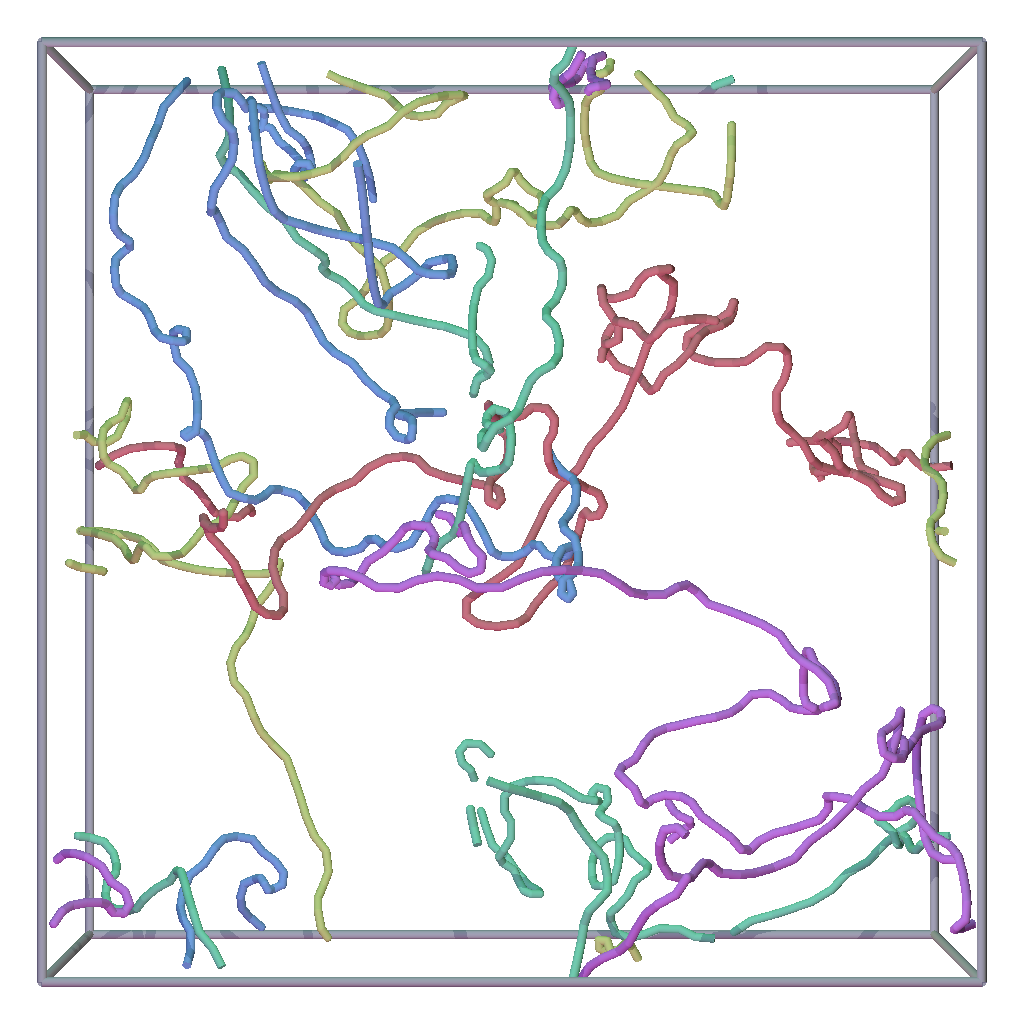}%
\includegraphics[width=0.2\textwidth]{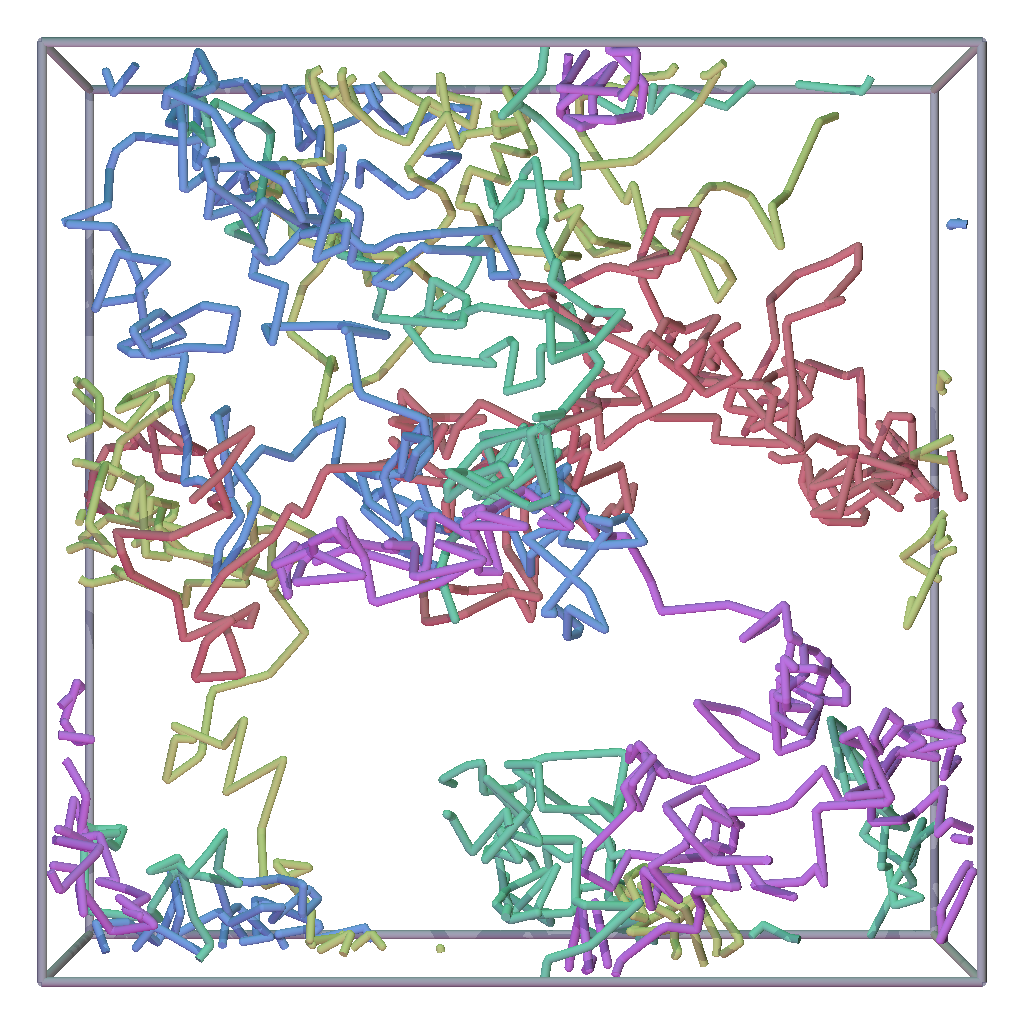}%
\includegraphics[width=0.2\textwidth]{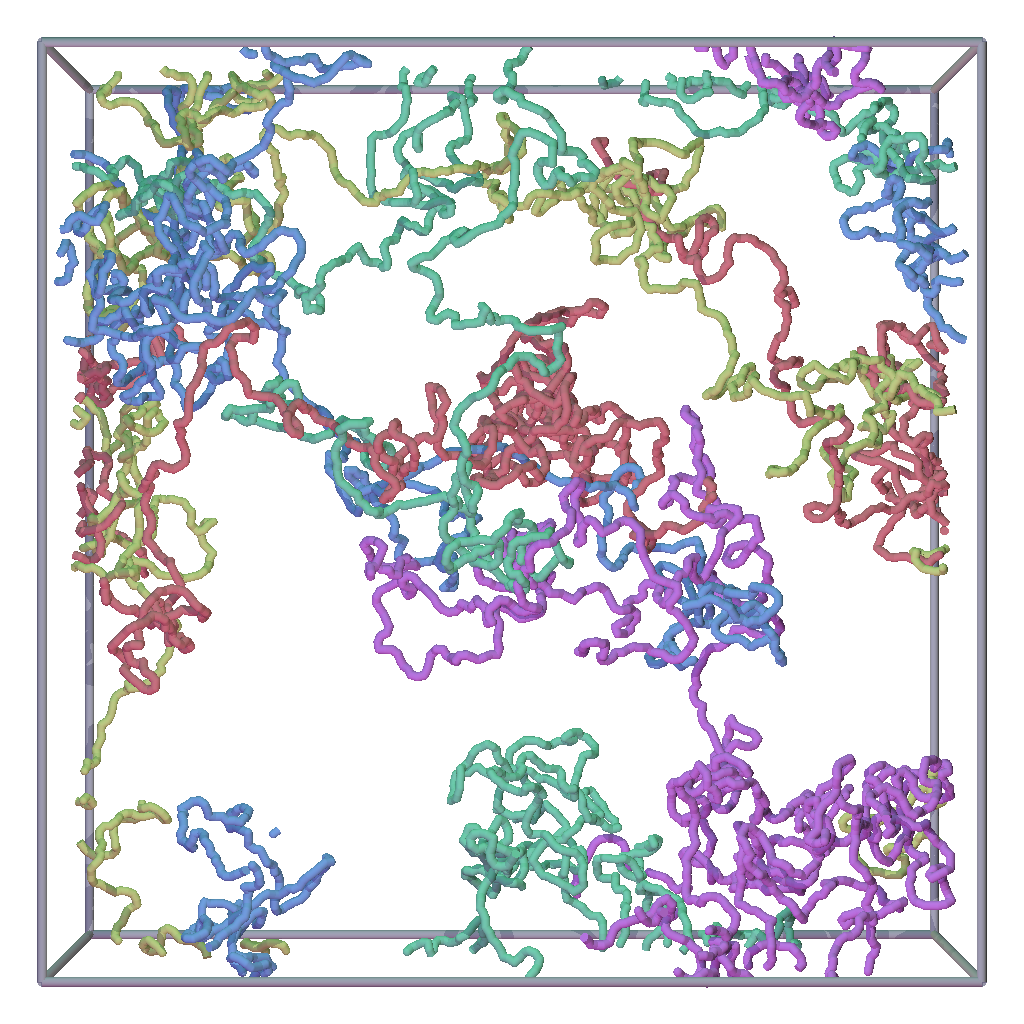}%
\includegraphics[width=0.2\textwidth]{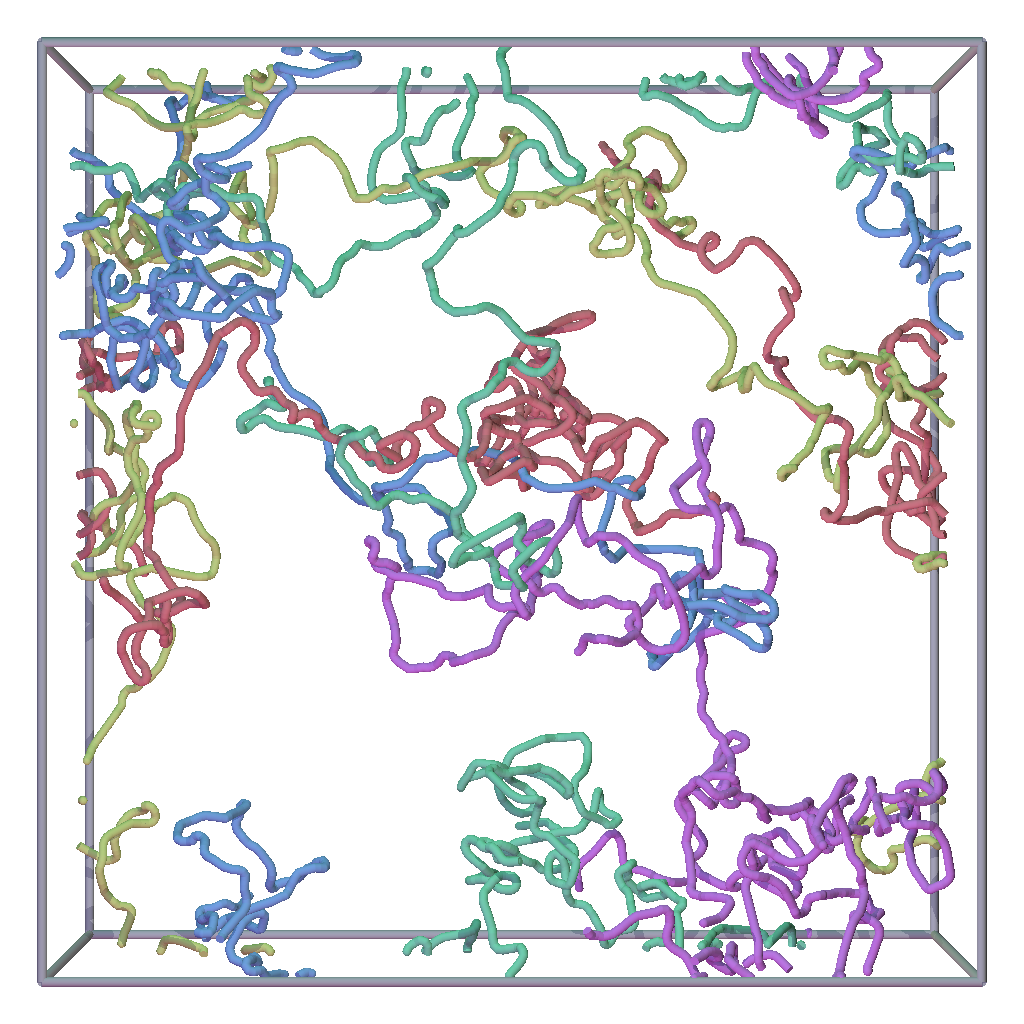}%

\includegraphics[width=0.2\textwidth]{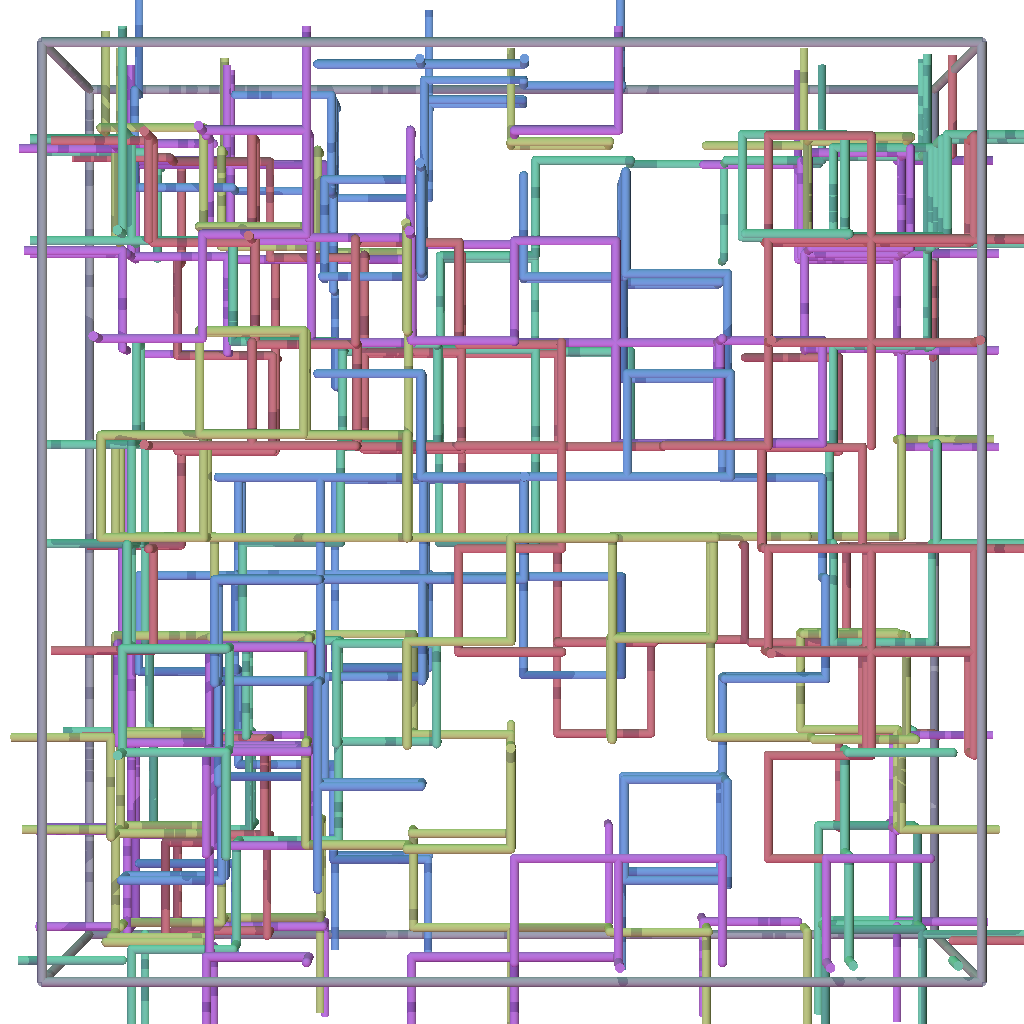}%
\includegraphics[width=0.2\textwidth]{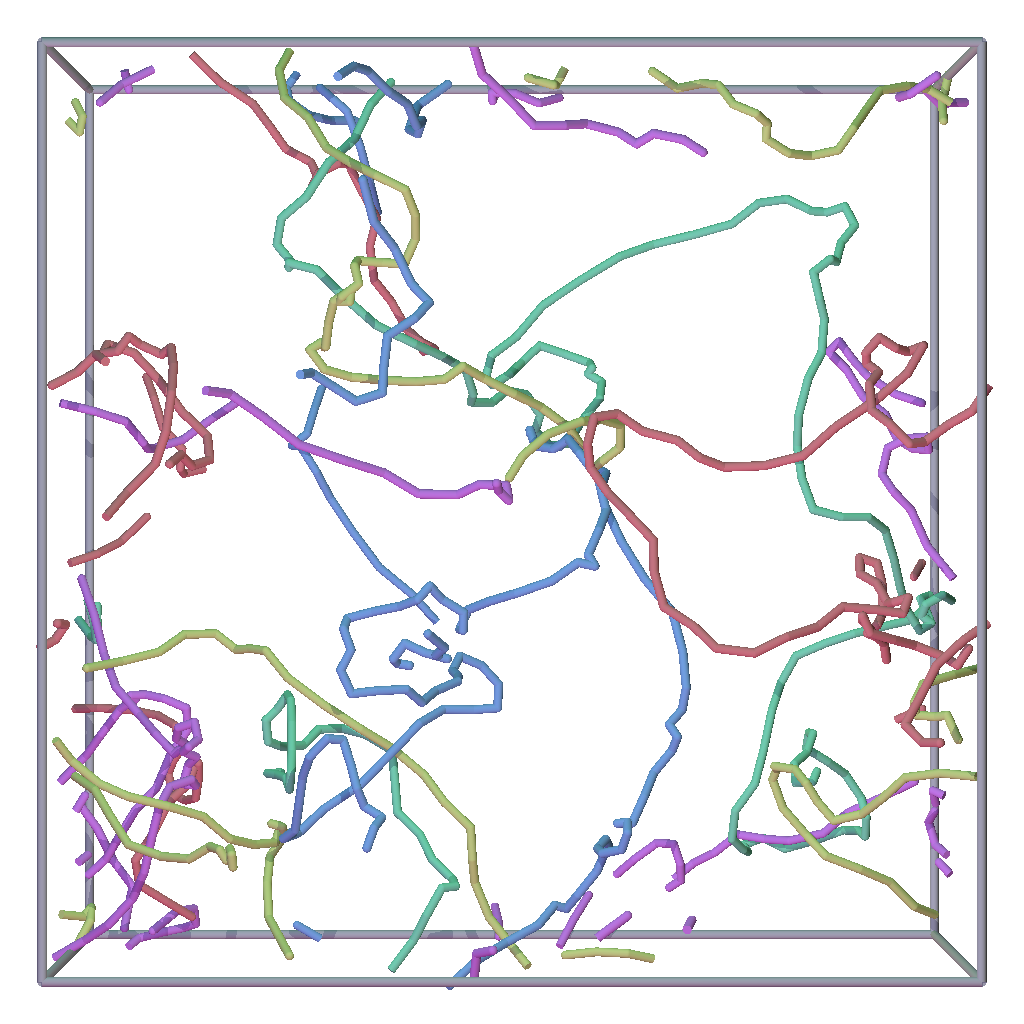}%
\includegraphics[width=0.2\textwidth]{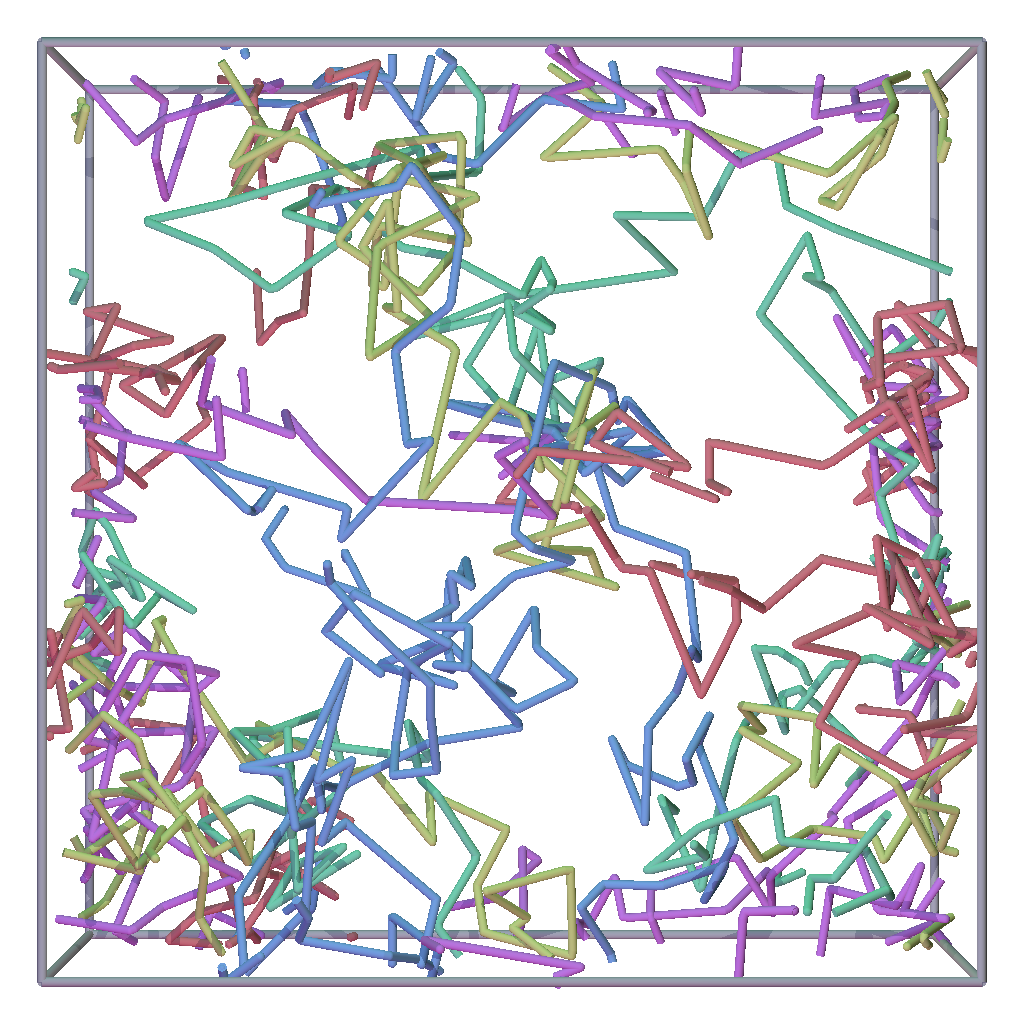}%
\includegraphics[width=0.2\textwidth]{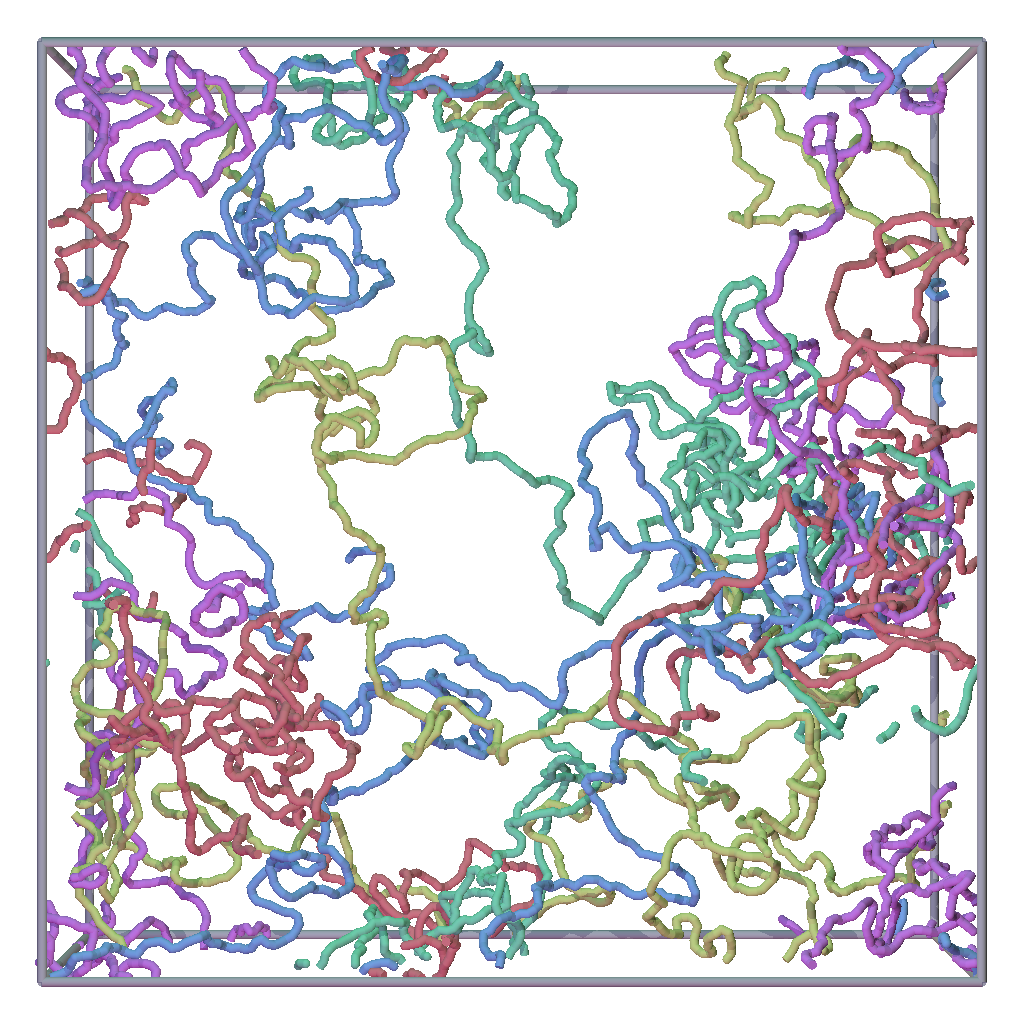}%
\includegraphics[width=0.2\textwidth]{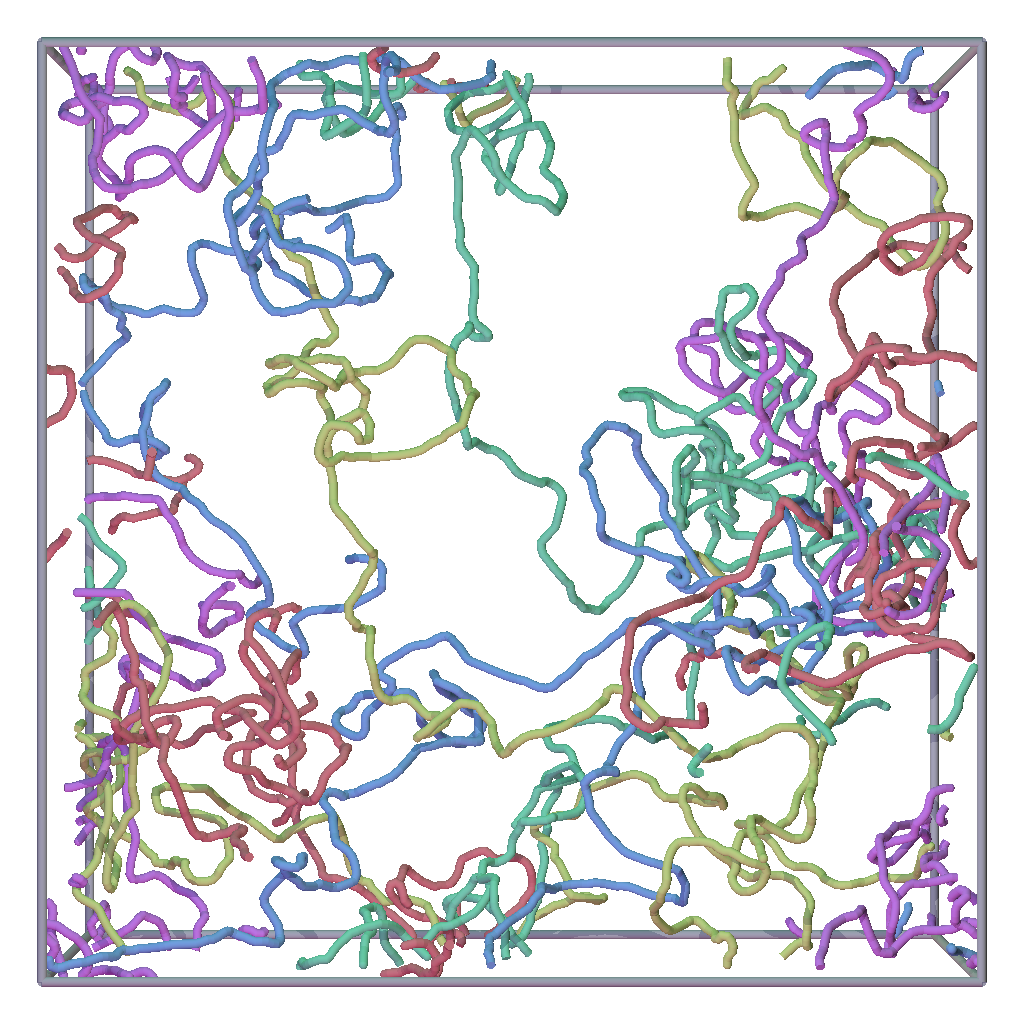}%

\caption{Visualization of five chains from a melt of $1000$ chains of $Z=200$ entanglements. Rows correspond to varying stiffness $\kappa=-2,0,2,4,6$
(top to bottom). Columns shows the lattice chain conformations before filtering, the low-pass filtered lattice conformations, the Kuhn scale
conformations with thermal fluctuations, the equilibrated KG bead-spring conformations, and finally for comparison the primitive path mesh
obtained from the equilibrated melt (left to right).\label{fig:Visualization}}
\end{figure*}

Figure~\ref{fig:Visualization} shows the evolution of $5$ randomly chosen chains (in melts composed of $1000$ chains, each with $200$ entanglements) in the course of the present multi-scale equilibration procedure. Columns illustrate the different stages of the algorithm, while the target stiffness of the KG chains increases from the top to the bottom row.

In our previous equilibration protocol\citep{SvaneborgEquilibration2016}, we used a computationally efficient lattice model~\citep{wang2009studying} to equilibrate the large scale structure of the polymer melt. The lattice constant was chosen to match the Kuhn length of the PPA~\cite{PPA}-inferred primitive paths, and multiple entanglement segments could occupy the same node in the lattice. We retain this lattice simulation, see leftmost column in Fig.~\ref{fig:Visualization}.
In a second step, we performed a lengthy push-off simulation starting from the lattice melt state using force-capped interactions (fourth column in Fig.~\ref{fig:Visualization}). The purpose of this second step was to introduce thermal fluctuations below the lattice scale while ensuring a homogeneous
spatial density of beads.

The present work introduces two major improvements, which smoothen the transitions between the lattice and bead scales:
Firstly, we use theoretical insight from a class of ``tube'' or random localization models~\citep{WarnerEdwards,read1997lozenge,everaers1998constrained,mergell2001tube} to switch from the lattice model to a chain representation on the Kuhn or bead scale. 
Constrained mode models represent chain conformations as the sum of a quenched mean conformation and annealed thermal fluctuation. 
In the original context, the mean conformation can be tentatively identified with the primitive paths, while the amplitude of the fluctuations defines the "tube" or "sausage" diameter.
When applied in a static sense, the formalism can be used to coarse-grain chains to an arbitrary length scale.
In the present context, we use it
(i) to reduce lattice artifacts in the primitive chain conformations (column 2 in Fig.~\ref{fig:Visualization}))
and (ii) to introduce the bead degrees of freedom together with random walk conformational statistics all the way down to the Kuhn scale of the chains (column 3 in Fig.~\ref{fig:Visualization})).
The resulting initial bead-spring ``melt'' conformations match the expected chain statistics already at the Kuhn scale and not, as in our original algorithm~\citep{SvaneborgEquilibration2016}, only at the tube scale.
The operation does not affect the large-scale spatial density distribution, which remains homogeneous above the tube scale. 

Secondly, we have carefully parameterized the intrinsic bending stiffness for a family of force-capped KG models such that we can tune models
with different levels of force-capping to generate chains with the same target Kuhn length. 
This iso-configurational push-off effectively avoids introducing distortions of the chain statistics during push-off, when we
build up the local chain structure and the liquid-like bead packing. The result is equilibrated starting states for
Molecular Dynamics simulations with the full KG potential  (column 4 in Fig.~\ref{fig:Visualization})).


For comparison, the rightmost column in Fig.~\ref{fig:Visualization} shows the outcome of the primitive path analysis\citep{PPA}, which we performed on the final configuration. The figure will be discussed in more detail at the end of the paper.


\subsection{Building equilibrated melt structures for higher resolution models}

In the present study we limit our efforts to building well-equilibrated melt configurations on the level of resolution of generic or Kuhn-mapped KG models. 
While these would be natural intermediate targets for building higher resolution models such as atomic polymer models, additional (and probably model specific) work would be needed to insert chemical details into the present description. For early work on equilibrating atomistic polymer models see e.g. Refs. \citep{
theodorou1985detailed,theodorou1986atomistic} and for later work utilizing multi-scale equilibration methods see e.g. Refs. \citep{kotelyanskii1996building,harmandaris2009dynamics,carbone2010fine,zhang2019hierarchical}.

\section{Background, Models and Methods \label{sec:Background}}

Below we summarize known models, theories and methods that form the background of the present work. 
In sub-sections \ref{subsec:KG-polymer-model} and \ref{subsec:Push-off-KG-model}, we introduce the KG polymer model and define the force-capped variant,
which forms the basis of the present work. Kuhn units are introduced in sub-section \ref{subsec:Kuhn-description}. Entanglements and primitive path analysis taking chain stiffness into account is introduced in sub-sections \ref{subsec:entanglements} and \ref{subsec:Primitive-path-analysis}. 
Sub-sections \ref{subsec:Lattice-melt-model} and \ref{subsec:Lattice-compatible-systems} introduce the lattice model of primitive-path chains, and how we generate bead-spring models based on lattice melt conformations.
Finally, we introduce the Rouse and the constrained mode localization models in sub-sections \ref{subsec:Rouse model} and \ref{subsec:Constrained-mode-models}.
While the present section contains all necessary information for reproducing our work, expert readers may prefer to skip it on first reading.
Note that the presentation of the background is structured in the coarse-graining sense from small to large scales, while the actual equilibration proceeds in the fine-graining direction from large to small scales.

\subsection{The Kremer-Grest model\label{subsec:KG-polymer-model}}

The KG polymer model\citep{grest1986molecular} is a bead-spring model where beads interact via 
a Lennard-Jones potential truncated at an outer cutoff radius of $r_c=2^{1/6}\sigma$ located at the minimum of the potential. 
This Weaks-Chandler-Andersen\citep{weeks1971role} (WCA) potential has the functional form

\begin{equation}
U_{WCA}(r)=\begin{cases}
4\epsilon\left[\left(\frac{\sigma}{r}\right)^{12}-\left(\frac{\sigma}{r}\right)^{6}+\frac{1}{4}\right] & r<2^{1/6}\sigma\\
0 & r\geq2^{1/6}\sigma
\end{cases}\label{eq:KG_WCA}
\end{equation}
Bonds are described by the FENE potential

\begin{equation}
U_{FENE}(r)=-\frac{kR^{2}}{2}\ln\left[1-\frac{r^{2}}{R^{2}}\right].\label{eq:KG_FENE}
\end{equation}
With the standard choices of $\epsilon=k_BT$, $R=1.5\sigma$ and $k=30\epsilon/\sigma^{2}$
the average bond length is given by $l_{b}=0.965\sigma$.

Following Faller and Müller-Plathe\citep{faller1999local,faller2000local},
we augment the model by a potential 

\begin{equation}
U(\Theta)=\kappa \,\epsilon \left(1-\cos\Theta\right),\label{eq:KG_COSINE}
\end{equation}
for the bond angle $\Theta$, which allows to control the chain stiffness.

KG simulations are performed at a standard bead density of $\rho_{b}=0.85\sigma^{-3}$ and at a standard temperature of $k_B T=1\epsilon$. 
\footnote{We have retained the standard KG notation, where $\epsilon$ denotes the energy scale of interactions and where the temperature is defined in units of $\epsilon/k_B$. In contrast, in our parameterization of the KG model for commodity polymer melts~\citep{Everaers2020Mapping} 
we have set the energy scale of all interactions to the thermal energy for the experimental target system, $\epsilon=k_B T_{exp}$, and thus rendered the model athermal.
This has the advantage that the predicted chain structures become independent of temperature in agreement with experimental observations.
For more details, we refer the reader to Ref.~\citep{Everaers2020Mapping}.}
Here and below we use subscript ``b'' to denote bead specific parameters.
We use a standard Langevin thermostat with friction $\Gamma_{b}=0.5m_{b}\tau^{-1}$
where $\tau=\sigma\sqrt{m_{b}/\epsilon}$ defines the simulation unit
of time, and $m_{b}$ denotes the bead mass. We integrate the dynamics
with a time step of $\Delta t=0.01\tau$ using the Farago/Grønbech-Jensen
integrator\citep{gronbech2013simple,gronbech2014application} implemented
in the Large Atomic Molecular Massively Parallel Simulator (LAMMPS)
code.\citep{PlimptonLAMMPS,PlimptonLAMMPS2}

\subsection{Force-capped KG models\label{subsec:Push-off-KG-model}}

Since the WCA potential, Eq.~(\ref{eq:KG_WCA}), diverges for small inter-particle distances, the full KG model can only be applied to melt conformations displaying the proper liquid-like local bead packing with inter-particle distances larger than $\sigma$ and bond lengths of the order of $l_b$. 

During push-off, we represent non-bonded interactions via a force-capped version of the WCA potential~\citep{auhl2003equilibration}:
\[
U_{WCA,cap}(r;r_{ci})=
\]

\[
\begin{cases}
U'_{WCA}(r_{ci})(r-r_{ci})+U_{WCA}(r_{ci}) & \mbox{for \quad}r<r_{ci}\\
U_{WCA}(r) & \mbox{for \quad}r_{ci}\leq r
\end{cases},
\]
As the inner cutoff distance $r_{ci}$ approaches the minimum of the LJ-Potential at $r_{c}=2^{1/6}\sigma$,
the pair potential for overlapping beads $U_{WCA,cut}(0;r_{ci})\rightarrow0$,
while for $r_{ci}\rightarrow0$ the potential $U_{WCA,cap}(0;r_{ci})\rightarrow\infty$
corresponding to the original KG model. Hence by changing the inner
cut-off we can gradually go from non-interacting beads to the full
KG model. Any choice of $r_{ci}$ corresponds to a force-cap on the
repulsive WCA interaction. We regard the inner cutoff distance as
a function of energy at overlap $U_{0}(r_{ci})=U_{WCA,cap}(0;r_{ci})$,
which we refer to below as the force cap. Inverting this relation,
we obtain $r_{ci}(U_{0})=26^{1/6}\sigma/(7+\sqrt{36+13U_{0}})^{1/6}.$

For bonded monomers the FENE+WCA potentials Eqs. (\ref{eq:KG_WCA} and \ref{eq:KG_FENE})
are replaced by the Taylor expansion
\begin{eqnarray}
U_{bond}(r) & = & 20.2026+490.628(r-l_{0})^{2}\nonumber \\
 &  & -2256.76(r-l_{0})^{3}+9685.31(r-l_{0})^{4},\label{eq:bond_poly05}
\end{eqnarray}
with $l_{0}=0.960897\sigma$ being the minimum of the potential. This
expression provides an excellent approximation for the FENE+WCA bond
potential close to the ground scale, while avoiding numerical problems
due to the divergence of the WCA potential at $r=0$ and the divergence of the FENE potential at $r=R$. 

There is no need to modify the functional form of the angular potential, Eq. \ref{eq:KG_COSINE}. 

%
%

\subsection{Kuhn model and units \label{subsec:Kuhn-description}}

Kuhn\citep{Kuhn} recognized that the large-scale conformational properties
of a polymer can be mapped to an equivalent freely jointed chain
model (FJC). Choosing the FJC step length as the Kuhn length 
\begin{equation}
l_{K}=\frac{\langle R^{2}\rangle}{L},\label{eq:kuhn_length_definition}
\end{equation}
the model reproduces the mean-square end-to-end distance $\langle R^{2}\rangle$
as well as the contour length $L$ of the target chains of long chains, $L\gg l_K$. 
The corresponding number of Kuhn segments along a chain is

\begin{equation}
N_{K}=\frac{\langle R^{2}\rangle}{l_{K}^{2}}=\frac{L^{2}}{\langle R^{2}\rangle}.\label{eq:chain_length}
\end{equation}
Qualitatively, the Kuhn length is a measure of stiffness with $c_{\infty}=l_{K}/l_{b}$ denoting the number of beads per Kuhn segment.
In particular, $l_K$ characterizes the contour length up to which chain sections are essentially straight.
Theories~\cite{DoiEdwards86} based on Gaussian chain models only apply above the Kuhn scale. 

Here and below the subscript ``K'' denotes quantities defined at the Kuhn scale. 
For example, $\rho_{K}$ denotes the number density of Kuhn segments and the Kuhn time, $\tau_{K}$,  the time it takes a Kuhn segment to diffuse over a distance of $l_K$ \citep{Svaneborg2020Characterization}.
While LJ units $(\sigma,\epsilon,\tau,m_{b},\dots)$ define
a natural set of units for reporting simulation results for the KG model,
universality suggests that choosing Kuhn units $(l_K,k_BT,\tau_K)$
facilitates the comparison between simulation results, theoretical predictions
and experimental results.\citep{Everaers2020Mapping}
In particular, Kuhn units provide direct access to the small number of dimensionless parameters, which control universal polymeric behavior:
the number of Kuhn segments, $N_K=L/l_K$ as a measure of chain length;
the Kuhn number
\begin{equation}
n_{K}\equiv\rho_{K}l_{K}^{3}=\rho_{b}l_{b}l_{K}^{2},\label{eq:kuhn_number}
\end{equation}
as a convenient measure of (contour) density;
 the Flory number, 
\begin{equation}
n_{F}=n_{K}N_{K}^{1/2}\ ,\label{eq:flory_number}
\end{equation}
as a measure of the degree of interpenetration of different polymers in a melt; and
the invariant degree of polymerization, 
\begin{equation}
N_p=n_{K}^{2}N_{K} = n_F^2\ ,\label{eq:invariant N}
\end{equation}
as dimensionless measure of chain length {\em and} interpenetration, which we denote with a subscript ``p'', because it measures chain extension in units of the diameter
\begin{equation}
p \equiv \sqrt{\langle R^{2}\rangle/N_p} =l_{K}/n_{K}\ \label{eq:p}
\end{equation}
of non-overlapping ``packing'' blobs with a reduced density of
\begin{equation}
n_p = \rho_p p^3 
= 1 \ .\label{eq:np}
\end{equation}
In melts, the contour length, $L_p = p N_p$ of the packing blob chain exceeds the physical chain contour length by a factor of $L_p/L = p N_p/(l_K N_K) = n_K$, because the formulations extrapolates the asymptotic random walk behavior to length scales below the Kuhn length.

\subsection{Entanglements \label{subsec:entanglements}}

In polymer melts, topological constraints dominate the dynamics of interpenetrating chains beyond a characteristic scale. 
Following the packing argument\citep{lin87,kavassalis87}
for flexible chains, the number of Kuhn segments between entanglements
can be well described by
\begin{equation}
N_{eK}\equiv\frac{\rho_{K}k_{B}T}{G_{e}}=\frac{\alpha^{2}}{n_{K}^{2}},\label{eq:packing_NeK}
\end{equation}
where $G_{e}=\frac{5}{4}G_{N}$ is the entanglement modulus, and $G_{N}$
denotes the melt plateau modulus.
The right-most expression follows via Eq.~(\ref{eq:invariant N}) from the packing argument\citep{lin87,kavassalis87}
\begin{equation}
N_{ep}\equiv\alpha^{2},\label{eq:packing_Nep}
\end{equation}
according to which the number of packing blobs per entanglement is a universal constant
and is only valid in the limit of flexible chains e.g. $n_{K}<10$ ($\kappa<2.5)$.
The entanglement Flory number $\alpha$ denotes the number of chains per entanglement volume. 
Experimental results~\cite{fetters2007chain}, simulation data~\cite{Svaneborg2020Characterization} and a geometric argument~\cite{rosa2014ring} for the local pairwise~\cite{everaers2012topological} entanglement of Gaussian chains suggest $\alpha = \frac{d_T}{p} = 18\pm2$.

Uchida et al.\citep{Uchida_jcp_08} developed an empirical
expression for $N_{eK}$ that also captures the behavior in the limit
of stiff chains. For $\alpha=18.0$, the number of Kuhn segments between
entanglements can be stated concisely as\citep{Svaneborg2020Characterization}

\begin{equation}
N_{eK}(\kappa)=x^{\frac{2}{5}}\left(1+x^{\frac{2}{5}}+x^{2}\right)^{\frac{4}{5}}\label{eq:Nek_Uchida}
\end{equation}
with $x=\alpha/n_{K}(\kappa)$.

Previously we performed a detailed Kuhn scale characterization of
the KG model\citep{Svaneborg2020Characterization}, in particular
we provided empirical expressions for the functions $l_{K}(\kappa),$
$n_{K}(\kappa)$, and $N_{eK}(\kappa)$ valid for stiffnesses $-1\leq\kappa\leq2.5$.
We will return to this point in the Results section below, where we
provide updated expressions for these functions valid for the range
$-2\leq\kappa\leq6$. 

\subsection{Primitive-path analysis\label{subsec:Primitive-path-analysis}}

To estimate the entanglement length $N_{eK}$ from a KG melt, we apply
Primitive Path Analysis\citep{PPA,sukumaran2005identifying}
with corrections to account for stiff chains\citep{Svaneborg2020Characterization}.
The PPA algorithm proceeds as follows: 1) the ends of all chains are
pinned in space, 2) local intra molecular pair interactions are disabled,
and 3) the energy is minimized. Minimization pulls the chains taught
to minimize the bond energy thus producing a unique primitive-path
mesh. Since we retain inter molecular pair interactions different
chains are unable to pass through each other, thus preserving entanglements.
In the original PPA algorithm all intra molecular interactions were
disabled, however, this removes self-entanglements between segments
far removed along a chain. It also removed entanglements of a chain
with images of itself due to periodic boundary conditions. Hence we
only disable intra molecular interactions up chemical distances of
$20$ bonds along the chain. This allows the chain to shrink to its
PPA contour length, while preserving distant self-entanglements. Other
PPA algorithms have also been proposed see e.g. \citep{kroger2005shortest,tzoumanekas2006topological}.
These algorithms have been shown to measure a topological entanglement
length which differs from the rheological entanglement length by 
about a factor of two.\citep{everaers2012topological}

To obtain the entanglement length in Kuhn units $N_{eK}$ from the
PPA analysis, we note the contour length of the chains before PPA
is $L=l_{K}N_{K}=l_{K}ZN_{eK}$, while after PPA the contour length
has shrunk to $L_{pp}=l_{e}Z$ where $l_{e}$ denotes the average
spatial distance between entanglements. To account for chain stiffness,
we assume the spatial distance is described by a WLC model\citep{KratkyPorod}
$l_{e}^{2}=\langle R(N_{eK};l_{K})^{2}\rangle\approx l_{K}^{2}N_{eK}(1-[1-\exp(-2N_{eK})]/[2N_{eK}])$.\citep{Svaneborg2020Characterization}.
The WLC expression reduce to the random walk limit $l_{e}^{2}=l_{K}^{2}N_{eK}$
for $N_{eK}\gg1$ corresponding to loosely entangled flexible chains,
while in the limit $N_{eK}\ll1$ it converges to the Semenov expression
for tightly entangled chains.\citep{Uchida_jcp_08} Hence the relation
between the number of Kuhn segments between entanglements and the
PPA observable $L_{pp}$ is given by

\begin{equation}
\left(\frac{L_{pp}}{L}\right)^{2}=\frac{2N_{eK}+\exp(-2N_{eK})-1}{2N_{eK}^{2}}.\label{eq:PPA_contraction_ratio}
\end{equation}

We numerically solve this equation for a given $L_{pp}$ to estimate
the number of Kuhn segments between entanglements $N_{eK}$, for loosely
entangled chains it reduces to the classical estimator of the entanglement
length.\citep{PPA} Noting that the mean-square end-to-end
vector is $\langle R^{2}\rangle=l_{K}L=a_{pp}L_{pp}$, we obtain
an expression for the Kuhn length of the primitive path mesh as

\begin{equation}
a_{pp}(\kappa)=l_{K}N_{eK}^{1/2}\left[1-\frac{1}{2N_{eK}}\left(1-e^{-2N_{eK}}\right)\right]^{-1/2}.\label{eq:app}
\end{equation}

We note that the standard random walk expression $a_{pp}^{2}=l_{K}^{2}N_{eK}$
is only valid for $N_{eK}\gg1$ corresponding to loosely entangled
chains. In the limit of stiff chains, $a_pp\sim l_K$ and $N_{eK}\sim1$, chains
are tightly entangled. 

\subsection{A lattice model for a melt of primitive chains\label{subsec:Lattice-melt-model}}

As a starting point for our fine-graining procedure in Ref. \citep{SvaneborgEquilibration2016} we chose a simple lattice model \citep{wang2009studying}, 
\begin{equation}
H=\frac{1}{2c\langle\rho\rangle}\sum_{s}\left(\rho_{s}-\langle\rho\rangle\right)^{2}\label{eq:original lattice_hamiltonian}
\end{equation}
which penalizes fluctuations in the occupation numbers $\rho_{s}$ of different sites $s$ of a simple cubic lattice to equilibrate the melt structure above the entanglement scale. 
Identifying the lattice parameter with the PPA Kuhn length is somewhat arbitrary but intuitive as we essentially simulate melts of freely jointed primitive chains. Each primitive chain segment corresponds to $N_{eK}$ Kuhn segments and in the packing limit we expect on average $\langle\rho\rangle=\alpha\approx18$ entanglement segments per site.
In particular, we expected nearly unperturbed random walk statistics from the screening of soft excluded volume interactions in a model with multiple occupied sites.

When we used simulated annealing to minimize density fluctuations, we observed that chains nevertheless had a weak tendency to stretch. 
To counter this effect we added the second term to the Hamiltonian

\begin{equation}
H=\frac{1}{2c\alpha}\sum_{s}\left(\rho_{s}-\langle\rho\rangle\right)^{2}+d(N_{FB}+N_{FF})\label{eq:lattice_hamiltonian}
\end{equation}
which provides an energy penalty for consecutive pairs of steps in the same
direction (FF - ``Forward-Forward'') or in opposite directions (FB
- ``Forward-Backwards'') and thus ensures random walk chain statistics
at large scales as the temperature is reduced. The compressibility
is denoted $c$, and we set $c=d=1$ during our simulations.

Simulated annealing is performed by reducing the temperature from
$T=100$ to $0.001$ in $20$ stages. For each stage we attempt $50$
moves per degree of freedom, that is one sweep comprises $50M(Z+1)$
attempts. As Monte Carlo moves, we use pivot\citep{madras1988pivot},
crank shaft, reptation, double bridging, and randomize moves with
equal probability. The three first are standard lattice MC moves and
are accepted with a Metropolis criterion. 
In contrast to off-lattice models, double bridging moves do not change the
energy and can always be accepted. 
The randomize move builds
a completely new random chain conformation accounting for the bias against subsequent steps along the same axis.
Hence it is accepted
with a Metropolis criterion taking into account only the energy change
due to density fluctuations.

\subsection{Choosing lattice-compatible chain and system sizes \label{subsec:Lattice-compatible-systems}}

Having identified the lattice constant of our coarse grain lattice model with the Kuhn length of the primitive paths, Eqs (\ref{eq:app}, \ref{eq:Kuhn_length_KG},
and \ref{eq:Nek_Uchida}), the linear dimension of our simulation box is necessarily an integer multiple of $a_{pp}$:
\begin{equation*}
L_{box} = n a_{pp} \ \ \ \ \mathrm{ with }\   n\in \mathbb{N}^+
\end{equation*}
A a consequence, the total volume of the simulation box has to fulfill the relation

\[
V\equiv\left(n a_{pp}\right)^{3}=\frac{N_{b,tot}}{\rho_{b}},
\]
with $N_{b,tot}=c_{\infty}(\kappa)N_{eK}(\kappa)ZM$ being the total number of beads in the melt. 
Given a target system size defined in terms of a target number $M_{target}$ of chains, we obtain from this relation the best integer approximation of the lattice size $n$. 
To avoid variations in the final bead density, we then choose the actual number of chains and beads as
\begin{eqnarray}
M &=& \mbox{Round}\left[\frac{a_{pp}^{3}n^{3}\rho_{b}}{c_{\infty}N_{eK}Z}\right]\\
N_{b}&=&\frac{\rho_{b}(n a_{pp})^{3}}{M}.
\end{eqnarray}
Here we have generated $33$ systems from $\kappa=-2$ to $6$ in steps of $0.25$, Tab. \ref{tab:systems} shows the details of the KG models
and systems for integer values of the stiffness. For all systems, we observed a variation $M/M_{target}$ in the range of $88-113\%$ and
$N_b/N_{b,target}$ in the range of $88-109\%$.\\

%
%

\subsection{The Rouse model\label{subsec:Rouse model}}

In a melt the chain statistics beyond the Kuhn scale can be described using a Gaussian chain model~\cite{rouse1953theory,DoiEdwards86}

\begin{equation}
\label{eq:Rouse model}
H=\frac{1}{2}k\sum_{\langle i,j\rangle}({\bf r}_{i}-{\bf r}{}_{j})^{2}
\end{equation}
where ${\bf r}_{i}$ denotes the instantaneous monomer positions and $\langle i,j\rangle$ denotes unique connected pairs of monomers. 
For a formulation on the Kuhn scale with $N_K$ segments, the strength of the harmonic springs connecting nearest-neighbor
beads along the chain is given by 
\begin{equation}\label{eq:k_K}
k_K =  \frac{3 N_K k_BT}{\langle R^2 \rangle} =   \frac{3 k_BT}{l_K^2} \ . 
\end{equation}

The Rouse model is easily solved in Fourier space. Defining the Rouse modes $u_{p\alpha}$ where the mode index $p=0,\dots,N_K$
and the Cartesian index $\alpha=x,y,z$. The transformation to Rouse modes is a matrix product using the standard Fourier convention, 
\begin{eqnarray}
\label{eq:Rouse transformation}
u_{p\alpha} &=& Q_{pi}r_{i\alpha}\\
Q_{pi} &=& \sqrt{\frac{2-\delta_{p0}}{N_{K}}}\cos\left(\left[i+\frac{1}{2}\right]\frac{\pi p}{N_{K}}\right)\ , \nonumber
\end{eqnarray}
 the back transformation to a real space node coordinates is given by the transposed matrix 
\begin{eqnarray}
\label{eq:inverse Rouse transformation}
r_{i\alpha} &=& Q_{ip}u_{p\alpha}\\
Q_{ip} &=& Q_{pi}\ . \nonumber
\end{eqnarray}

In contrast to the node degrees of freedom, Rouse modes are decoupled. Writing $H=\sum_{p}\sum_{\alpha}H_{p\alpha}$, the 1D mode-space Hamiltonian has the simple form
\[
H_{p\alpha}=\frac{1}{2}k_{p}u{}_{p\alpha}^{2}
\]
with the spring constant of the $p$'th Rouse mode 
%
\begin{eqnarray}
\label{eq:k_p}
k_{p} 
&=&4k_K\sin^{2}\left(\frac{\pi p}{2N_{K}}\right) \\
\label{eq:k_p approx}
&\approx& \pi^2  k_K \left(\frac{p}{N_{K}}\right)^2
\end{eqnarray} 
and mean-square mode amplitudes given by equipartition:
\begin{eqnarray}
 \langle u_{p\alpha}^{2}\rangle &=&\frac{k_{B}T}{k_{p}} \ .
\end{eqnarray}

%
%

\subsection{Constrained mode ``tube'' models\label{subsec:Constrained-mode-models}}

Variants of the Warner-Edwards tube model \citep{WarnerEdwards,heinrich1988rubber,read1997lozenge,rubinstein1997nonaffine,everaers1998constrained,mergell2001tube} are based on a real space Hamiltonian 

\[
H=\frac{1}{2}k\sum_{\langle i,j\rangle}({\bf r}_{i}-{\bf r}{}_{j})^{2}+\frac{1}{2}\ell \sum({\bf r}_{i}-{\bf R}{}_{i})^{2}\ .
\]
The additional term describes springs of strength $\ell$ connecting each segment to a ``randomly quenched''  pinning position ${\bf R}_{i}$ and causing chains to become localized in space.
The chain mean positions can then be identified with an instantaneous ``tube'' axis or the primitive chain conformation, while the amplitude of the fluctuations around the mean positions define a ``tube'' diameter. 
Transformation to Fourier/Rouse modes \citep{read1997lozenge,everaers1998constrained,mergell2001tube},
yields again a simple 1D Hamiltonian for decoupled modes,

\[
H_{p\alpha}=\frac{1}{2}k_{p}u{}_{p\alpha}^{2}+\frac{1}{2}\ell (u_{p\alpha}-v_{p\alpha})^{2}\ ,
\]
where $v_{p\alpha}$ denotes the Rouse/Fourier-transformed pinning positions ${\bf R}_{i}$.
Completing the square, the mode Hamiltonian is

\[
H_{p\alpha}=\frac{1}{2}(k_{p}+\ell )\delta u_{p\alpha}^{2}+\frac{1}{2}\frac{k_{p}(k_{p}+\ell )}{\ell }U_{p\alpha}^{2},
\]
where the mean mode excitations $U_{p\alpha}=[\ell /(k_{p}+\ell )] \ v_{p\alpha}$ and the fluctuations
$\delta u_{p\alpha}=u_{p\alpha}-U_{p\alpha}$ around the mean excitations. In particular,
\begin{eqnarray}
 \langle U_{p\alpha}^{2} \rangle &=& \frac{k_BT \ell }{k_{p}(k_{p}+\ell ) } = \gamma_p \frac{k_{B}T}{k_{p}}
 \label{eq:Up^2}\\
 \langle \delta u_{p\alpha}^{2} \rangle &=&\frac {k_BT }{k_{p}+\ell} = (1-\gamma_p) \frac{k_{B}T}{k_{p}}
 \label{eq:delta up^2}\\
 \langle u_{p\alpha}^{2}\rangle &=& \langle U_{p\alpha}^{2}\rangle+\langle\delta u_{p\alpha}^{2}\rangle = \frac{k_{B}T}{k_{p}} 
 \label{eq:up^2}
\end{eqnarray}
with
\begin{equation}
0 \le \gamma_{p}=\frac{\ell}{k_{p}+\ell} \le 1
\label{eq:gammap}
\end{equation}
so that the ensemble average of the chain statistics remains equal to the ``annealed'' statistics of  free Rouse chains.
In the weak localization limit, $\ell \rightarrow0$ and $\gamma_p \rightarrow0$, the amplitudes of the quenched mean excitations vanish, $ \langle U_{p\alpha}^{2} \rangle \rightarrow 0$, with the fluctuations $\langle \delta u_{p\alpha}^{2} \rangle \rightarrow k_BT / k_{p}$ becoming equal to those of free chains. 
In the strong localization limit, $\ell \rightarrow\infty$ and $\gamma_p \rightarrow1$, the amplitude of the remaining thermal fluctuations drops to zero, $\langle \delta u_{p\alpha}^{2} \rangle \rightarrow 0$, as the chains become rigidly quenched in mean conformations, whose statistics is equal to the statistics of free chains, $ \langle U_{p\alpha}^{2} \rangle \rightarrow k_BT / k_{p}$.

Defining the (1D) tube diameter $d_{T\alpha}^{2}$via the fluctuations
of a 1D chain as

\[
d_{T\alpha}^{2}\equiv\frac{1}{N}\sum_{p=1}^{N}\langle\delta u_{p\alpha}^{2}\rangle \ ,
\]
using Eq.~(\ref{eq:k_p approx}) and replacing the sum by an integral in the continuum limit $N_K\gg1$, one finds \citep{WarnerEdwards,read1997lozenge,mergell2001tube}
\begin{eqnarray}
d_{T\alpha}^{2} &=& \frac{1}{N}\int_{0}^{\infty}\mbox{d}p\frac{k_{B}T}{k_{p}+\ell}=\frac{k_{B}T}{2\sqrt{k\ell} \ ,}\end{eqnarray}
and hence for the 3D tube diameter
\begin{eqnarray}\label{eq:dT(l)}
d_{T}^{2} &=& 3d_{T\alpha}^{2}= \frac{3 k_{B}T}{2\sqrt{k\ell}} \ ,
\end{eqnarray}
since fluctuations along the three Cartesian directions are uncorrelated.

\section{Method development I: Fine-graining from the tube to the Kuhn scale \label{sec:CMM finegraining}}


\begin{figure}
\includegraphics[width=0.95\columnwidth]{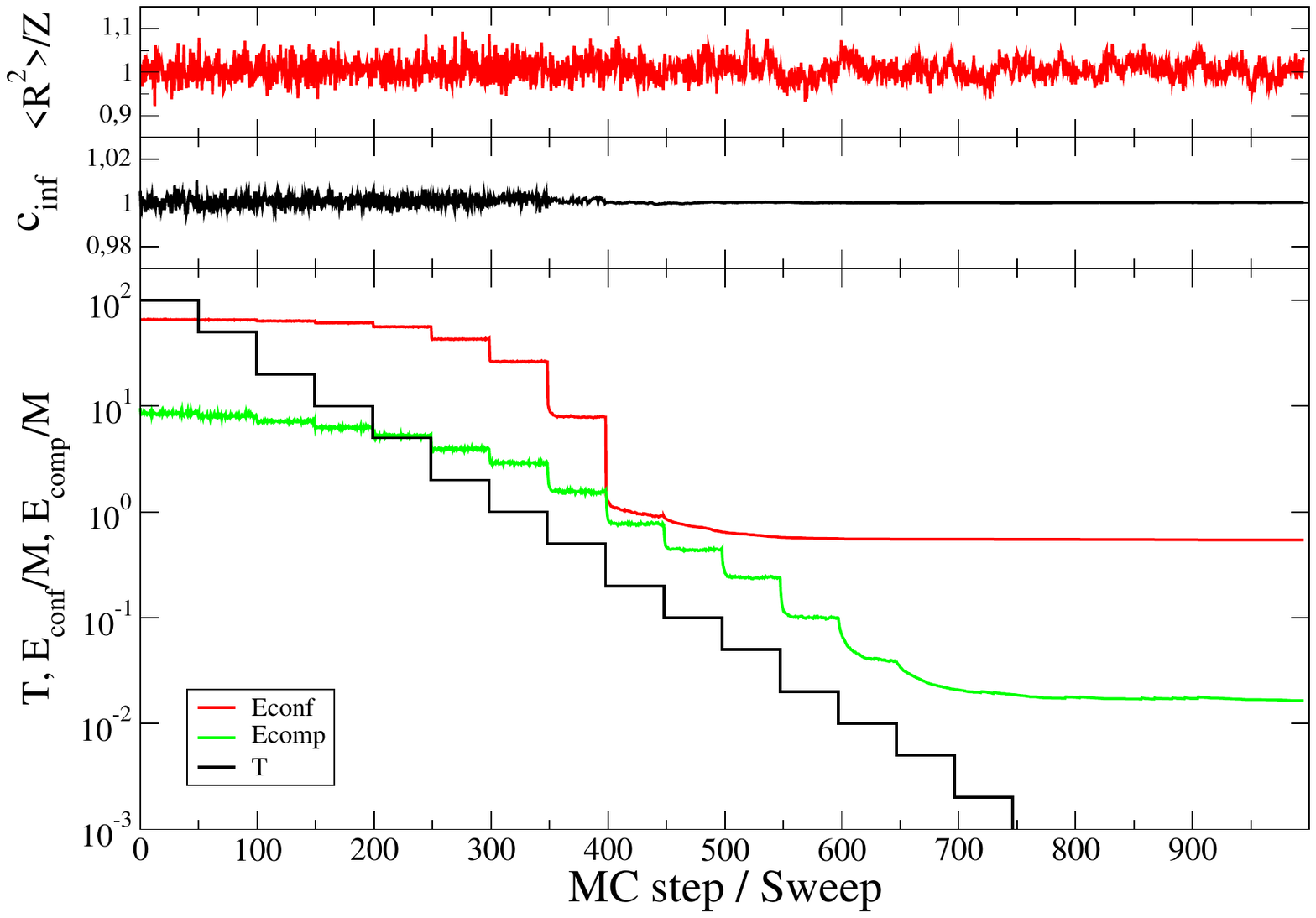}

\caption{\label{fig:Evolution-during-simulated}
Characterization of lattice melts during Monte Carlo simulated annealing.
$E_{comp}$ denotes Eq. (\ref{eq:original lattice_hamiltonian}), and $E_{conf}$ denotes the second term in Eq. (\ref{eq:lattice_hamiltonian})
}
\end{figure}


To generate the target KG melts, we start by randomly placing $M$ primitive chains of $Z$ entanglement segments on the lattice followed by simulated annealing simulations as described in Sect. \ref{subsec:Lattice-melt-model}. 
The temperature profile
and energies are shown for a melt with $Z=200$ and $M=1000$ chains
in Fig. \ref{fig:Evolution-during-simulated}. During the annealing
process both the conformation and density energy contributions drop
by nearly two orders of magnitude.

During the annealing process we monitor the large-scale chain statistics
via the mean square end-to-end vector: $\langle R^{2}\rangle=a_{pp}^{2}Z$.
The local chain statistics is characterized by the average angle between
subsequent bonds $\langle\cos\theta\rangle=(N_{FF}-N_{FB})/(Z-2)$,
represented in terms of Flory's $c_{\infty}\equiv(1+\langle\cos\theta\rangle)/(1-\langle\cos\theta\rangle)=1$
for a random walk. We observe that both observables stay at the expected random walk target values. 
The chain configurations resulting from the process are illustrated in the leftmost column in Fig.~\ref{fig:Visualization}). 

What is missing in the lattice melt conformations are the fluctuations of the microscopic chain degrees of freedom around the instantaneous primitive chain conformations.\citep{Edwards_procphyssoc_67} 
Below we show how to systematically add these fluctuations using the constrained mode formulation~\citep{everaers1998constrained,mergell2001tube}
of the Warner-Edwards ``tube'' localization model~\citep{WarnerEdwards,heinrich1988rubber,read1997lozenge}, which we have briefly summarized in Sec.~\ref{subsec:Constrained-mode-models}.
The idea is 1) to identify the instantaneous primitive chain conformations from the lattice model with quenched mean conformations from the localization model, 2) to infer the stiffness, $\ell$, of the localizing springs in Eq.~(\ref{eq:dT(l)}), and 3) to generate the missing fluctuations from  Eq.~(\ref{eq:delta up^2}) when fine-graining from the tube to the Kuhn scale.

\subsection{Statistics of the mean conformations in the constrained mode localisation model}
Due to linearity of the Rouse transformation and the additivity of the Rouse mode amplitudes $U_p$ for the primitive chain and $\delta u_p$ for the fluctuations around it, 
the real space coordinates of each chain can also be split along equivalent lines: 
\begin{eqnarray}
\label{eq:inverse Rouse transformation R}
R_{i\alpha} &=& Q_{ip}U_{p\alpha}\\
\delta r_{i\alpha} &=& Q_{ip} \delta u_{p\alpha}\\
r_{i\alpha} &=& R_{i\alpha} + \delta r_{i\alpha} \\
  &=& Q_{ip}\left( U_{p\alpha}+\delta u_{p\alpha} \right)  \nonumber\\
  &=& Q_{ip}u_{p\alpha}  \nonumber
\end{eqnarray}
By construction, the conformations of a Gaussian chain model are random walks. Assuming discretization on the Kuhn scale, the mean-square-internal
distances are given by
\begin{equation}\label{eq:msid}
\left\langle (r_{i}-r_{j})^{2}\right\rangle =l_{K}^{2} n,
\end{equation}
where $n=|i-j|$ is the chemical distance between the $i$'th and $j$'th Kuhn segment. 
To derive the conformational statistics of the primitive chain, we note that mean modes are uncoupled and given by the constrained mode model as
\[
\langle U_{p\alpha}U_{q\alpha}\rangle=\frac{k_{B}T}{k_{p}}\times\frac{\ell}{k_{p}+\ell}\delta_{pq},
\]
hence the 1D chain statistics is can be expressed using Eq. (\ref{eq:inverse Rouse transformation R}) and expressed as
\[
\left\langle (R_{i\alpha}-R_{j\alpha})^{2}\right\rangle =\int_{0}^{\infty}\mbox{d}p\langle U_{p\alpha}^{2}\rangle\left(Q_{ip}^{2}+Q_{jp}^{2}-2Q_{ip}Q_{jp}\right).
\]
After simplifications, the 3D mean chain conformations are described
by
\begin{equation}\label{eq:msid_mean}
\frac{\left\langle (R_{i}-R_{j})^{2}\right\rangle }{l_{K}^{2}n}=1-f(y),
\end{equation}
where $f(y)=\left[1-e^{-y}\right]/y$ and $y(n)=l_{K}^{2}n/(2d_{T}^{2})$ is the reduced chemical distance. 
In the limit of large chemical distances, $y_{n}\gg1$, the second term can be neglected and the chain statistics is dominated by the quenched part.
In the opposite limit, the primitive chain conformations are locally rod-like with a
step length of 
\[
\left\langle (R_{i}-R_{j})^{2}\right\rangle \approx\frac{l_{K}^{4}n^{2}}{4d_{T}^{2}}\equiv b_{pp}^{2}n^{2}.
\]

Using the definition $L_{pp}=b_{pp}N_K$ and the relation $a_{pp}L_{pp}=l_{K}^{2}N_K$,
we obtain the relation between the Kuhn length of the mean chain conformations and the tube diameter
as 
\begin{equation}
\label{eq:app vs dT CMM}
a_{pp}=2d_{T}
\end{equation}

Finally, we note that as a corollary of Eqs. (\ref{eq:msid}) and (\ref{eq:msid_mean}), the mean-square fluctuations in the distance around the primitive chain mean distance are given by
\begin{equation}\label{eq:msid_delta}
\frac{\left\langle (\delta r_{i}-\delta r{}_{j})^{2}\right\rangle }{l_{K}^{2}n}=f(y) \ .
\end{equation}

\subsection{Constrained mode localisation models for KG melts}

In the case of KG melts, the spring constant, Eq.~(\ref{eq:k_K}), of a Gaussian chain model resolved at the Kuhn scale is 
\begin{equation}
k_K(\kappa) = k_K\left(l_K(\kappa)\strut\right)=\frac{3k_{B}T}{l_{K}^{2}(\kappa)} \ .
\label{eq:springconnectivity}
\end{equation}
Combining Eqs.~(\ref{eq:dT(l)}) and (\ref{eq:app vs dT CMM}) the strength of the confinement spring given by

\begin{equation}
\ell_K(\kappa) = \ell_K\left(l_K(\kappa), a_{pp}(\kappa)\strut\right)
=12\frac{k_{B}Tl_{K}^{2}(\kappa)}{a_{pp}^{4}(\kappa)}\ .
\label{eq:springlocalization}
\end{equation}

The wave-length dependent degree of
confinement of the Rouse modes can be expressed as

\[
\gamma_{p}(\kappa)=\left(1+\frac{\pi^{2}a_{pp}^{4}(\kappa)p^{2}}{4l_{K}^{4}(\kappa)N_{K}^{2}}\right)^{-1}\approx\frac{4Z^{2}}{\pi^{2}p^{2}+4Z^{2}},
\]
where the latter expression is derived using the approximation $a_{pp}^{2}=l_{K}^{2}N_{eK}$
and hence is only valid in the flexible chain limit. The cross-over
mode number $\gamma_{p}=0.5$ depends on the number of entanglement
segments as $p\approx0.64Z$. Since we know $l_{K}(\kappa)$ and $a_{pp}(\kappa)$
for KG models, we can use Eqs. (\ref{eq:springconnectivity}) and (\ref{eq:springlocalization})
to derive a constrained mode model that exactly matches the conformational
properties on and above the Kuhn scale of a target KG model. 

%
%
%
%
%

\subsection{Analysis of the primitive chain conformations from the lattice model}

\begin{figure}
\includegraphics[width=0.99\columnwidth]{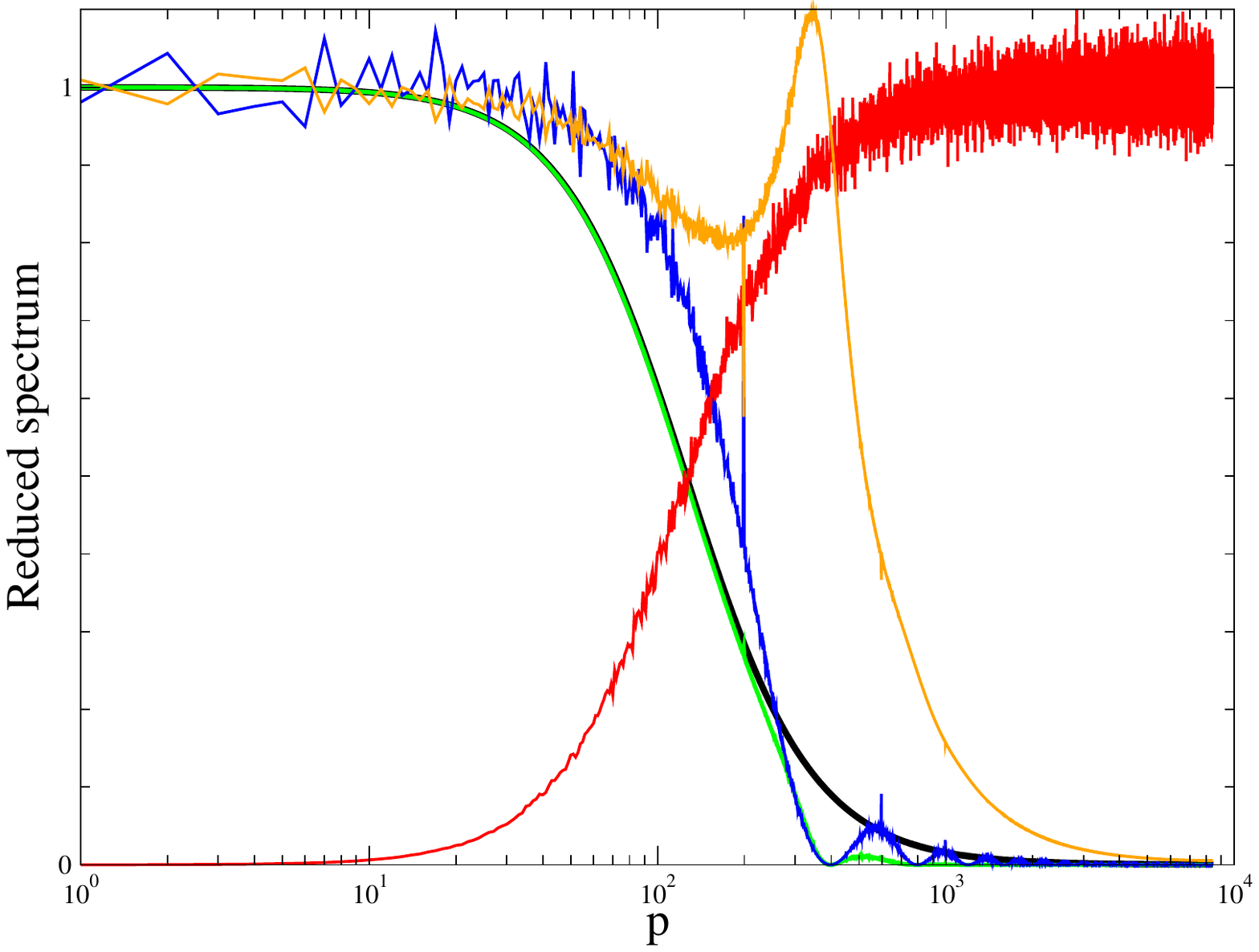}
\includegraphics[width=0.99\columnwidth]{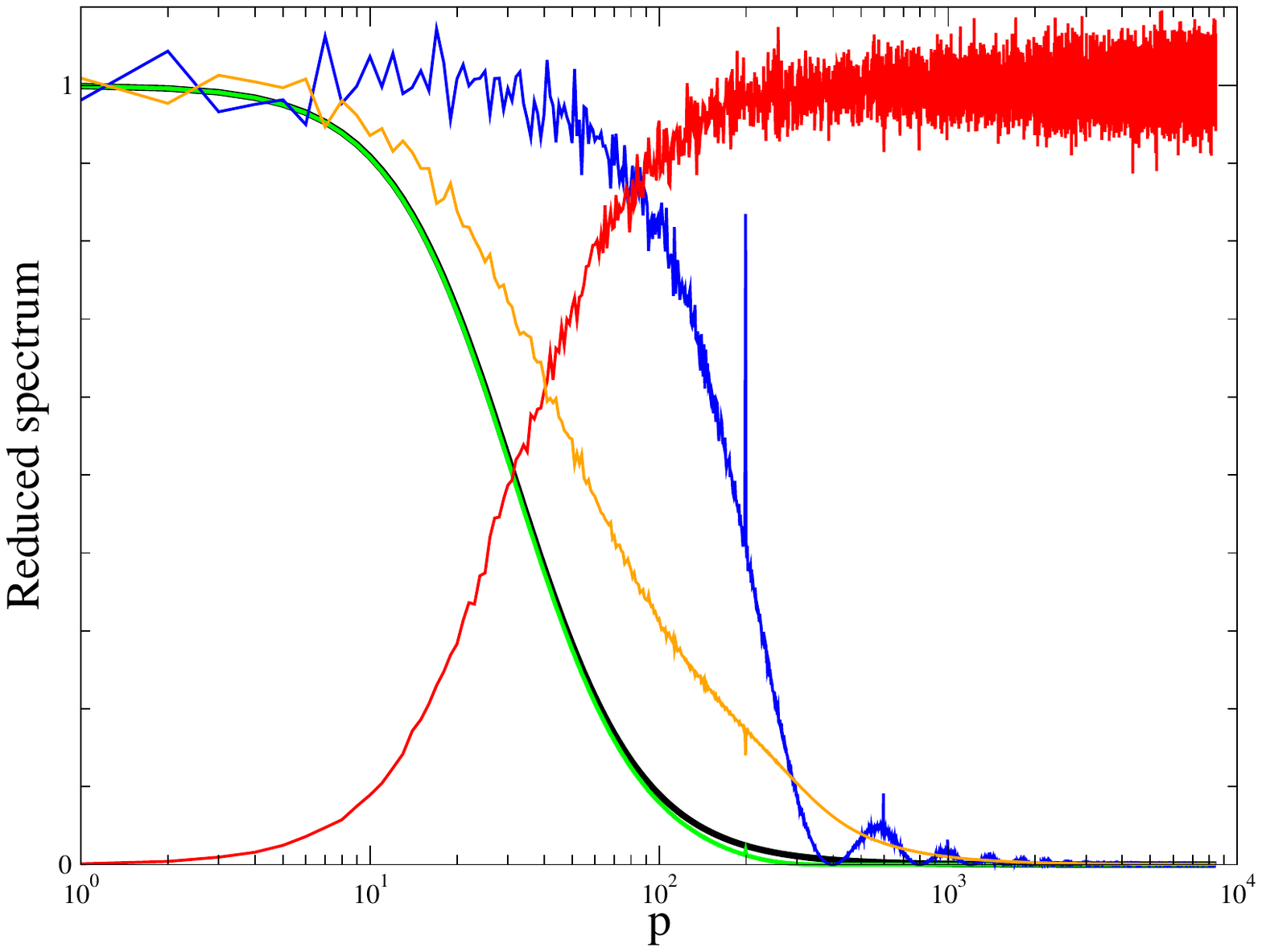}

\caption{\label{fig:Reduced-mode-spectrum}Reduced mode spectrum for the melt
with $\kappa=0$ with the strength, Eq.~(\ref{eq:springlocalization}), of the confinement springs chosen as
$\ell_K\left(l_K(\kappa), a_{pp}(\kappa)\strut\right)$ (top) and 
$\ell_K\left(l_K(\kappa), 2 a_{pp}(\kappa)\strut\right)$ (bottom). 
Lattice mean-square amplitude $K_{p}\langle\left(U_{p\alpha}^\mathrm{on}\right)^{2}\rangle$
(blue line), off-lattice mean-mean amplitude $K_{p}\langle\left(U_{p\alpha}^\mathrm{off}\right)^{2}\rangle$
(green line), target mean-square amplitude $\gamma_{P}$ (black line). Also
shown is the fluctuation spectrum $K_{p}\langle\delta u{}_{p\alpha}^{2}\rangle$
(red line) and the filter function $f_p$ Eq. (\ref{eq:rescale lattice Rouse modes}) (orange line).
The mode specific scaling constant is $K_{p}=k_{p}/k_{B}T$.
}
\end{figure}

\begin{figure}
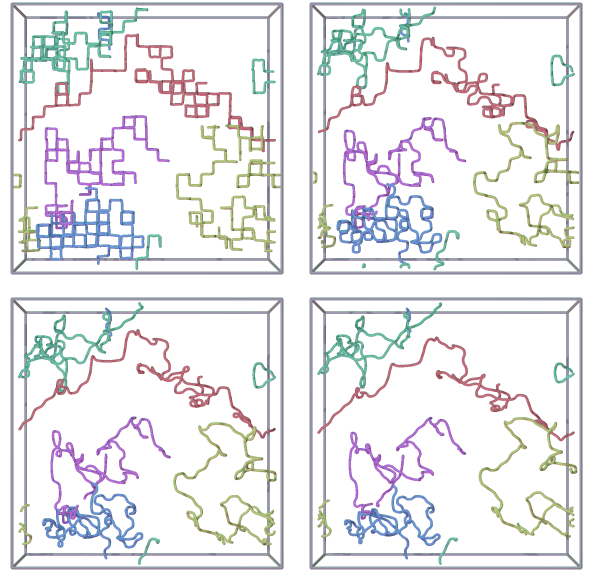

\includegraphics[width=0.45\columnwidth]{Fig4_row1_col1}
\includegraphics[width=0.45\columnwidth]{Fig4_row1_col2}

\includegraphics[width=0.45\columnwidth]{Fig4_row2_col1}
\includegraphics[width=0.45\columnwidth]{Fig4_row2_col2}

\caption{\label{fig:app_filter}Visualizations of the filtered primitive chain conformations
for $\kappa=0$  for $\ell_K\left(l_K(\kappa), \lambda a_{pp}(\kappa)\strut\right)$ with
$\lambda=1, 1.5, 2, 2.5$ (top left, top right, bottom left, bottom right).
}
\end{figure}

\begin{figure}
\includegraphics[width=0.95\columnwidth]{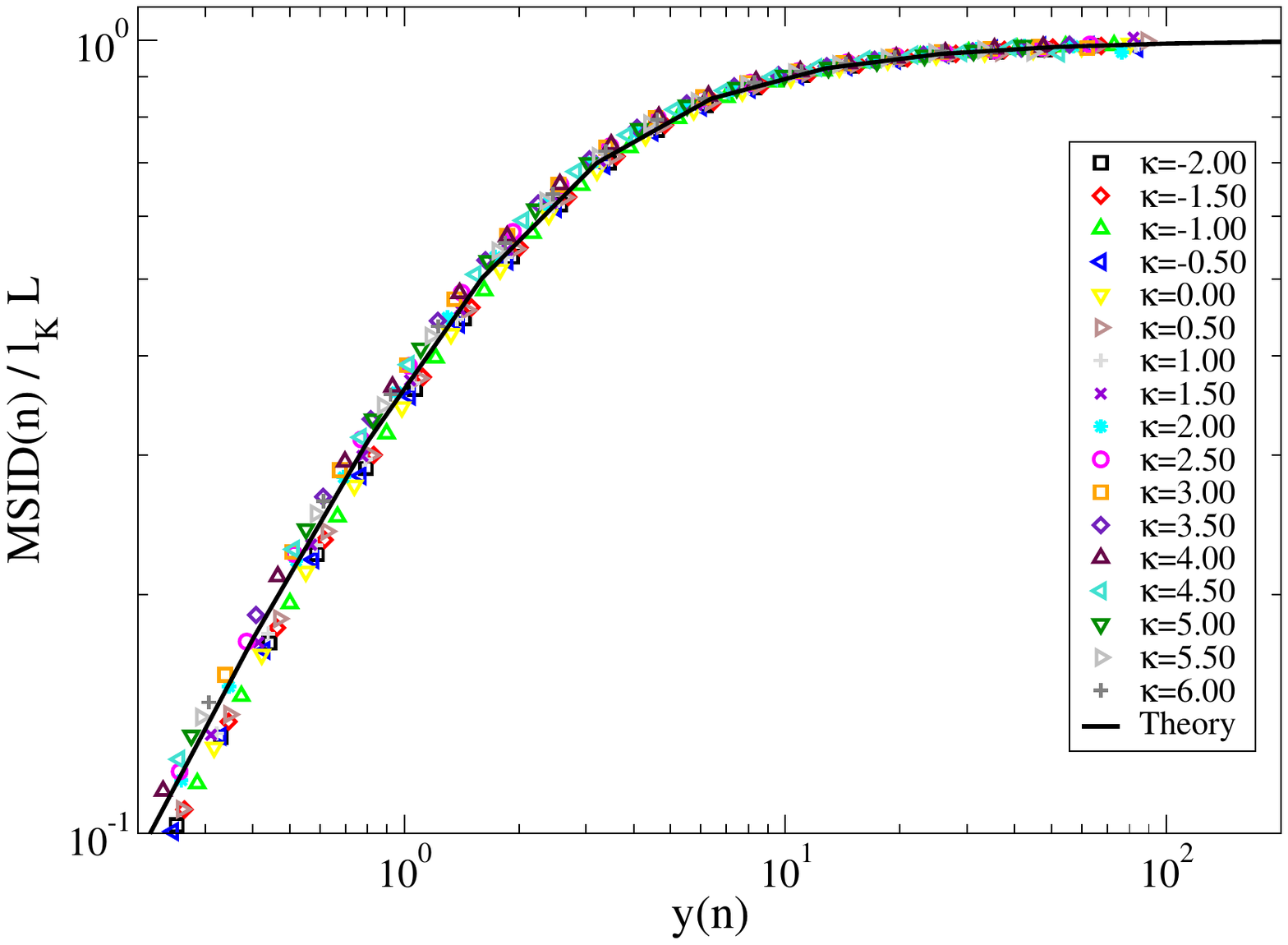}

\caption{\label{fig:MSIDmean}Mean-square internal distances, $\left\langle (R_{i}^\mathrm{off}-R_{j}^\mathrm{off})^{2}\right\rangle$, for the off-lattice primitive conformations after
rescaling, Eq.~(\ref{eq:rescale lattice Rouse modes}), for various chain stiffness (symbols) and theoretical
expectation Eq. (\ref{eq:msid_mean}) (solid black line).
}
\end{figure}

We uniformly decorate each lattice chain conformation with $N_{K}=N_{eK}Z$
Kuhn segments. Let $R_{i\alpha}^\mathrm{on}$ denote the on-lattice position of the
$i$'th Kuhn segment, and $\alpha$ the Cartesian index of a lattice
conformation, then the corresponding Rouse amplitudes of the lattice
chain conformation are given by

\[
U_{p\alpha}^\mathrm{on}=Q_{pi}R_{i\alpha}^\mathrm{on}\ .
\]
The blue symbols in Figure \ref{fig:Reduced-mode-spectrum} show the reduced variances of the lattice amplitudes $\frac{k_p}{k_BT}\langle\left(U_{p\alpha}^\mathrm{on}\right)^{2}\rangle$
(averaged over all $M$ chains and three Cartesian directions).
They are in good qualitative agreement with the theoretical expectation from the constrained mode ``tube'' model (Eq.~(\ref{eq:Up^2}) with $\ell_K(a_{pp}(\kappa))$ from Eq.~(\ref{eq:springlocalization}), black line in the top panel). In particular, the suppression of fluctuations relative to the full Rouse spectrum occurs on the expected length scale. 

To reduce lattice and decoration artifacts at short wavelengths, we filter the lattice modes by rescaling them as
\begin{eqnarray}\label{eq:rescale lattice Rouse modes}
U_{p\alpha}^\mathrm{off} = f_p U_{p\alpha}^\mathrm{on}  \\
f_p = \sqrt{\frac{\langle U_{p\alpha}^{2}\rangle}{\langle\left(U_{p\alpha}^\mathrm{on}\right)^{2}\rangle+c}}
\end{eqnarray}
with a mode cutoff $c=\langle U_{p^\ast\alpha}^{2}\rangle$ for a weakly constrained mode with $\gamma_{p^\ast}=0.01$.

When choosing the theoretical $\langle U_{p\alpha}^{2}\rangle$ in Eq.~(\ref{eq:rescale lattice Rouse modes}) using the confinement springs
$\ell_K\left(l_K(\kappa), a_{pp}(\kappa)\strut\right)$ corresponding to the Kuhn length of the primitive paths, Eq.~(\ref{eq:springlocalization}), the filtered ensemble of off-lattice primitive chain conformations is obviously in perfect agreement with the target statistics
with only minor deviations being visible for $p>100$ where fluctuations dominate the spectrum
(top panel in Fig.~\ref{fig:Reduced-mode-spectrum}). However, visual inspection of the primitive chain conformations (top left panel in Fig.~\ref{fig:app_filter}) shows that the original lattice is still clearly visible.

Choosing a larger scale (i.e. using weaker localization springs $\ell_K\left(l_K(\kappa), \lambda a_{pp}(\kappa)\strut\right)$ with $\lambda>1$) removes lattice artifacts more thoroughly (Fig.~\ref{fig:app_filter}), while preserving perfect agreement of the mode spectrum with the redefined target statistics (bottom panel in Fig.~\ref{fig:Reduced-mode-spectrum}). However this obviously comes at the expense of reintroducing density fluctuations on scales, on which they had already been suppressed in the lattice model. Here we have opted to proceed with a smoothing factor of $\lambda=2$ to minimize the effect of lattice artifacts.

We also note in passing that our choices of $a_{pp}$ as the lattice parameter and a decoration $N_K=Z N_{eK}$ are inconsistent for stiff chains. This
can be shown by calculating the mean-square distance $\langle R^2 \rangle = a_{pp}^2 Z = l_K^2 N_K /\left[1-\frac{1}{2N_{eK}}\left(1-e^{-2N_{eK}}\right)\right]$. In the limit of loosely entangled chains $N_{eK}\gg1$ the denominator is one and hence the Kuhn conformations match the
target chain statistics. However in the limit of tightly entangled chains $N_{eK}\ll1$, the denominator is $\sim N_{eK}$ and hence the decorated conformations are effectively stretched. We could fix this by choosing $l_K^2 N_{eK}$ as lattice spacing, however, we retained the choice since rescaling by Eq. (\ref{eq:rescale lattice Rouse modes}) effectively compresses the stiff chain conformations to match the target statistics.

As a final validation, Fig. \ref{fig:MSIDmean} compares the mean-square internal distances for the rescaled primitive chain conformations to the theoretical expectation, Eq. \ref{eq:msid_mean}. The agreement is observed to be excellent. In future work, it might be worthwhile to explore if we have erred on the side of caution in the sense that the removal of the lattice artifacts could be trusted to the subsequent push-off phase of the algorithm, and furthermore to explore the optimal design of the lattice model and -parameter for removal of density fluctuations at shorter scales than the tube scale.

\subsection{Introducing fluctuations at the Kuhn scale}
To obtain chain conformations with proper fluctuations all the way down to the Kuhn scale we add fluctuations $\delta u_{p\alpha}$ to the primitive chain Rouse modes,
\begin{equation}\label{eq:add CMM noise to lattice Rouse modes}
u_{p\alpha}^\mathrm{off}=U_{p\alpha}^\mathrm{off}+\delta u_{p\alpha},
\end{equation}
where the fluctuation amplitudes $\delta u_{p\alpha}$ are chosen
from a Gaussian distribution with zero mean and a variance given by Eq.~(\ref{eq:delta up^2}).
Note that the chosen strength of the localization springs, $\ell$, has to be consistent between rescaling, Eq.~(\ref{eq:rescale lattice Rouse modes}), and the addition of noise, Eq.~(\ref{eq:add CMM noise to lattice Rouse modes}). 

The center-of-mass mode ($p=0$) is treated separately: $u_{0\alpha}^{\mathrm{off}}=U_{o\alpha}^{\mathrm{on}}+\xi_{\alpha}$,
where $\xi_{\alpha}$ is an uniform random number from $[-a_{pp}/2,a_{pp}/2]$
to corresponding to a random shift of the lattice cell.

Fig. \ref{fig:Reduced-mode-spectrum} shows the reduced spectra of the fluctuations around the primitive chain,
$\frac{k_p}{k_BT}\langle\delta u_{p\alpha}^{2}\rangle$, for the above example. They are negligible at small mode numbers / large scales
($p<35$ for $\kappa=0$), where the chain fluctuations are dominated by the primitive chain component.
At large mode numbers / small scales the additional fluctuations provide the required complement to reach Gaussian statistics on all scales down to the Kuhn scale.

Finally the chain Rouse modes from Eq.~(\ref{eq:add CMM noise to lattice Rouse modes}) are converted back to real space
conformation 
\[
r_{i\alpha}^\mathrm{off}=Q_{ip}u_{p\alpha}^\mathrm{off}\ .
\]
The resulting
Kuhn scale conformations match the primitive chain conformations from the lattice model at scales
above the tube scale, while having random walk statistics all the way down to Kuhn scales.
They are thus much closer to the target statistics than the bead-decorated lattice chains from which we used to initiate the push-off in our previous equilibration protocol (Fig. 10 in Ref.~\cite{SvaneborgEquilibration2016}).

\section{Method development II: A quasi-static push-off procedure for introducing the local chain structure and bead packing \label{sec:Push-off}}
In a second step we perform a push-off with a sequence of force-capped KG force fields. 
Ref.~\cite{SvaneborgEquilibration2016} describes how establishing the local bead packing reduces the remaining density fluctuations at the tube scale.
Here we address the challenge of how to adjust the value of the intrinsic bending stiffness, Eq.~(\ref{eq:KG_COSINE}), for changing levels of force-capping such that the emerging average chain structure always remains on target.
To fully control the conformational fluctuations, we perform the push-off at constant $k_BT=1\epsilon$ and make no use of energy minimization. 


\subsection{The Kuhn length of KG models\label{subsec:Kuhn-length-KG-model}}

\begin{figure}
\includegraphics[width=0.95\columnwidth]{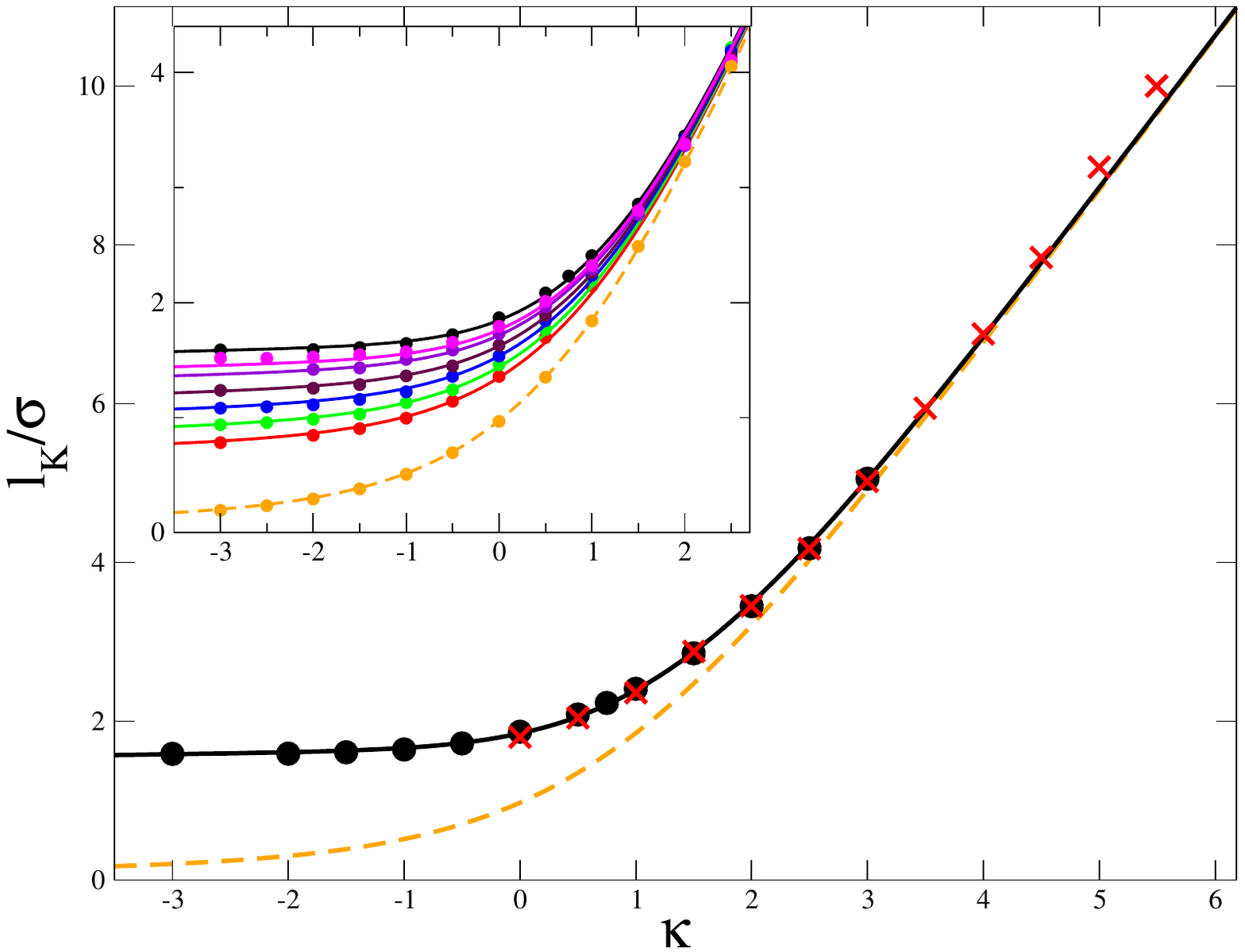}

\caption{Kuhn length vs. chain stiffness. Our simulation data (filled black
$\circ$) extended by data from Dietz and Hoy\citep{dietz2022facile}
(red $\times$), theoretical prediction for Kuhn length Eq. (\ref{eq:lk0})
(dashed orange line), and empirical expression for the Kuhn length
Eqs. (\ref{eq:lk0} and \ref{eq:msid_mean}) (solid black line). The
inset shows Kuhn lengths obtained from simulations (symbols) compared
to the empirical expression (line) for force caps $U_{0}/\epsilon=0,0.25,0.5,1,2,5,10,\infty$
(bottom to top) .\label{fig:Kuhn-length-vs.chainstiffness}}
\end{figure}

For stiff chains with $\kappa>2$ the Kuhn length of Kremer-Grest models is essentially controlled by the intrinsic bending potential, Eq.~(5) of Ref.  \citep{auhl2003equilibration}:
\begin{equation}
l_{K}^{(0)}(\kappa)=l_{b}\begin{cases}
\frac{2\kappa+e^{-2\kappa}-1}{1-e^{-2\kappa}(2\kappa+1)} & \mbox{if\quad}\kappa\neq0\\
1 & \mbox{if\quad}\kappa=0
\end{cases},\label{eq:lk0}
\end{equation}
which asymptotically reduces to the simple relation $l_{K}=l_{b}(2\kappa-1)$  (Fig. \ref{fig:Kuhn-length-vs.chainstiffness}).
For more flexible chains, there are important corrections due to the excluded volume interactions between non-bonded beads. 
Because the bending angle is limited by the repulsive WCA interaction between the next-nearest neighbor beads along the chain, $l_K$ never decreases below 
$1.58\sigma$ even when negative values of $\kappa$ would favor back-folding,

Accounting for the direct interactions between next-nearest neighbor beads along the chain \citep{auhl2003equilibration} is not enough for a quantitative description, since there are also indirect effects from the liquid-like bead packing. We have previously characterized the dependence of the Kuhn length of KG chains on the intrinsic stiffness parameter $\kappa$ via the empirical expression\citep{Svaneborg2020Characterization}:
\begin{equation}
l_{K}=l_{K}^{(0)}(\kappa)+\delta l_{K}
\label{eq:Kuhn_length_KG}
\end{equation}
with
\begin{eqnarray}
\lefteqn{\delta l_{K}(\kappa)  =  0.7276\sigma} \label{eq:lk_correction_KG} \\
 &  & \ \ \ \ \times\left(1+\tanh\left(0.002\kappa^{3}-0.469\kappa+0.214\right)\right) \nonumber
\end{eqnarray}


For validation consider Fig. \ref{fig:Kuhn-length-vs.chainstiffness}, where we show our own data from Ref.~\citep{Svaneborg2020Characterization} together with recent results from Dietz and Hoy~\citep{dietz2022facile}, who studied KG melts of constant chain length $N_{b}=400$ and stiffness varying from $\kappa=0$
up to $5.5$. We observe excellent agreement of their results
with Eqs.~(\ref{eq:lk0})--(\ref{eq:lk_correction_KG}) for all $\kappa$-values up to the isotropic-nematic transition. Faller et al.\citep{faller2000local} estimated that the transition is located at $l_K\approx \gg10\sigma$ corresponding to $\kappa\gg 4.7$. We have performed simulations of oligomer melts
suggesting the transition is located at $\kappa=5.25$.

\subsection{The Kuhn length of force-capped KG models\label{subsec:Kuhn-length-pushoff-KG-model}}

Here we have performed corresponding extensive brute force hybrid MD/MC simulations of
force-capped KG model systems with $U_{0}/\epsilon=0,0.25,1,2,5,10,\infty$
for $\kappa$ values in the range from $-3$ up to $3$ for a range
of chain lengths up to $N_{K}=80$. During the MD simulations we performed
double bridging moves to accelerate the equilibration.\citep{sides2004effect}
In contrast to the full KG model~\citep{dietz2022facile}, double bridging remains efficient in our force-capped models.
%
The computational effort invested in these simulations is about 52
core years ($0.45$M core hours).

We have characterized the resulting Kuhn length using the methods
described in Ref. \citep{Svaneborg2020Characterization}. 
They are shown in the inset of Fig. \ref{fig:Kuhn-length-vs.chainstiffness}.
Not surprisingly, force-capped excluded volume interactions have weaker but qualitatively similar effects compared to the full WCA interactions in the KG model.

For numerical ease, we have generalized Eq.~(\ref{eq:lk_correction_KG})
using yet another auxiliary force cap parameter, $W(U_{0})=U_{0}^{1/6}/[1+U_{0}^{1/6}],$
which maps the force cap range $[0:\infty[$ onto the interval $[0:1[$.
As shown by the excellent agreement between all simulation data and the correspondingly colored lines in Fig. \ref{fig:Kuhn-length-vs.chainstiffness},
$\delta l_{K}$ is well described by the empirical expression

\begin{eqnarray}
\lefteqn{\delta l_{K}(\kappa,W(U_{0}))  =  0.3638\sigma} \label{eq:lk_correction} \\
& & \ \ \ \ \times \left(1-\tanh\left(-5.481W^{3}-3.870W+2.318\right)\right)\nonumber \\
&  & \ \ \ \ \times\left(1+\tanh\left(0.002\kappa^{3}-0.469\kappa+0.214\right)\right) \nonumber \ .
\end{eqnarray}

Having an empirical expression for the Kuhn length, we also have produced
an equation for the number of beads per Kuhn length $c_{\infty}(\kappa)=[l_{K}^{(0)}(\kappa)+\delta l_{K}(\kappa,1)]/l_{b}$
of the full KG model. 

\subsection{Renormalized bending stiffness for force-capped KG models\label{subsec:renormalized kappa}}
In particular, we can numerically solve the equation 
\begin{equation}\label{eq:renormalized_kappa_for_force_capping}
l_{K}(\kappa,1)=l_{K}(\kappa',W(U_{0})) \ ,
\end{equation}
which defines the renormalized stiffness $\kappa'(\kappa,U_{0})$
required for any force-capped KG model to match the Kuhn length and large-scale chain statistics of
a full KG model with stiffness $\kappa$. This relation enables us to perform an iso-configurational
push-off.

\subsection{Starting configurations for simulations at the bead scale}

\begin{figure}
\includegraphics[width=0.95\columnwidth]{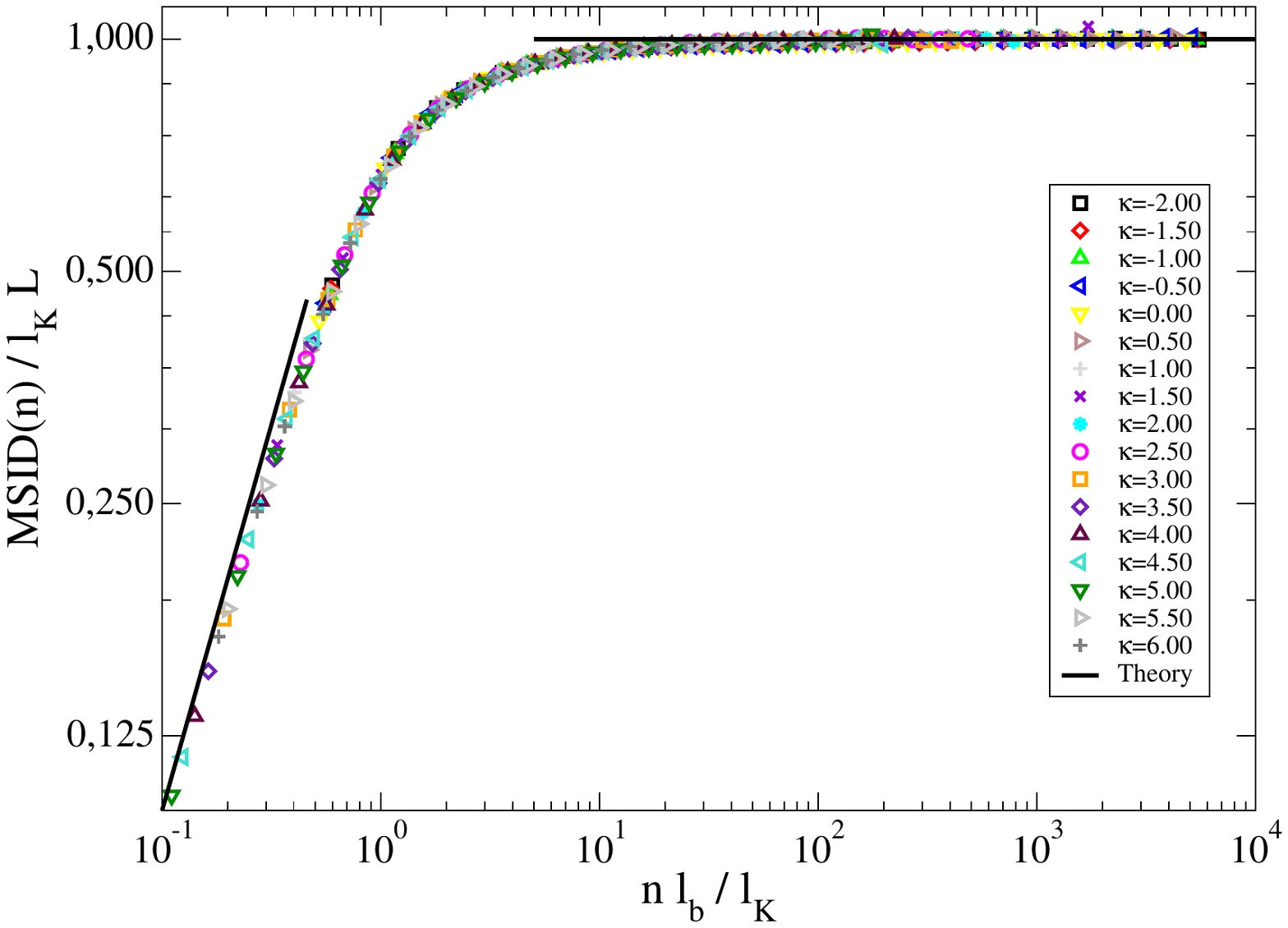}

\caption{\label{fig:MSIDKuhn}Chain statistics of bead-decorated Kuhn chain
conformations (symbols) compared to expectation (solid black lines).} 
\end{figure}

To introduce the Kremer-Grest force field into fluctuation augmented melt configurations derived from our lattice model, we place $N_{b}$ beads along the chain with a uniform bead density per Kuhn segment. 
Being derived from a Gaussian chain model discretized at the Kuhn scale, the bead-decorated  conformations do {\em not} display a realistic KG chain structure at the bead scale. By construction, the bonds lengths strongly fluctuate and there are kinks every $c_{\infty}$ beads between rod-like chain sections.
Nevertheless, the chains follow the target statistics {\em on average}.
Fig. \ref{fig:MSIDKuhn} shows 
the mean-square spatial distances, $\langle \left(\vec r_i - \vec r_j \right)^2 \rangle$, between monomers as a function of their contour distance, $l_b |i-j|$  for
the resulting bead-decorated conformations, where we have properly rescaled the data to collapse results for chains with different intrinsic stiffness. 
At large length scales
($L\gg l_{K}$) we observe random walk statistics with the desired
Kuhn lengths in agreement with our expected target statistics.
Locally ($L\ll l_{K}$) the conformations display rod-like
chain statistics due to the decoration of Kuhn segments with several beads. Comparing Fig.
\ref{fig:MSIDKuhn} and \ref{fig:MSIDmean}, we note that the cross
over from rod-like to random-like behavior occurs at $L\sim l_{K}$
for the Kuhn scale chain conformations while it occurs at $L\sim a_{pp}$
for the mean chain conformations.

\subsection{Reducing density fluctuations with strongly force-capped WCA interactions}

\begin{figure}
\includegraphics[width=0.99\columnwidth]{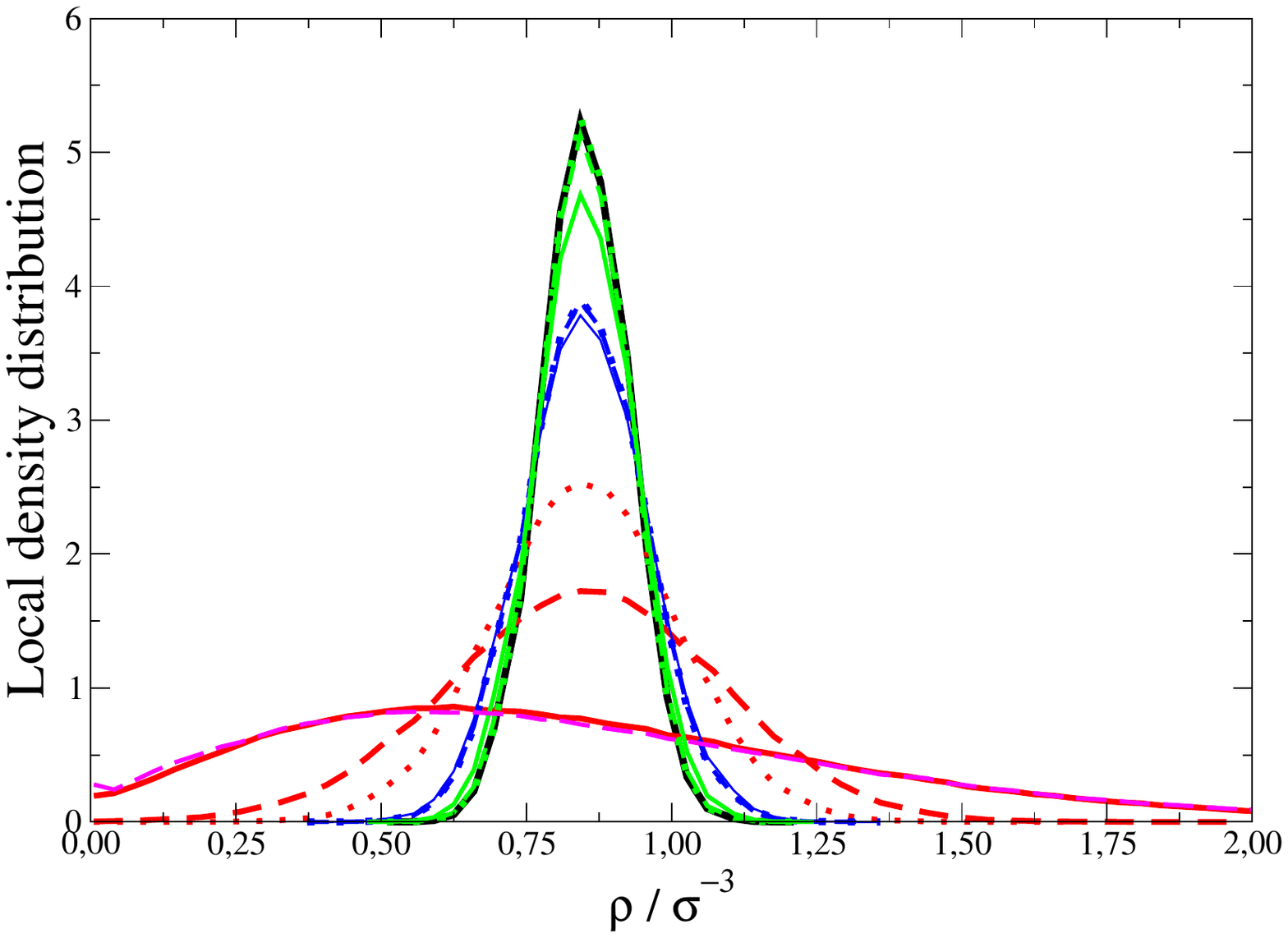}
\includegraphics[width=0.99\columnwidth]{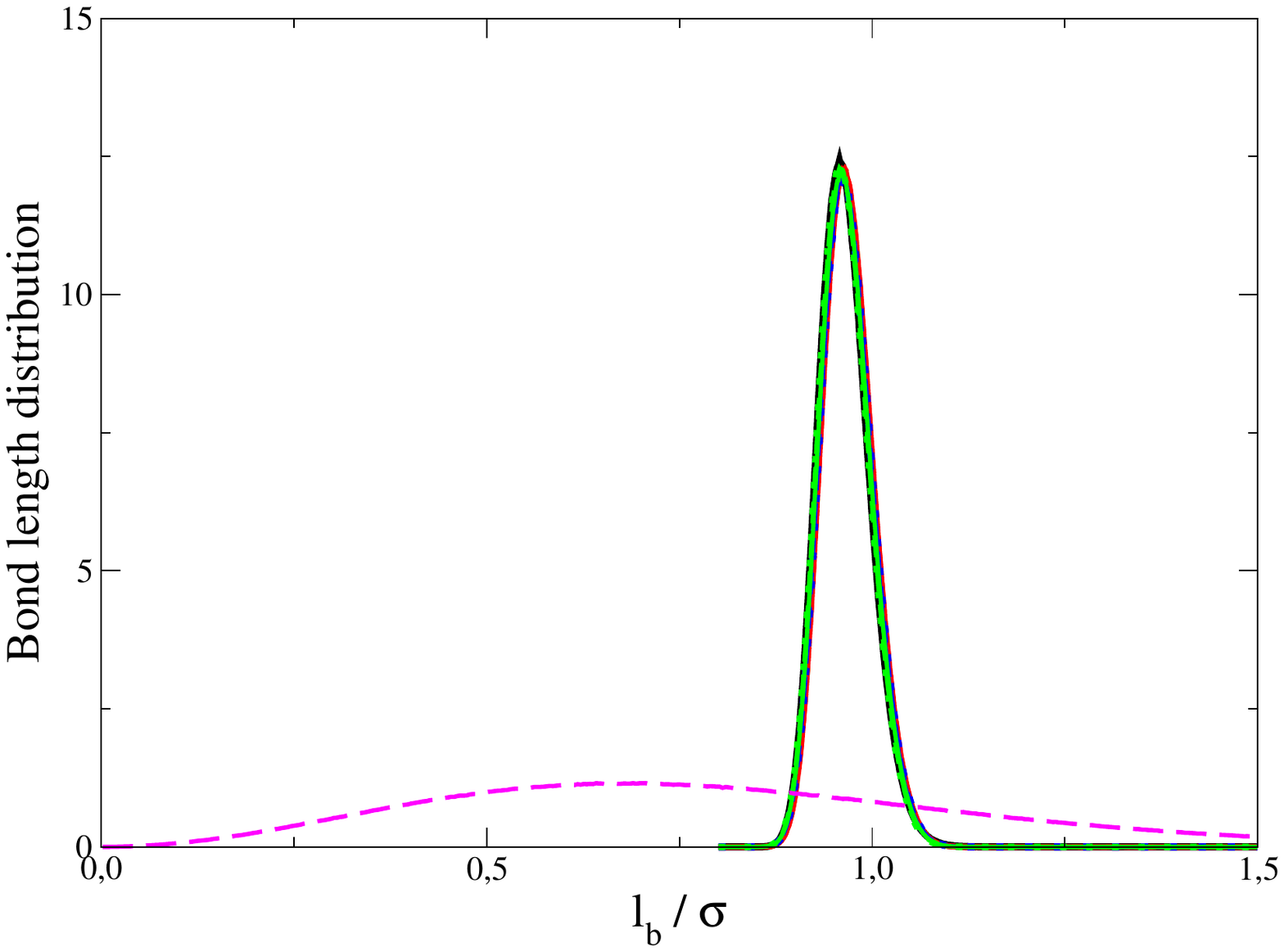}
\includegraphics[width=0.99\columnwidth]{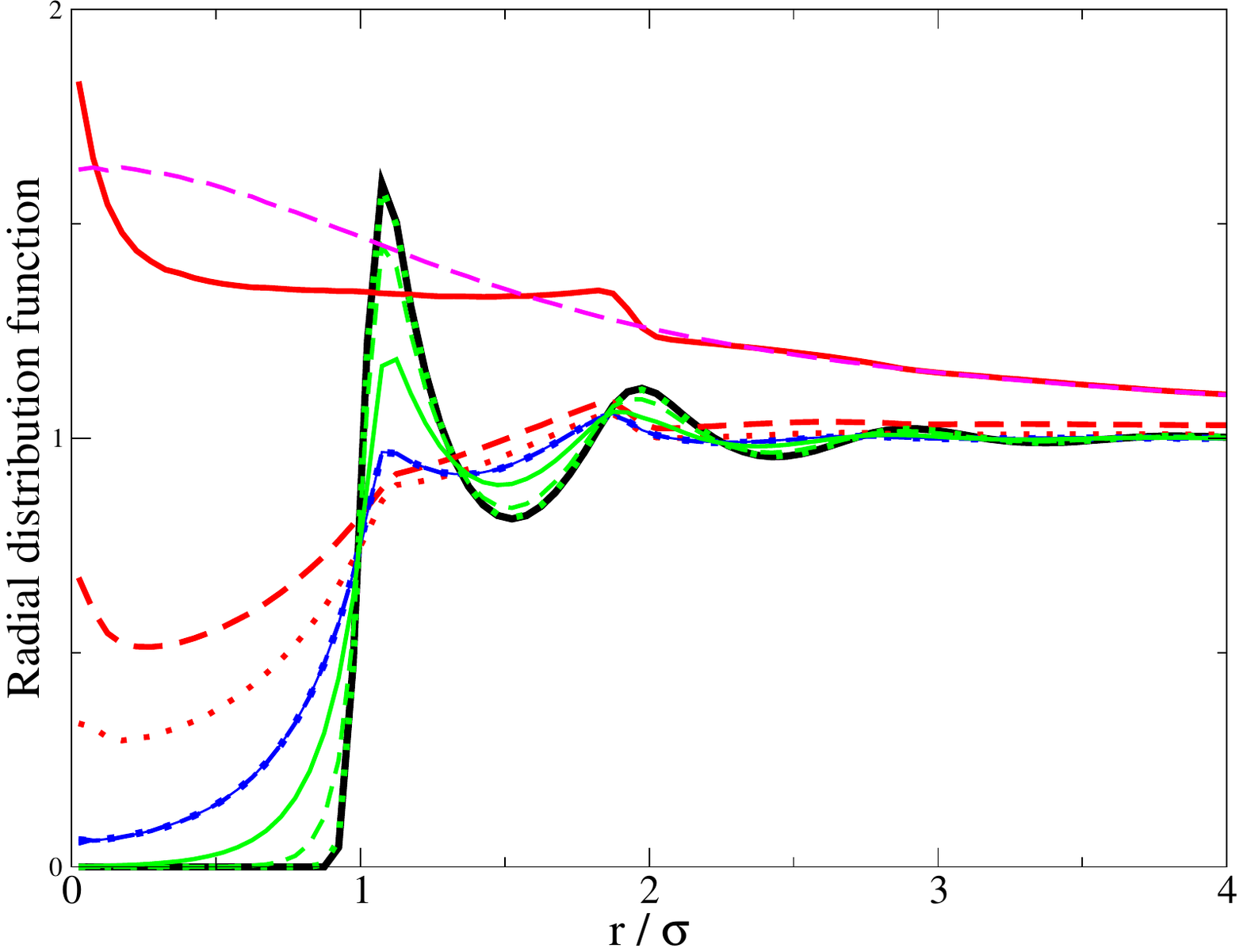}

\caption{\label{fig:rho_lb_rdf}
Structural evolution during the push-off for the $\kappa=0$ melt:
(a) Probability distributions of the density in subsystems with a volume of $\left(3\sigma\right)^3$,
(b) Probability distributions for the bond length, and
(c) Radial distribution functions for non-bonded pairs of monomers.
The initial bead-decorated Kuhn chains are shown by a dashed magenta line.
Intermediate results from the first push-off iterations with $U_0/\epsilon=0,1,2$ are shown with red solid, dashed, dotted lines, respectively.
Data from the  equilibration simulation with $U_0/\epsilon=5$ and $t'=0.02,0.25,2$ are shown with blue solid, dashed, dotted lines, respectively,
Results from the third push-off phase with $U_0/\epsilon=10,20,50$ are shown with green solid, dashed, dotted lines, respectively.
Data for the final KG equilibrium conformation are shown with a thick black line.}
\end{figure}

%

Our starting states are characterized by frequent and strong overlaps between beads. 
To remove these, we perform push-off simulations with our force-capped KG models at increasing levels of force-capping.
The push-off procedure is divided into four distinct phases of different duration.

The three panels of Figure~\ref{fig:rho_lb_rdf} monitor the evolution of the distribution of local densities and of the bond lengths during the push-off. Furthermore, we follow the emergence of liquid structure via the radial distribution functions for non-bonded pairs of beads.
Our starting states (dashed magenta lines) exhibit enormous density fluctuations and the typical correlations of Gaussian chains.

The first phase of the push-off consists of a sequence of short runs with strongly force-capped (i.e. relatively soft) WCA interactions.
Each stage is run for $10^{4}$ steps with a friction of $\Gamma=5m_{b}\tau^{-1}$, and with the appropriately readjusted 
stiffness $\kappa'(\kappa,U_{0})$ to preserve the large-scale chain statistics. 

The first push-off iteration with $U_0=0\epsilon$ (red line),  locally rearranges bead positions to establish the typical bond lengths around $l_b=0.97\sigma$ of the KG model. Note that this distribution remains essentially unchanged during the further stages of the push-off procedure and subsequent runs with the full KG force field.
In the absence of interactions between non-bonded beads there is no effect on the local density fluctuations. The pair-correlation function $g(r)$ is weakly affected via the modification of distances to second and third nearest-neighbor beads along the chains.

The subsequent push-off runs with force caps of $U_{0}=1,2,5\epsilon$ drastically reduce the local density fluctuations. For  $U_{0}=5\epsilon$ the probability of two beads to approach each other to a distance $r<\sigma$ is already significantly reduced. 
However, $g(r)$ does not yet display the typical first neighbor shell peak of liquid-like bead packing.

\subsection{Re-equilibrating the chain statistics at an intermediate level of force-capping}

\begin{figure}
\includegraphics[width=0.95\columnwidth]{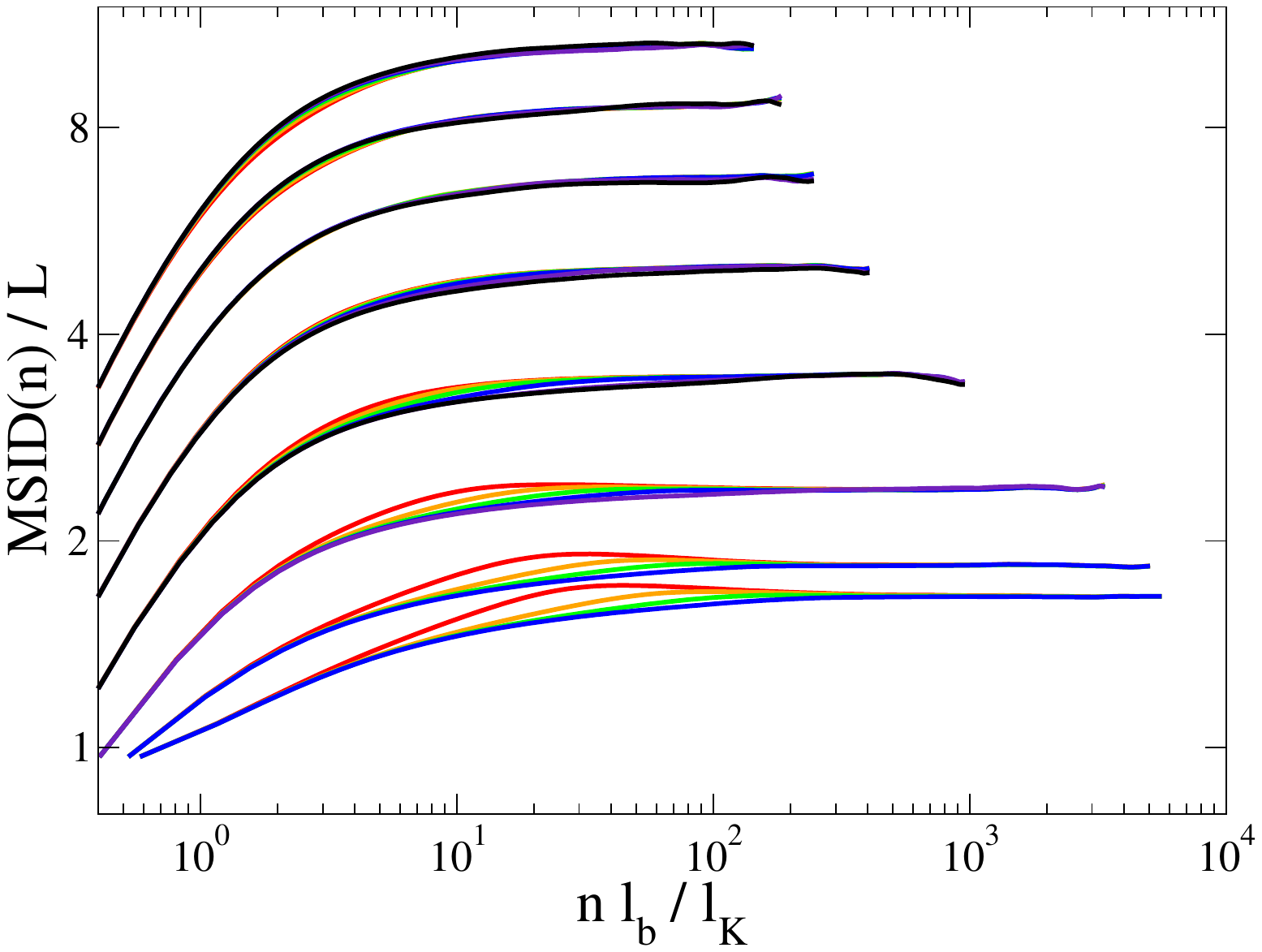}

\caption{Evolution of the chain statistics during push-off with force cap $U_{0}=5\epsilon$
as function of parametric time $t'=g_1(t)/a_{pp}^2$. colored lines denoting $t'=$$0.1$ (red), $0.25$ (orange), $0.5$
(green), $1$ (blue), 4 (indigo), $8$ (black) ) for $\kappa=-1,\dots,6$
(bottom to top).\label{fig:pushoff_MSID}}
\end{figure}

\begin{figure}
\includegraphics[width=0.95\columnwidth]{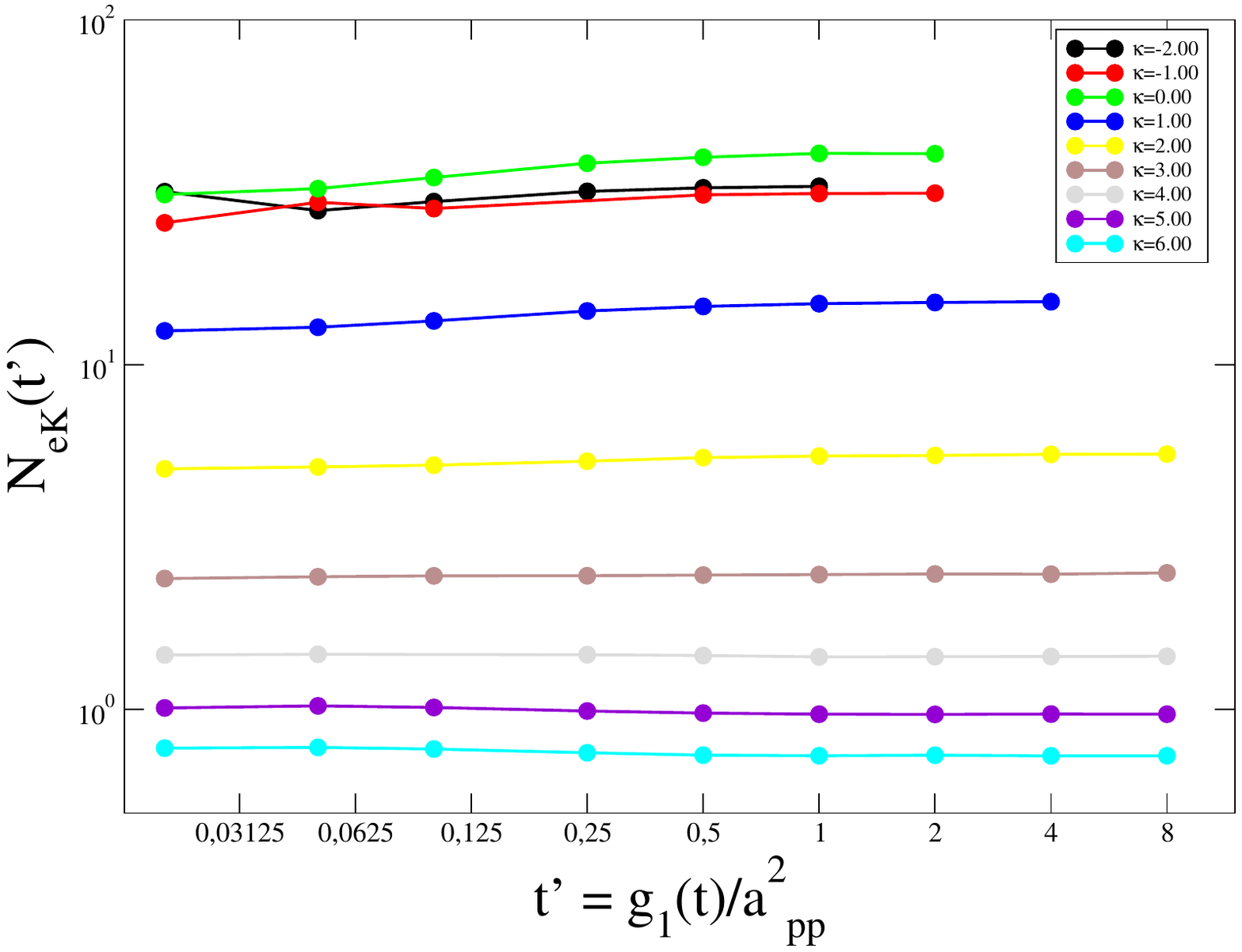}

\caption{Evolution of entanglement length $N_{eK}(t')$ during push-off simulation with $U_0=5\epsilon$.\label{fig:pushoff_NEK} 
}
\end{figure}

The red lines in Fig.~\ref{fig:pushoff_MSID} 
show the mean-square internal distances, $\langle \left(\vec r_i - \vec r_j \right)^2 \rangle$, of monomers as a function of their contour distance, $l_b |i-j|$,
after the first phase of the push-off.
The chain statistics at large scales remains unaffected and locally the chains can equilibrate to the target statistics.
However, a small amount of chain swelling develops at intermediate scales\citep{auhl2003equilibration}.

We chose to reequilibrate the chain and melt structure at $U_{0}=5\epsilon$, because
(i) local density fluctuations are sufficiently suppressed,
(ii) the intrinsic bead friction is low in the absence of caging within a well-established first neighbor shell, and
(iii) the chains can still freely pass through each other exhibiting Rouse dynamics~\cite{SvaneborgEquilibration2016}.

The equilibration simulations were run at friction $\Gamma=0.5m_{b}\tau^{-1}$.
To monitor the progress of the evolution of target properties during push-off we prefer a parametric representation of time
in terms of monomer mean-square displacements, $g_1(t)=\langle\left({\bf r}_{i}(t)-{\bf r}_{i}(0)\right)^{2}\rangle$, relative to the wavelength of the remaining density inhomogeneities on the tube scale, $a_{pp}^2$.
As shown in Fig.~\ref{fig:pushoff_MSID} the perturbations at intermediate scales equilibrate, when monomers diffuse over the entanglement scale.
In particular, this {\em local} reequilibration is sufficient to reestablish the target statistics on {\em all} length scales.

As the entanglement length is another sensitive
measure of the equilibration of polymer melts \citep{PhysRevE.72.061802}, we show in Fig. \ref{fig:pushoff_NEK}
the evolution of $N_{eK}$ during push-off.
To perform PPA in these systems we first have to
quench the topological structure by introducing the full KG force field via a quick push-off.
Compared to the corresponding Fig.~11 for our previous equilibration protocol\citep{SvaneborgEquilibration2016}, the inferred entanglement lengths now remain nearly constant throughout the push-off. We take this as further evidence for the utility of our improved fine-graining from the tube to the Kuhn scale. 
On closer inspection, the entanglement length nevertheless grows by approximately $30\%$ for standard KG melts with $\kappa=0$, while we observe essentially no evolution during push-off for the stiff chain melts with $\kappa\geq2$.
This observation is in good qualitative agreement with the larger disturbances of the chain statistics at intermediate scales reported for flexible chains in Fig. \ref{fig:pushoff_MSID} and the observations in Ref.  \citep{PhysRevE.72.061802}.
For all melts, we observe again that the equilibrium entanglement length appears
to be established, when monomers have diffused over the entanglement scale.

We conclude that push-off simulations reaching $g_1(t) = a_{pp}^2$ are sufficient to establish equilibrium chain statistics and entanglement length, when the systems were previously equilibrated down to the entanglement scale using the lattice model. 
In terms of MD steps, this corresponds to relatively modest simulations of $2.4,1.1,0.16,0.24,0.72\times10^{6}$
MD steps, respectively, for $\kappa=-2,0,2,4,6$. The non-monotonous
behavior is surprising, since $a_{pp}$ depends only weakly on
the chain stiffness. We attribute this to the effective friction that
depends on the bead interactions and local melt structure. These data
suggests that the friction is minimal around $\kappa\sim2$ for the
force capped model, while previously it was observed that dynamics for the
full KG model accelerates with increasing chain stiffness.\citep{faller2001chain,Svaneborg2020Characterization}
At large stiffnesses a transition from isotropic to nematic order is expected,
and this might explain why the friction increases for the stiffest melts.

\subsection{Establishing the local bead packing with weakly force-capped WCA interactions}

In the third step of the push-off protocol we rapidly raise the force cap in five additional stages as $U_{0}=10,$
$20,$ $50,$ $75,$ $100\epsilon$ during a sequence of runs of $10^{4}$
MD steps each and at a friction of $\Gamma=5m_{b}\tau^{-1}$. Again
we use a force-cap readjusted chain stiffness $\kappa'(\kappa,U_{0})$
that keeps the Kuhn length constant during the push-off process. 
As shown in Fig.~\ref{fig:rho_lb_rdf} this further reduces the local density fluctuations,
while generating the typical liquid-like local bead packing of KG models
In particular, the minimal distance between any bead pair increases from zero up to approximately
$0.8\sigma$ for the largest force cap.

\subsection{Switching to the full KG model}

During the final phase of the push-off we switch to the full KG force field.

In a first step we replace the $U_{0}=100\epsilon$ force-capped
interaction with the full WCA interactions, but retain the polynomial
bonding potential Eq. (\ref{eq:bond_poly05}). We perform a simulation
of $10^{4}$ MD steps with maximal displacement spatial displacement
of $0.05$ during a time step. Within $100$ steps we observe that
the minimal distance between any bead pair increases to approximately
$0.86\sigma$ as expected for the KG force field. 
In a second step,  we replace
the polynomial bond potential with the FENE bond potential, and perform
an additional $5\times10^{4}$ MD steps. The heat produced by changing
the force-cap is removed by the Langevin thermostat. The maximal temperature
increase observed during the entire push-off process is less than
$10\%$.


\begin{table*}
\caption{\label{tab:systems}
The Kuhn number $n_K$ was calculated using Eq. (\ref{eq:kuhn_number}),
the Kuhn length $l_K$ was estimated using Eqs. (\ref{eq:lk0},\ref{eq:lk_correction_KG}),
the Kuhn length of the primitive paths $a_{pp}$ was calculated using Eq. (\ref{eq:app}),
and the entanglement length $N_{eK}$ was estimated using Eq. (\ref{eq:Nek_Uchida}).
The Kuhn and entanglement times are calculated using Eqs. (25,33,63) of Ref. \cite{Everaers2020Mapping}. 
Both time scales depends on an empirical relation for the bead friction valid within
$-1\leq \kappa \leq 2$ by $^*$ we denote extrapolated values. The radius of gyration is
given by $R_g^2 = l_K^2 N_K/6$
}
\begin{ruledtabular}
\begin{tabular}{|cc|cccc|cc|cccccc|}
$\kappa$  & $n_K$   &
$l_K/\sigma$ & $c_\infty$  &  $a_{pp}/\sigma$  &   $N_{eK}$        &
$\tau_K/\tau$  &  $\tau_e/\tau$  &
$M/M_{target}$  & $Z/Z_{target}$ & $N_b/10^{3}$ & $M\times N_{b}/10^{6}$ & $L_{box}/R_g$ & $n$  \\
\hline 
$-2$ & $2.13$ & $1.61$  & $1.67$  & $13.93$ & $72.6$  & $4.95^*$  & $27400^*$ & $0.98$ & $1.06$  & $25.61$ & $25.22$ & $3.80$  & 22 \\
$-1$ & $2.27$ & $1.66$  & $1.72$  & $13.52$ & $62.0$  & $5.36$    & $23000$   & $0.99$ & $1.09$  & $23.24$ & $22.96$ & $3.80$  & 22 \\
$0$  & $2.80$ & $1.85$  & $1.91$  & $12.31$ & $37.8$  & $6.94$    & $13400$   & $1.00$ & $1.16$  & $16.73$ & $16.80$ & $3.84$  & 22 \\
$1$  & $4.69$ & $2.39$  & $2.48$  & $9.97$  & $15.0$  & $13.6$    & $3870$    & $0.93$ & $1.11$  & $8.30$ & $7.76$ & $3.70$    & 21 \\
$2$  & $9.99$ & $3.49$  & $3.62$  & $8.19$  & $5.42$  & $38.9$    & $954$     & $1.04$ & $0.93$  & $3.63$ & $3.76$ & $3.64$    & 20 \\
$3$  & $20.9$ & $5.05$  & $5.23$  & $8.33$  & $2.46$  & $113^*$   & $484^*$   & $1.09$ & $0.85$  & $2.19$ & $2.39$ & $3.35$    & 17 \\
$4$  & $38.4$ & $6.84$  & $7.09$  & $9.65$  & $1.44$  & $275^*$   & $439^*$   & $0.92$ & $0.88$  & $1.80$ & $1.66$ & $2.81$    & 13 \\
$5$  & $62.4$ & $8.72$  & $9.04$  & $11.37$ & $0.98$  & $565^*$   & $484^*$   & $0.99$ & $0.95$  & $1.68$ & $1.65$ & $2.57$    & 11 \\
$6$  & $92.8$ & $10.63$ & $11.02$ & $13.23$ & $0.75$  & $1020^*$  & $557^*$   & $0.88$ & $0.99$  & $1.63$ & $1.43$ & $2.25$    & 9  \\
\end{tabular}
\end{ruledtabular}
\end{table*}

\section{Results \label{sec:results-and-discussion}}

Below we analyze the conformational properties of highly entangled KG bead-spring melts, which we have generated with our revised multiscale equilibration procedure composed of three steps:
1) a Monte Carlo equilibration of a simple and computationally efficient lattice model for the melt structure above the entanglement scale (Sec.~\ref{subsec:Lattice-melt-model}),
2) the application of a new random localization theory-based procedure for reducing lattice artifacts and adding fluctuations all the way down to the Kuhn scale (Sec.~\ref{sec:CMM finegraining}), and
3) the execution of a series of Molecular Dynamics simulations for bead-spring chains with force-capped WCA interactions and carefully parameterized intrinsic bending rigidities to slowly introduce the liquid-like bead packing characteristic of Kremer-Grest melts (Sec.~\ref{sec:Push-off}).

In Fig.~\ref{fig:Visualization} we document the evolution of the conformation of randomly chosen chains during this procedure.
Each row of the figure corresponds to a different value of the bending parameter $\kappa$. The chosen values cover the entire range from experimentally relevant bending rigidities\citep{Everaers2020Mapping} up to and beyond the limit of the isotropic-nematic transition\citep{faller2000local}.
We generated melts of approximately $M_{target}=1000$ chains with effective chain lengths of $Z_{target}=200$ entanglements.
Since  the number of beads per entanglement segment varying from $149$ for the most flexible melts ($\kappa=-2)$ down to $10$ for the stiffest melts ($\kappa=6$),
this correspond to $N_b=25000$-$2000$ beads per chain and to total system sizes of $24.4$M
down to $1.4$M beads. 
The precise characteristics of the generated melts are listed in Table~\ref{tab:systems}, the choices for the number of chains in the systems and the exact number of beads per chain follow from the considerations in Sec.~\ref{subsec:Lattice-compatible-systems}.

The left-most column of Fig.~\ref{fig:Visualization} shows the lattice melts conformations
where each chain is shifted by a random cell vector. We observe interpenetrating
random walk chains as expected for a melt. 
The density of beads is kept constant, which effectively
results in lattices with a smaller number of nodes for increasing chain
stiffness. The second column shows the mean conformations produced
after mode rescaling of the lattice conformations. Artifacts due
to the lattice structure are completely removed. The
third column shows bead-spring chains after
adding thermal fluctuations all the way down to the Kuhn scale. 
The fluctuations strongly contribute to the conformations
for the flexible melts (top), while for the stiff melts the Kuhn scale
chain conformations are similar to the mean conformations. The fourth
column shows the KG melt conformations after the push-off, which only affects the local packing. 

In the following, we compare measures of the chain statistics in the generated melts to target values from the literature,
check their invariance during longer simulations with the full KG force field.

\subsection{Mean-square internal distances\label{subsec:MSID}}

\begin{figure}
\includegraphics[width=0.95\columnwidth]{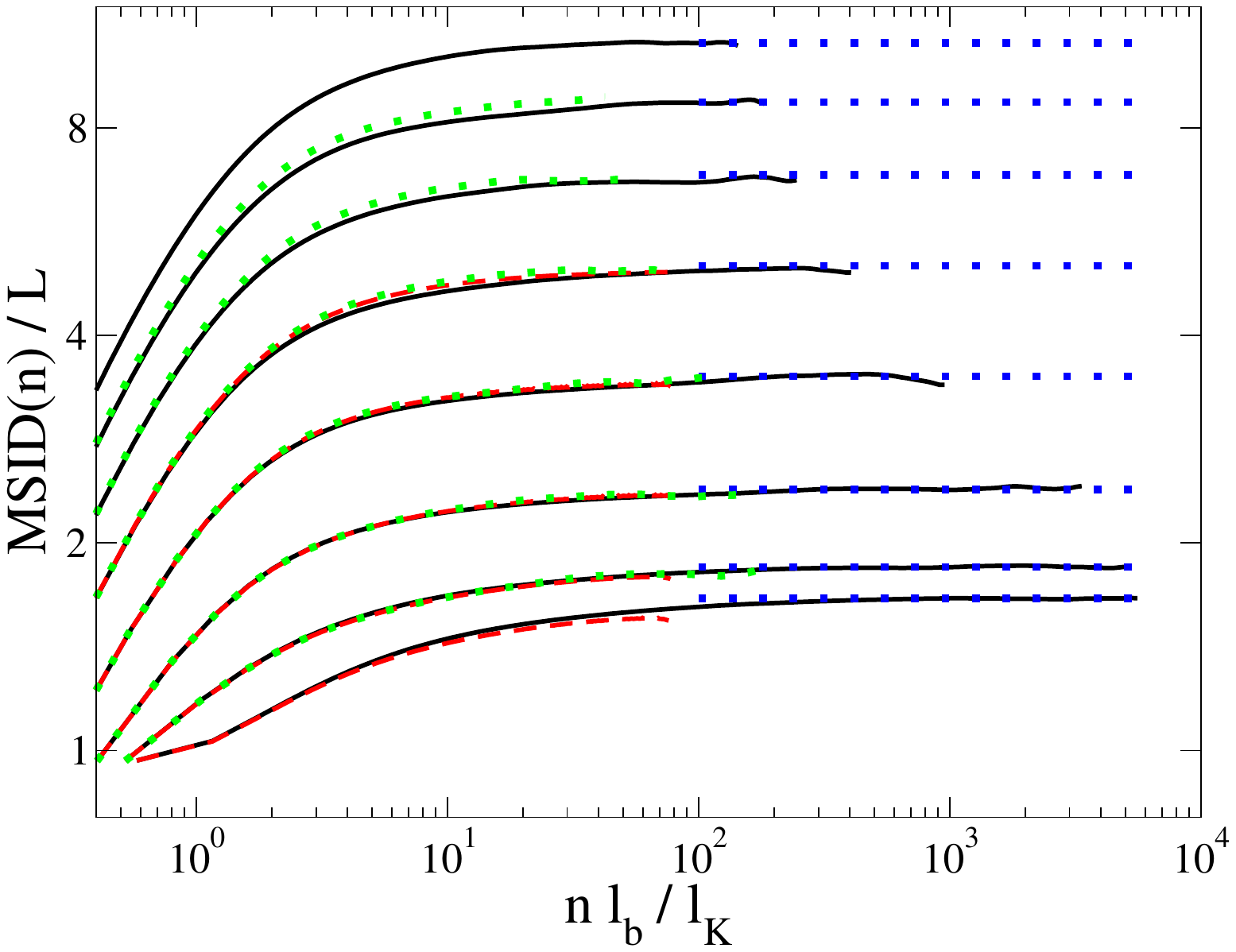}

\caption{Comparison of chain statistics of the equilibrated melts (black line)
with the target asymptotic Kuhn lengths (blue dotted line), with conformations
equilibrated with the algorithm in Ref. \citep{SvaneborgEquilibration2016}
(red dashed line), and with brute force equilibrated conformations
generated by Dietz and Hoy\citep{dietz2022facile} (green dotted line)
for stiffnesses $\kappa=-1,\dots6$. (bottom to top) \label{fig:KGMSID_comparison} }
\end{figure}

\begin{figure}
\includegraphics[width=0.95\columnwidth]{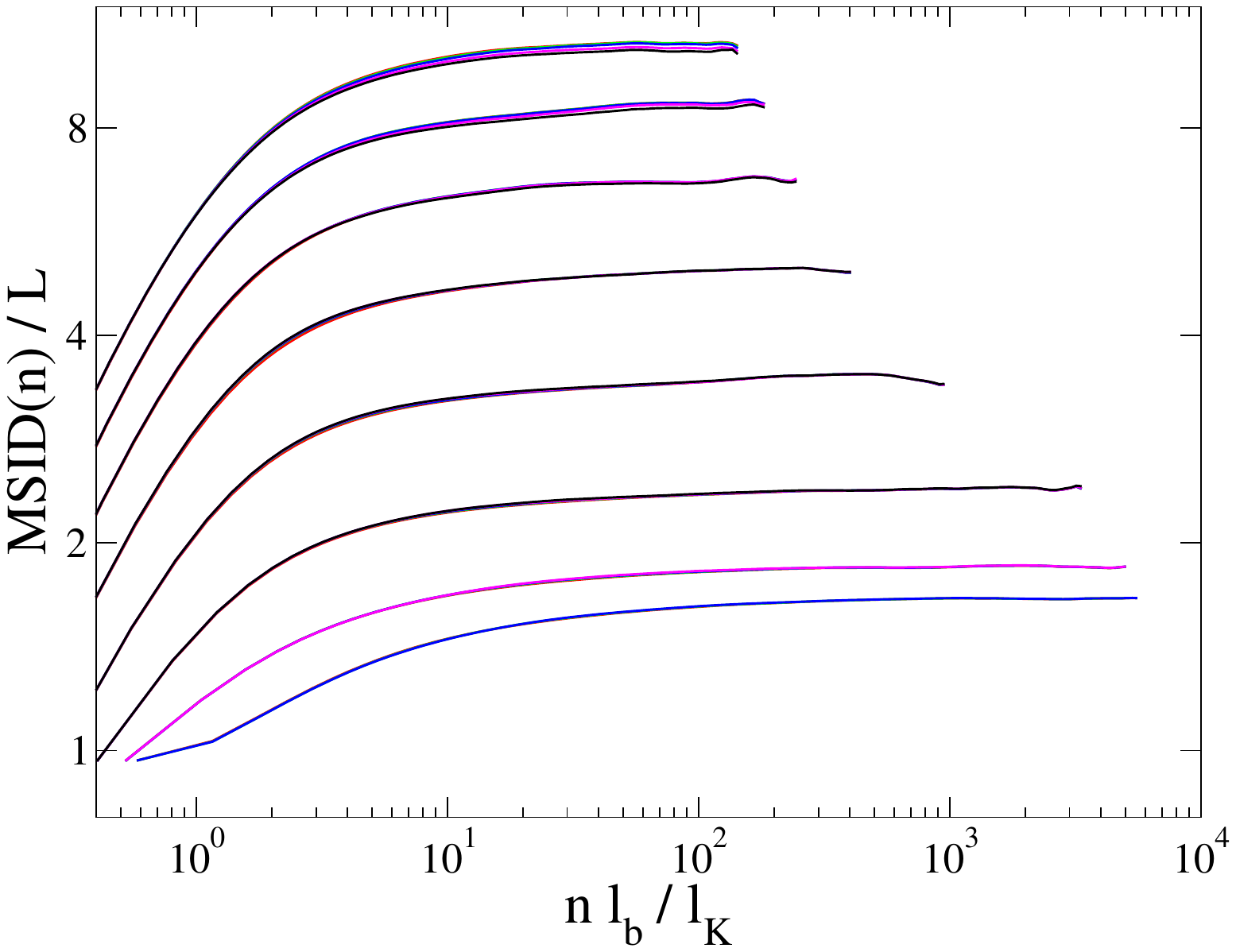}

\caption{Evolution of mean-square internal distances during an equilibrium
simulation (colored lines denoting $t'=0.05$ (red), $0.1$ (green),
$0.2$ (blue), $0.5$ (magenta), and $1$ (black) for $\kappa=-1,\dots,6$.
\label{fig:evolution_eq_msid}
}
\end{figure}

\begin{figure}
\includegraphics[width=0.95\columnwidth]{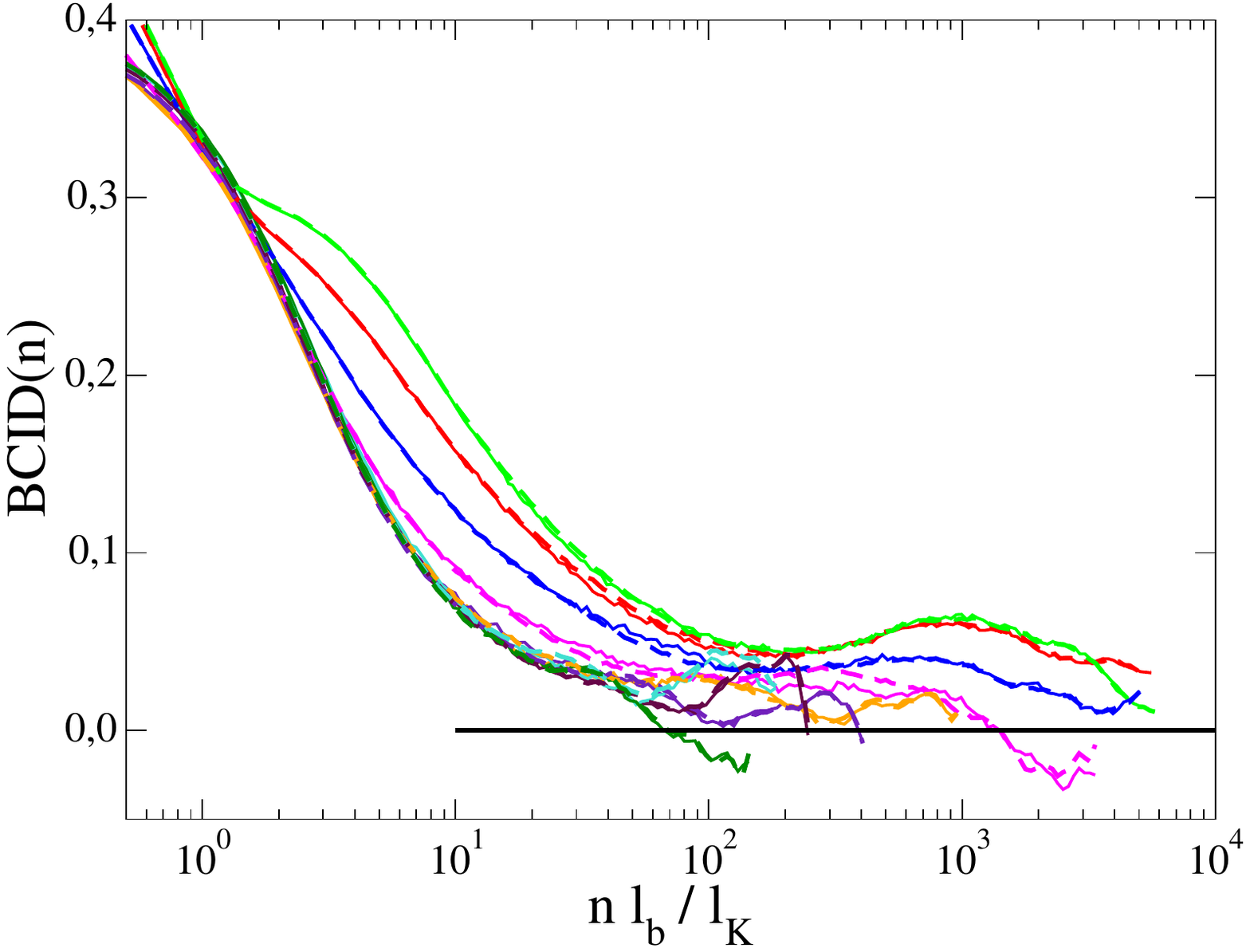}

\caption{Evolution of binder cumulants for the equilibrated melts with $\kappa=-2,-1,0,1,2,3,4,5,6$
(solid lines colored green, red, blue, magenta, orange, indigo, maroon,
turquoise, dark green, respectively) compared to columants sampled
at the end of the equilibrium simulation shown in Fig. \ref{fig:evolution_eq_msid}
(dashed lines with matching colors). \label{fig:BinderCumulant} 
}
\end{figure}

Figure \ref{fig:KGMSID_comparison} shows the mean-square internal
distances of the chains in our generated melts compared to literature results. 
This is an ideal way to check if we reproduce the target statistics at all length scales. 
We observe excellent agreement both with our own previous melt conformations
as well as those generated by Dietz and Hoy. At large length scales
the melt statistics are also in excellent agreement with
the target Kuhn lengths\citep{Svaneborg2020Characterization}.

To check the equilibration of the present melts we performed
equilibrium simulations up to the entanglement time to look for signs of changes
in the melt structure. Fig. \ref{fig:evolution_eq_msid} shows the
evolution of the mean-square internal distances. 
%
%
We observe essentially no change, except for $\kappa=5$ and $6$ where there is a small drop in the chain statistics in the asymptotic plateau,
which perhaps is due to the onset of local nematic ordering. 

\subsection{Beyond second moments \label{subsec:Beyond second moments}}

The equilibration procedure as well as most of our quality control were based on the analysis of second moments of the internal distance distributions.
Having Gaussian chain statistics well above the Kuhn length is obviously of primary importance for claiming our melts are equilibrated. 
However, the related example of building melts of crumpled rings \cite{schram2019local} should serve as a warning, that the second moment
is only a {\em necessary}, but not a {\em sufficient} condition for concluding that we have the desired distribution of equilibrium chain statistics. 
Short of inspecting the complete internal distance distribution functions, we have analyzed the Binder Cumulant\citep{binder1981finite} as a
measure of their deviation from Gaussian statistics. The sampled Binder Cumulant for internal distances (BCID) is defined as
\[
BCID(n)=1-\frac{\langle R_{\alpha}^{4}(n)\rangle}{3\langle R_{\alpha}^{2}(n)\rangle^{2}}
\]
where the averages are over all three Cartesian components $R_{\alpha}$
of the internal distances between beads for fixed chemical distance.
For Gaussian distributed stochastic variable $X$, $\langle X^{4}\rangle=3\langle X^{2}\rangle^{2}$,
hence the Binder Cumulant is zero in this case. 

Figure~\ref{fig:BinderCumulant} shows that our conformations obey Gaussian chain statistics above the Kuhn scale. At local scales we
observe deviations. This is to be expected since due to the local rod-like chain structure and liquid packing of beads of the KG model.
We note that while virtually all theories are based on Gaussian chain models\citep{DoiEdwards86}, real polymers are made up of monomers
with a finite size and with a local liquid-like packing. In this sense, the KG model is a more realistic representation of real chemical
polymers than most theoretical models.\cite{SGE_epl_05,Everaers2020Mapping} Finally a point to note for our present purposes is that Figure~\ref{fig:BinderCumulant} shows no difference between results at the beginning and the end of the equilibrium runs with the full KG potential.

%

\subsection{Primitive path analysis of the generated melts \label{subsec:PPA of generated melts}}

\begin{figure}
\includegraphics[width=0.95\columnwidth]{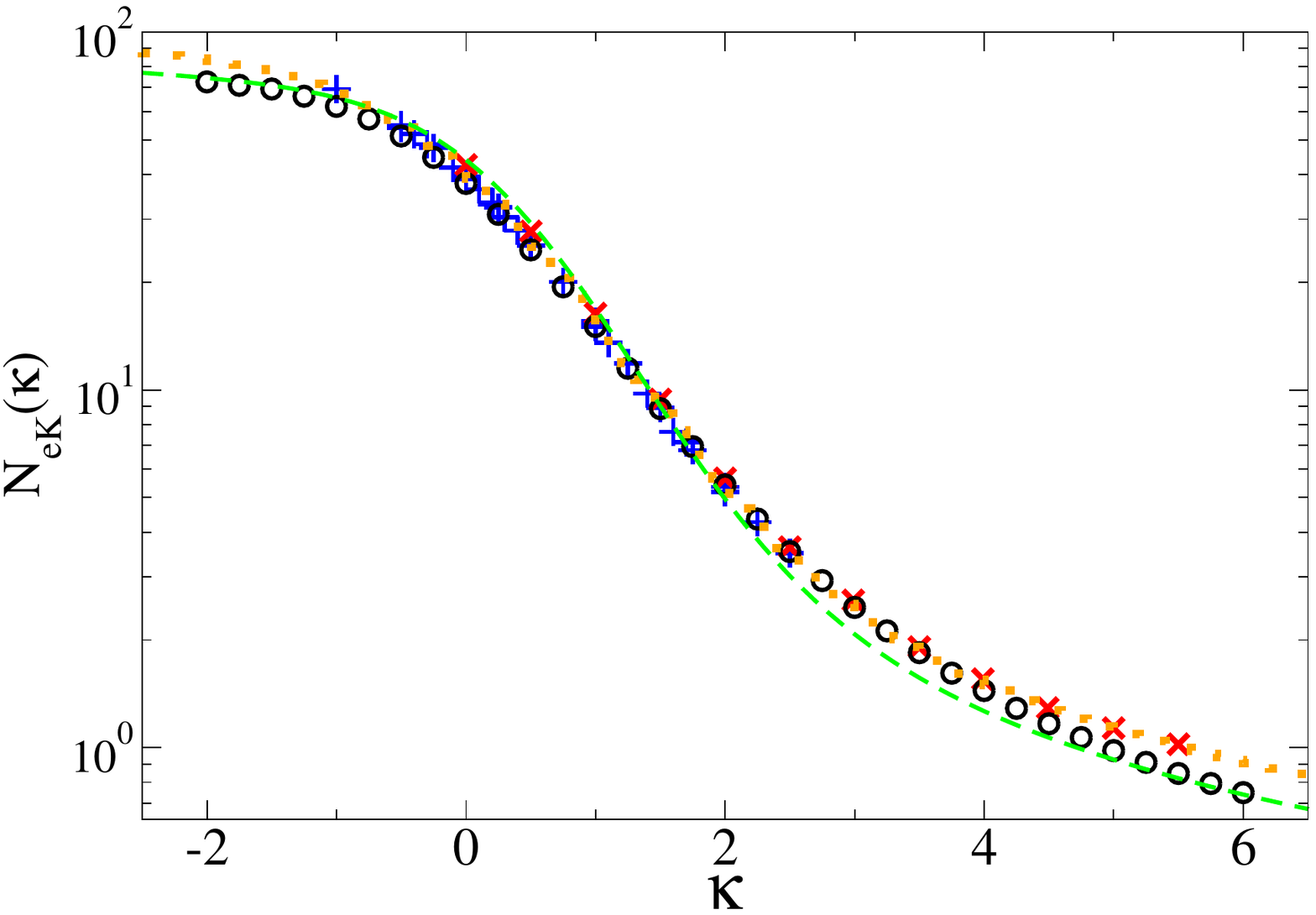}

\caption{Comparison of entanglement lengths for melts equilibrated with the
present method (black $\circ$), melts with $Z\geq100$ equilibrated
with the method in Ref. \citep{SvaneborgEquilibration2016} (blue
$+$), and brute force equilibrated melts with $N_{b}=400$ from Ref.
\citep{dietz2022facile} (red $\times$). For comparison the entanglement
lengths given by eq. (\ref{eq:Nek_Uchida}) (green dashed curve), and
Eq. (26) of Ref. \citep{dietz2022facile} (orange dotted curve).\label{fig:Nek_comparison}}
\end{figure}

So far we have focused on single chain observables characterizing the structure of individual chains. 
The microscopic topological state of entangled polymer systems sensitively depends on how different chains mutually interpenetrate 
and its characterisation provide a sensitive measure of chain equilibration~\citep{PhysRevE.72.061802,tubiana2021comparing}.
The right-most column in Fig. \ref{fig:Visualization} shows the primitive
path mesh obtained from primitive path analysis of the equilibrated melt. 
The stiffer the chains, the smaller the effect of applying primitive path analysis, since the entanglement
length decreases with increasing $\kappa$ (Fig.~\ref{fig:Nek_comparison}).
 
Fig. \ref{fig:Nek_comparison} compares the entanglements lengths of our present melts to
Literature data. We observe excellent agreement with our previous data and those from Dietz and Hoy for $\kappa\leq4$. 
Similarly to the values for the Kuhn length shown in Fig.~\ref{fig:Kuhn-length-vs.chainstiffness}
we observe some small deviations for the stiffest melts close to the isotropic to nematic transition at $\kappa=5.25$
There is no notable evolution of the entanglement length during the equilibrium simulation (no shown). 

It might be worthwhile to explore to what extent the primitive path chain conformations
match the primitive chain conformations shown in Fig. \ref{fig:Visualization} for flexible
chains, and how to generalize the constrained mode model to include bending modes relevant
for describing fluctuations of stiff chains. This would be a further step towards
generating equilibrium primitive path meshes ab-initio. However, this is beyond the scope
of the present paper.

\section{Discussion\label{sec:Discussion}}

Having described the method and having demonstrated its effectiveness in generating plausibly equilibrated samples of highly entangled isotropic model polymer melts, we now turn to the required computational effort and the question, how to put our data and methods to best use. 

\subsection{Computational effort}
To help the reader appreciate the utility of the method, 
we estimate\citep{Svaneborg2020Characterization} the brute-force equilibration/disentanglement times for the $Z=200$ and $\kappa=0$ systems to be $2.6\times 10^{11}\tau$.
With a time step of $\delta t=0.01\tau$ and $M=1000$ chains this corresponds to $4.3 \times 10^{20}$ particle updates for the system we generate here.
Assuming the simulation is run on Joliot-Curie Skylake partition this corresponds to $9.2\times 10^6$ core-years of simulation.
For comparison, the computationally most expensive part of the present equilibration protocol are the long push-off runs at the intermediate
force-capping value of $U_{0}=5\epsilon$. We invested just $55$ (CPU) core years of simulation time ($0.48$M core hours) on performing push-offs for all the $33$ melts reported here.

The computational effort required by the preceding lattice simulation and subsequent
mode processing is very modest, and only requires a few days on a single core on a standard
work station. The major computational effort is the push-off process, which requires 
$0.8-2.4$M integration steps. The wall time required for such a simulation depends on
system size and the number of nodes used to perform the simulations. However, this is a
modest computational effort compared to previous methods. For comparison, Moreira et al.
used about $8.6$M steps\citep{moreira2015direct} of varying duration to equilibrate
$1000$ chains of length $2000$ matching the smallest systems equilibrated here. 
For comparison, double-bridging augmented brute-force simulations with softened Kremer-Grest models
call for substantially longer simulation times for much shorter chains.
Auhl et al. required about $5 \times10^{7}$ integration steps\citep{auhl2003equilibration} to
equilibrate $M=500$ chains of $N=500$  beads. Dietz and Hoy\citep{dietz2022facile} recently
 equilibrated melts $M=1000$ chains of $N=400$  beads for varying stiffnesses
using simulations up to $10^{8}$ MD steps. 

\subsection{Multiscale equilibration of long-chain melts}

Multiscale equilibration proceeds from large to small scales. 
For the largest scales, we have retained our original choice~\citep{wang2009studying,SvaneborgEquilibration2016} of a lattice model, while Zhang et al.\citep{zhang2014equilibration,zhang2015communication,zhang2019hierarchical} employ a hierarchical blob-based description.
For both approaches, the total computational effort for the multiscale equilibration of long-chain polymer melts is dominated by the equilibration of the {\em local} chain and melt structure during the last fine-graining step. In the present case, this is the isoconfigurational push-off which we terminate once beads have moved over a spatial distance $g_1(t) = a_{pp}^2$ of the order of the resolution of the lattice model. 

This observation has several implications:
Most importantly and in marked contrast to the physical dynamics or double-bridging~\cite{karayiannis2002novel,karayiannis2002atomistic,karayiannis2003advanced}, the necessary computer time for multiscale equilibration of long chain melts becomes {\em independent} of chain length.
Secondly and in spite of the overhead for treating the large scale models, the approach essentially becomes advantageous for the shortest chain lengths resolved by the coarse-grain descriptions (i.e., in the present case, beyond a few entanglement lengths).
Thirdly, efforts for speeding up the equilibration are probably best concentrated on the critical (and in our experience most delicate) final step.

\subsection{Fine-graining to the Kuhn scale}

In Ref.~\citep{SvaneborgEquilibration2016}, we generated the initial bead-spring conformations for the push-off by simply decorating the lattice conformations with equidistantly placed beads. While this procedure preserves the conformational statistics and the chain packing at large scales, the resulting bond lengths are extremely compressed when compared to the target system. As a consequence, the initial relaxation of these states by energy minimisation was prone to perturb the chain structure on intermediate scales.

Here we have shown how to generate lattice-derived Gaussian chain conformations, which are discretized at the Kuhn scale. This has the advantage that the resulting bead-spring conformations exhibit the proper average chain contour length in addition to preserving the large scale structure. 
We have achieved this by switching to mode space and adapting the constrained mode formalism for rubber-elastic systems~\citep{everaers1998constrained,mergell2001tube} to the present problem. 
In particular, we identify the quenched (mean) configurations with the primitive chain lattice conformations.
This allows us to (i) filter out lattice artifacts and (ii) add contour length and bending fluctuations down to the Kuhn scale.
Our results show that the resulting conformations can be used without further precautions as starting states for our push-off procedure for KG melts.
We believe that the procedure would be equally useful for the transition to other mildly coarse-grain or even atomistic models.

\subsection{Isoconfigurational force-capped KG models}


As a methodological improvement geared exclusively towards applications of the (stiffness modulated [Faller]) KG model~\citep{Everaers2020Mapping}, we have improved and extended the range of our parameterization~\citep{SvaneborgEquilibration2016} of ``isoconfigurational'' force-capped KG models with suitably renormalized bending stiffnesses, Eqs.~(\ref{eq:Kuhn_length_KG}), (\ref{eq:lk_correction}) and (\ref{eq:renormalized_kappa_for_force_capping}).
In particular, we now cover the entire range  of Kuhn numbers $n_K= \rho_K l_K^3 \in [2,100]$ for isotropic polymer melts ranging from experimentally relevant bending rigidities up to and beyond the limit of the isotropic- nematic transition.

Employing the present family of force-capped KG models in brute-force equilibration runs is always computationally more efficient than the use of the full KG model (Fig.~6 in Ref.~\citep{SvaneborgEquilibration2016}), since the computational effort for the final switch to the full KG potential is negligible. 
For chains below the entanglement length the speed-up of the order of two due to the smaller effective bead friction may be of little relevance, but for longer chains the possibility of chain crossing assures that the maximal relaxation time increases only like $N^2$ as in the Rouse model as opposed to $N^{3+x}$ for entangled KG chains.
Moreover, LAMMPS allows to accelerate the equilibration via double-bridging moves so that the present families of potentials can be used along the lines of the recent work of Dietz and Hoy\citep{dietz2022facile}.
LAMMPS scripts for performing the present push-off simulations will be made available on our website (http://polymer.zqex.dk/), but a systematic analysis of the effectiveness of the double-bridging algorithm as a function of chain length and at different levels of force-capping is beyond the scope of the present work.

\subsection{Fine-graining to the atomic scale}

As in our previous work~\citep{Everaers2020Mapping} we have focused on Kuhn-scale mapped KG models as a minimal microscopic description of commodity homopolymer melts in the spirit of standard textbooks of polymer physics\cite{DoiEdwards86,RubinsteinColby,deGennes,dealy2018structure}. 
We also think of the resulting melt configurations as suitable starting points for further fine-graining to an atomistic description.
Note, however, that  this remains to be shown in practice and that it is by no means necessary to include a generic description on the Kuhn scale into the employed hierarchy of models. Ref.~\citep{zhang2019hierarchical} demonstrates that multiscale equilibration can proceed all the way down to the atomic scale through more material-specific coarse-grain models.

\subsection{Choosing the length scales for the crossover between the different models}

Our lattice model and the blob model of Zhang et al.\citep{zhang2014equilibration,zhang2015communication,zhang2019hierarchical} are much simpler and computationally more efficient than the more microscopic target models.
This raises the question, down to which scales one should use these descriptions in multiscale equilibration schemes.
A full investigation is beyond the scope of the present work, but some insight can be gained by analyzing the issue on a scaling level.

Given that our models and methods are fully compatible with those from Refs.~\citep{zhang2014equilibration,zhang2015communication}, we stick to a lattice notation. 
Be $a$ the lattice constant and $\langle R^2 \rangle = a^2 N_a$ the mean-square end-to-end distance of a $N_a$-step lattice chain representing a target polymer. From the number density of lattice chain segments, $\rho_a = \rho_{ch} N_a$, follows the dimensionless number of chain segments per lattice site, $n_a = \rho_a a^3 =  \rho_{ch} \langle R^2 \rangle^{3/2} / N_a^{1/2} = \left( N_p/ N_a \right)^{1/2}$. 
In terms of available physical length scales, one can plausibly associate the lattice constant $a$ with (a multiple of) the Kuhn length~\cite{rosa2014ring,schram2019local}, the packing length~\citep{zhang2014equilibration,zhang2015communication}, and the entanglement length~\citep{SvaneborgEquilibration2016}. We have not done so here, but we note that one can easily extend the idea of multiscale equilibration to chains of arbitrary length by defining a hierarchy of more and more coarse-grain models indexed $n=0,1,2,...$ for any such choice of $a$ with lattice constants $a^{(n)} = 2^n a$, segment numbers $N_a^{(n)} = 2^{-2n} N_a$ and reduced densities $n_a^{(n)} = 2^n n_a$.

First consider a lattice (or blob) scale equal to a multiple of the Kuhn scale, $a=\alpha_K l_K$. 
In this case,  the average number of segments occupying a lattice site is proportional to the Kuhn number,  $n_a = \alpha_K n_K$, i.e. {\it a priori} we have to study a different (hierarchy of) lattice models for each target material.
In contrast, choosing the lattice (or blob) scale equal to a fixed multiple of the packing length, $a=\alpha_p p$, 
the average number of segments occupying a lattice site $n_a = \alpha_p n_p = \alpha_p$, i.e. {\em all} target materials can be studied using a universal (hierarchy of) coarse-grain models.~\citep{zhang2014equilibration,zhang2015communication}
For intrinsically flexible commodity polymers with $n_K \in [2,10]$~\citep{fetters2007chain,Everaers2020Mapping} our choice of equa\-ting the lattice scale with the entanglement length corresponds to $\alpha_p = \alpha$ and $\alpha_K = \alpha/n_K$ with $\alpha=18$ from the packing argument, Eqs.(\ref{eq:packing_NeK}) and (\ref{eq:packing_Nep}) and effectively coincides with the choice of Refs.~\citep{zhang2014equilibration,zhang2015communication}.

Choosing smaller values of $\alpha_K$ or $\alpha_p$ is promising, as one might hope to terminate the subsequent, computationally expensive push-off runs once beads have moved over the correspondingly reduced lattice scale, $g_1(t) = a^2$.
For more dilute target systems, Refs.~\cite{rosa2014ring,schram2019local} worked with a lattice scale $a=l_K$ equal to the Kuhn scale, which offers the added benefit that the lattice chains have the same (average) contour length as the target chains and can be conveniently decorated with the beads of the off-lattice model. 
In the present work we have nevertheless retained our original choice~\citep{wang2009studying,SvaneborgEquilibration2016} to define the lattice model on the entanglement scale.
The original reason for this choice was that excluded volume interactions are no longer fully screened,  \cite{Wittmer_Meyer_PRL04,wittmer2007polymer,wittmer2007intramolecular,beckrich2007intramolecular,semenov2010bond} when the number $n_a$ of overlapping chain segments per lattice site approaches unity. 
The  coarse-grain models loose much of their elegance and simplicity, if additional work is required to preserve the target relation $\langle R^2 \rangle = a^2 N_a$ 
or if one has to perform a parameterization of the relation between the asymptotic Kuhn length and the model parameters along the lines of Ref. \cite{Everaers2020Mapping}.
This seems inevitable on the packing scale, where $a=p$ and $n_a = n_p = 1$, and might still be an issue on the Kuhn scale, where  for commodity polymers $n_a =  n_K$ 
falls into the range of  $n_K\in [2,10]$~\citep{Everaers2020Mapping}. 

The second reason to continue to work on the entanglement scale was that we think it interesting~\cite{Uchida_jcp_08,dietz2022facile} to study isotropic polymer melts over the full range of intrinsic stiffnesses up to and beyond the onset of the isotropic- nematic transition around $n_K \approx 100$. For the present purposes it would probably be possible to employ a universal large scale model with a fixed resolution of $a = \alpha p$ along the lines of Refs.\citep{zhang2014equilibration,zhang2015communication}. 
However, one would need to test, if the push-off protocol can deal with the unphysical chain contour lengths and the random walk statistics of the lattice chains below the target Kuhn scale. Alternatively, one could remove these artifacts using the methods from Sec.~\ref{sec:CMM finegraining} at the price of reintroducing density fluctuations on scales up to the Kuhn scale. 
Our choice to identify the lattice scale with the Kuhn length of the primitive paths, $a=a_{pp}$, allowed as to avoid these complications, because it assures~\cite{Uchida_jcp_08} that the lattice constant reduces to the Kuhn length $l_K$ of the chains in an uncritical limit, where $n_K \gg 1$.

\section{Summary and Conclusion\label{sec:Conclusions}}

We have presented an effective, computationally efficient multi-scale equilibration process for long chain polymer melts resolved at the Kuhn scale using the Kremer-Grest model \cite{grest1986molecular,Everaers2020Mapping}.

Our method proceeds from large to small scales and uses three distinct polymer models to represent the chains. 
We start out by equilibrating our melts above the tube scale using a computationally very efficient Monte-Carlo lattice model. 
Each step of the primitive lattice chains corresponds to an entanglement segment and multiple segments are allowed to occupy the same lattice site. 
We use simulated annealing Monte Carlo to progressively penalize density fluctuations without introducing deviations from the target random walk chain statistics. 
To fine-grain the primitive chain lattice conformations to the Kuhn scale we switch to (Rouse) mode space.
In the context of tube models for rubber elastic systems long wavelength modes beyond the entanglement scale are quenched while short wavelength modes are
free to fluctuate. Here we identify the quenched (mean) configurations with the primitive chain lattice conformations and use a constrained mode formalism~\citep{everaers1998constrained,mergell2001tube} to (i) filter out lattice artifacts and (ii) add contour length and bending fluctuations down to the Kuhn scale. 
Finally, we switch to a bead-spring representation and implement a push-off process using a sequence of force-capped KG models
to arrive at viable starting conformations for simulations with the full Kremer-Grest polymer model\citep{grest1986molecular}. 
To avoid the introduction of artifacts in this stage, we have carefully parameterized the bending stiffness to be employed for each level of force-capping
so as to always reproduce the Kuhn length of the target KG model. 

We have used this method to generate sample melt configurations for chain stiffnesses $\kappa=-2$ up to $6$ ranging from values corresponding to the most flexible experimental systems~\cite{Everaers2020Mapping} to the onset of the nematic-isotropic transition. With the transition located around $\kappa\approx5.5$~\cite{dietz2022facile}  the conformations we generated for the largest $\kappa$ values should probably be regarded as metastable isotropic melt states and not true equilibrium melt states. 
All melts are composed of $M\approx1000$ chains of an effective chain length of $Z\approx200$ entanglements per chain. 
The number of beads per chain varies between $N=25000$ and $N=2000$ for the highest stiffness. 
The resulting melt structures show excellent agreement with literature values for shorter chains with respect to the chain mean-square internal distances and PPA-derived entanglement lengths. 
These statistical properties remain virtually constant over the entanglement time, when the systems are subsequently propagated with the full KG force field. 
We take these observations as evidence that the generated melt conformations can be plausibly regarded as equilibrium states.




In future work, it would be interesting to explore, if we one can preserve the simple form of the lattice model for resolutions approaching the Kuhn scale and to further reduce the CPU time for the isoconfigurational push-off. 
Furthermore, it is straightforward to generalize both the lattice and constrained mode models to (mixtures of) chains of arbitrary connectivities. We expect the present approach to be applicable with minor modifications in all cases where the systems reach thermal equilibrium and where topological constraints do not affect static properties. This may be particularly fruitful in the case of entangled star and branched polymers, where the physical relaxation times become exponentially longer than for linear chains.\cite{DoiEdwards86,dealy2018structure}
As a counterexample, we note that the multiscale generation of melts of non-concatenated ring polymer melts or solutions requires special methods~\cite{rosa2014ring,schram2019local}.

To end on a note of caution, no number of successful validations of a multiscale equilibration procedure precludes the discovery of deviations in the analysis of additional,  more refined observables (for a recent example see Refs.~\cite{rosa2014ring,schram2019local} on generating ring melts).
In the case of long-chain homopolymer melts suitable candidates for further comparisons might be the analysis of intra-chain knots~\citep{tubiana2021comparing} and of long-range correlations in the bond orientational correlation function~\cite{Wittmer_Meyer_PRL04,wittmer2007polymer,wittmer2007intramolecular,beckrich2007intramolecular,semenov2010bond}.
Given that brute-force MD equilibration for the present chain lengths requires of the order of $10^6$ more CPU time, we may have to accept this uncertainty for the foreseeable future and resign ourselves to revising the equilibration procedure whenever shortcomings are identified.

\subsection{Data Availability}
All the data that support the findings of this study are available from the corresponding author upon reasonable request.
Researchers interested in analyzing our long chain melts or in using them as starting conformations for their own simulations can find equilibrated melt conformations of $M=1000$ and $Z=200$ for integer and half-integer values of the stiffness parameter $\kappa$ at Zenodo at http://doi.org/10.5281/zenodo.7034881. We have generated a similar but smaller set of melts with $M=500$ and $Z=100$ which are also openly available at Zenodo at http://doi.org/10.5281/zenodo.7319837.

\begin{acknowledgments}
The simulations were performed by using the Large Atomic Massively
Parallel Simulator (LAMMPS) Molecular Dynamics software.\citep{PlimptonLAMMPS,PlimptonLAMMPS2}
We acknowledge that part of the results of this research have been
achieved using the PRACE Research Infrastructure resource Joliot-Curie
SKL based in France at GENCI@CEA. Computing facilities were provided
by the eScience Center at the University of Southern Denmark and financed
by the Faculty of Science. CS gratefully acknowledges conformational
data shared by R. Hoy and discussions with I.A. Gula.
\end{acknowledgments}

\newpage

\bibliography{melteq}

\begin{thebibliography}{117}%
\makeatletter
\providecommand \@ifxundefined [1]{%
 \@ifx{#1\undefined}
}%
\providecommand \@ifnum [1]{%
 \ifnum #1\expandafter \@firstoftwo
 \else \expandafter \@secondoftwo
 \fi
}%
\providecommand \@ifx [1]{%
 \ifx #1\expandafter \@firstoftwo
 \else \expandafter \@secondoftwo
 \fi
}%
\providecommand \natexlab [1]{#1}%
\providecommand \enquote  [1]{``#1''}%
\providecommand \bibnamefont  [1]{#1}%
\providecommand \bibfnamefont [1]{#1}%
\providecommand \citenamefont [1]{#1}%
\providecommand \href@noop [0]{\@secondoftwo}%
\providecommand \href [0]{\begingroup \@sanitize@url \@href}%
\providecommand \@href[1]{\@@startlink{#1}\@@href}%
\providecommand \@@href[1]{\endgroup#1\@@endlink}%
\providecommand \@sanitize@url [0]{\catcode `\\12\catcode `\$12\catcode
  `\&12\catcode `\#12\catcode `\^12\catcode `\_12\catcode `\%12\relax}%
\providecommand \@@startlink[1]{}%
\providecommand \@@endlink[0]{}%
\providecommand \url  [0]{\begingroup\@sanitize@url \@url }%
\providecommand \@url [1]{\endgroup\@href {#1}{\urlprefix }}%
\providecommand \urlprefix  [0]{URL }%
\providecommand \Eprint [0]{\href }%
\providecommand \doibase [0]{http://dx.doi.org/}%
\providecommand \selectlanguage [0]{\@gobble}%
\providecommand \bibinfo  [0]{\@secondoftwo}%
\providecommand \bibfield  [0]{\@secondoftwo}%
\providecommand \translation [1]{[#1]}%
\providecommand \BibitemOpen [0]{}%
\providecommand \bibitemStop [0]{}%
\providecommand \bibitemNoStop [0]{.\EOS\space}%
\providecommand \EOS [0]{\spacefactor3000\relax}%
\providecommand \BibitemShut  [1]{\csname bibitem#1\endcsname}%
\let\auto@bib@innerbib\@empty
\bibitem [{\citenamefont {Everaers}\ \emph {et~al.}(2020)\citenamefont
  {Everaers}, \citenamefont {Karimi-Varzaneh}, \citenamefont {Fleck},
  \citenamefont {Hojdis},\ and\ \citenamefont
  {Svaneborg}}]{Everaers2020Mapping}%
  \BibitemOpen
  \bibfield  {author} {\bibinfo {author} {\bibfnamefont {R.}~\bibnamefont
  {Everaers}}, \bibinfo {author} {\bibfnamefont {H.~A.}\ \bibnamefont
  {Karimi-Varzaneh}}, \bibinfo {author} {\bibfnamefont {F.}~\bibnamefont
  {Fleck}}, \bibinfo {author} {\bibfnamefont {N.}~\bibnamefont {Hojdis}}, \
  and\ \bibinfo {author} {\bibfnamefont {C.}~\bibnamefont {Svaneborg}},\
  }\bibfield  {title} {\enquote {\bibinfo {title} {Kremer–grest models for
  commodity polymer melts: Linking theory, experiment, and simulation at the
  kuhn scale},}\ }\href {\doibase
  https://dx.doi.org/10.1021/acs.macromol.9b02428} {\bibfield  {journal}
  {\bibinfo  {journal} {Macromolecules}\ }\textbf {\bibinfo {volume} {53}},\
  \bibinfo {pages} {1901} (\bibinfo {year} {2020})}\BibitemShut {NoStop}%
\bibitem [{\citenamefont {Doi}\ and\ \citenamefont
  {Edwards}(1986)}]{DoiEdwards86}%
  \BibitemOpen
  \bibfield  {author} {\bibinfo {author} {\bibfnamefont {M.}~\bibnamefont
  {Doi}}\ and\ \bibinfo {author} {\bibfnamefont {S.~F.}\ \bibnamefont
  {Edwards}},\ }\href@noop {} {\emph {\bibinfo {title} {The Theory of Polymer
  Dynamics}}}\ (\bibinfo  {publisher} {Clarendon},\ \bibinfo {address}
  {Oxford},\ \bibinfo {year} {1986})\BibitemShut {NoStop}%
\bibitem [{\citenamefont {Faller}(2007)}]{faller2007coarse}%
  \BibitemOpen
  \bibfield  {author} {\bibinfo {author} {\bibfnamefont {R.}~\bibnamefont
  {Faller}},\ }\bibfield  {title} {\enquote {\bibinfo {title} {Coarse-grain
  modeling of polymers},}\ }\href@noop {} {\bibfield  {journal} {\bibinfo
  {journal} {Rev. Comp. Ch.}\ }\textbf {\bibinfo {volume} {23}},\ \bibinfo
  {pages} {233} (\bibinfo {year} {2007})}\BibitemShut {NoStop}%
\bibitem [{\citenamefont {Peter}\ and\ \citenamefont
  {Kremer}(2010)}]{MultiscalePeterKremerFaradayDiss2010}%
  \BibitemOpen
  \bibfield  {author} {\bibinfo {author} {\bibfnamefont {C.}~\bibnamefont
  {Peter}}\ and\ \bibinfo {author} {\bibfnamefont {K.}~\bibnamefont {Kremer}},\
  }\bibfield  {title} {\enquote {\bibinfo {title} {Multiscale simulation of
  soft matter systems},}\ }\href@noop {} {\bibfield  {journal} {\bibinfo
  {journal} {Faraday Discuss.}\ }\textbf {\bibinfo {volume} {144}},\ \bibinfo
  {pages} {9} (\bibinfo {year} {2010})}\BibitemShut {NoStop}%
\bibitem [{\citenamefont {Gooneie}, \citenamefont {Schuschnigg},\ and\
  \citenamefont {Holzer}(2017)}]{gooneie2017review}%
  \BibitemOpen
  \bibfield  {author} {\bibinfo {author} {\bibfnamefont {A.}~\bibnamefont
  {Gooneie}}, \bibinfo {author} {\bibfnamefont {S.}~\bibnamefont
  {Schuschnigg}}, \ and\ \bibinfo {author} {\bibfnamefont {C.}~\bibnamefont
  {Holzer}},\ }\bibfield  {title} {\enquote {\bibinfo {title} {A review of
  multiscale computational methods in polymeric materials},}\ }\href@noop {}
  {\bibfield  {journal} {\bibinfo  {journal} {Polymers}\ }\textbf {\bibinfo
  {volume} {9}},\ \bibinfo {pages} {16} (\bibinfo {year} {2017})}\BibitemShut
  {NoStop}%
\bibitem [{\citenamefont {Karayiannis}, \citenamefont {Mavrantzas},\ and\
  \citenamefont {Theodorou}(2002)}]{karayiannis2002novel}%
  \BibitemOpen
  \bibfield  {author} {\bibinfo {author} {\bibfnamefont {N.~C.}\ \bibnamefont
  {Karayiannis}}, \bibinfo {author} {\bibfnamefont {V.~G.}\ \bibnamefont
  {Mavrantzas}}, \ and\ \bibinfo {author} {\bibfnamefont {D.~N.}\ \bibnamefont
  {Theodorou}},\ }\bibfield  {title} {\enquote {\bibinfo {title} {A novel monte
  carlo scheme for the rapid equilibration of atomistic model polymer systems
  of precisely defined molecular architecture},}\ }\href@noop {} {\bibfield
  {journal} {\bibinfo  {journal} {Phys. Rev. Lett.}\ }\textbf {\bibinfo
  {volume} {88}},\ \bibinfo {pages} {105503} (\bibinfo {year}
  {2002})}\BibitemShut {NoStop}%
\bibitem [{\citenamefont {Karayiannis}\ \emph {et~al.}(2002)\citenamefont
  {Karayiannis}, \citenamefont {Giannousaki}, \citenamefont {Mavrantzas},\ and\
  \citenamefont {Theodorou}}]{karayiannis2002atomistic}%
  \BibitemOpen
  \bibfield  {author} {\bibinfo {author} {\bibfnamefont {N.~C.}\ \bibnamefont
  {Karayiannis}}, \bibinfo {author} {\bibfnamefont {A.~E.}\ \bibnamefont
  {Giannousaki}}, \bibinfo {author} {\bibfnamefont {V.~G.}\ \bibnamefont
  {Mavrantzas}}, \ and\ \bibinfo {author} {\bibfnamefont {D.~N.}\ \bibnamefont
  {Theodorou}},\ }\bibfield  {title} {\enquote {\bibinfo {title} {Atomistic
  monte carlo simulation of strictly monodisperse long polyethylene melts
  through a generalized chain bridging algorithm},}\ }\href@noop {} {\bibfield
  {journal} {\bibinfo  {journal} {J. Chem. Phys.}\ }\textbf {\bibinfo {volume}
  {117}},\ \bibinfo {pages} {5465} (\bibinfo {year} {2002})}\BibitemShut
  {NoStop}%
\bibitem [{\citenamefont {Karayiannis}, \citenamefont {Giannousaki},\ and\
  \citenamefont {Mavrantzas}(2003)}]{karayiannis2003advanced}%
  \BibitemOpen
  \bibfield  {author} {\bibinfo {author} {\bibfnamefont {N.~C.}\ \bibnamefont
  {Karayiannis}}, \bibinfo {author} {\bibfnamefont {A.~E.}\ \bibnamefont
  {Giannousaki}}, \ and\ \bibinfo {author} {\bibfnamefont {V.~G.}\ \bibnamefont
  {Mavrantzas}},\ }\bibfield  {title} {\enquote {\bibinfo {title} {An advanced
  monte carlo method for the equilibration of model long-chain branched
  polymers with a well-defined molecular architecture: Detailed atomistic
  simulation of an h-shaped polyethylene melt},}\ }\href@noop {} {\bibfield
  {journal} {\bibinfo  {journal} {J. Chem. Phys.}\ }\textbf {\bibinfo {volume}
  {118}},\ \bibinfo {pages} {2451} (\bibinfo {year} {2003})}\BibitemShut
  {NoStop}%
\bibitem [{\citenamefont {Auhl}\ \emph {et~al.}(2003)\citenamefont {Auhl},
  \citenamefont {Everaers}, \citenamefont {Grest}, \citenamefont {Kremer},\
  and\ \citenamefont {Plimpton}}]{auhl2003equilibration}%
  \BibitemOpen
  \bibfield  {author} {\bibinfo {author} {\bibfnamefont {R.}~\bibnamefont
  {Auhl}}, \bibinfo {author} {\bibfnamefont {R.}~\bibnamefont {Everaers}},
  \bibinfo {author} {\bibfnamefont {G.~S.}\ \bibnamefont {Grest}}, \bibinfo
  {author} {\bibfnamefont {K.}~\bibnamefont {Kremer}}, \ and\ \bibinfo {author}
  {\bibfnamefont {S.~J.}\ \bibnamefont {Plimpton}},\ }\bibfield  {title}
  {\enquote {\bibinfo {title} {Equilibration of long chain polymer melts in
  computer simulations},}\ }\href@noop {} {\bibfield  {journal} {\bibinfo
  {journal} {J. Chem. Phys.}\ }\textbf {\bibinfo {volume} {119}},\ \bibinfo
  {pages} {12718} (\bibinfo {year} {2003})}\BibitemShut {NoStop}%
\bibitem [{\citenamefont {Dietz}\ and\ \citenamefont
  {Hoy}(2022)}]{dietz2022facile}%
  \BibitemOpen
  \bibfield  {author} {\bibinfo {author} {\bibfnamefont {J.~D.}\ \bibnamefont
  {Dietz}}\ and\ \bibinfo {author} {\bibfnamefont {R.~S.}\ \bibnamefont
  {Hoy}},\ }\bibfield  {title} {\enquote {\bibinfo {title} {Facile
  equilibration of well-entangled semiflexible bead--spring polymer melts},}\
  }\href@noop {} {\bibfield  {journal} {\bibinfo  {journal} {J. Chem. Phys.}\
  }\textbf {\bibinfo {volume} {156}},\ \bibinfo {pages} {014103} (\bibinfo
  {year} {2022})}\BibitemShut {NoStop}%
\bibitem [{\citenamefont {Madras}\ and\ \citenamefont
  {Sokal}(1988)}]{madras1988pivot}%
  \BibitemOpen
  \bibfield  {author} {\bibinfo {author} {\bibfnamefont {N.}~\bibnamefont
  {Madras}}\ and\ \bibinfo {author} {\bibfnamefont {A.~D.}\ \bibnamefont
  {Sokal}},\ }\bibfield  {title} {\enquote {\bibinfo {title} {The pivot
  algorithm: a highly efficient monte carlo method for the self-avoiding
  walk},}\ }\href@noop {} {\bibfield  {journal} {\bibinfo  {journal} {J. Stat.
  Phys.}\ }\textbf {\bibinfo {volume} {50}},\ \bibinfo {pages} {109} (\bibinfo
  {year} {1988})}\BibitemShut {NoStop}%
\bibitem [{\citenamefont {Flory}(1949)}]{flory49}%
  \BibitemOpen
  \bibfield  {author} {\bibinfo {author} {\bibfnamefont {P.~J.}\ \bibnamefont
  {Flory}},\ }\bibfield  {title} {\enquote {\bibinfo {title} {The configuration
  of real polymer chains},}\ }\href@noop {} {\bibfield  {journal} {\bibinfo
  {journal} {J. Chem. Phys.}\ }\textbf {\bibinfo {volume} {17}},\ \bibinfo
  {pages} {303} (\bibinfo {year} {1949})}\BibitemShut {NoStop}%
\bibitem [{\citenamefont {Kremer}\ and\ \citenamefont
  {Grest}(1990{\natexlab{a}})}]{kremer90}%
  \BibitemOpen
  \bibfield  {author} {\bibinfo {author} {\bibfnamefont {K.}~\bibnamefont
  {Kremer}}\ and\ \bibinfo {author} {\bibfnamefont {G.~S.}\ \bibnamefont
  {Grest}},\ }\bibfield  {title} {\enquote {\bibinfo {title} {Dynamics of
  entangled linear polymer melts: A molecular dynamics simulation},}\
  }\href@noop {} {\bibfield  {journal} {\bibinfo  {journal} {J. Chem. Phys.}\
  }\textbf {\bibinfo {volume} {92}},\ \bibinfo {pages} {5057} (\bibinfo {year}
  {1990}{\natexlab{a}})}\BibitemShut {NoStop}%
\bibitem [{\citenamefont {Carbone}, \citenamefont {Karimi-Varzaneh},\ and\
  \citenamefont {M{\"u}ller-Plathe}(2010)}]{carbone2010fine}%
  \BibitemOpen
  \bibfield  {author} {\bibinfo {author} {\bibfnamefont {P.}~\bibnamefont
  {Carbone}}, \bibinfo {author} {\bibfnamefont {H.~A.}\ \bibnamefont
  {Karimi-Varzaneh}}, \ and\ \bibinfo {author} {\bibfnamefont {F.}~\bibnamefont
  {M{\"u}ller-Plathe}},\ }\bibfield  {title} {\enquote {\bibinfo {title}
  {Fine-graining without coarse-graining: An easy and fast way to equilibrate
  dense polymer melts},}\ }\href@noop {} {\bibfield  {journal} {\bibinfo
  {journal} {Faraday Discuss.}\ }\textbf {\bibinfo {volume} {144}},\ \bibinfo
  {pages} {25} (\bibinfo {year} {2010})}\BibitemShut {NoStop}%
\bibitem [{\citenamefont {Moreira}\ \emph {et~al.}(2015)\citenamefont
  {Moreira}, \citenamefont {Zhang}, \citenamefont {M{\"u}ller}, \citenamefont
  {Stuehn},\ and\ \citenamefont {Kremer}}]{moreira2015direct}%
  \BibitemOpen
  \bibfield  {author} {\bibinfo {author} {\bibfnamefont {L.~A.}\ \bibnamefont
  {Moreira}}, \bibinfo {author} {\bibfnamefont {G.}~\bibnamefont {Zhang}},
  \bibinfo {author} {\bibfnamefont {F.}~\bibnamefont {M{\"u}ller}}, \bibinfo
  {author} {\bibfnamefont {T.}~\bibnamefont {Stuehn}}, \ and\ \bibinfo {author}
  {\bibfnamefont {K.}~\bibnamefont {Kremer}},\ }\bibfield  {title} {\enquote
  {\bibinfo {title} {Direct equilibration and characterization of polymer melts
  for computer simulations},}\ }\href@noop {} {\bibfield  {journal} {\bibinfo
  {journal} {Macromol. Theory Simul.}\ }\textbf {\bibinfo {volume} {24}},\
  \bibinfo {pages} {419} (\bibinfo {year} {2015})}\BibitemShut {NoStop}%
\bibitem [{\citenamefont {Svaneborg}\ \emph {et~al.}(2016)\citenamefont
  {Svaneborg}, \citenamefont {Karimi-Varzaneh}, \citenamefont {Hojdis},
  \citenamefont {Fleck},\ and\ \citenamefont
  {Everaers}}]{SvaneborgEquilibration2016}%
  \BibitemOpen
  \bibfield  {author} {\bibinfo {author} {\bibfnamefont {C.}~\bibnamefont
  {Svaneborg}}, \bibinfo {author} {\bibfnamefont {H.~A.}\ \bibnamefont
  {Karimi-Varzaneh}}, \bibinfo {author} {\bibfnamefont {N.}~\bibnamefont
  {Hojdis}}, \bibinfo {author} {\bibfnamefont {F.}~\bibnamefont {Fleck}}, \
  and\ \bibinfo {author} {\bibfnamefont {R.}~\bibnamefont {Everaers}},\
  }\bibfield  {title} {\enquote {\bibinfo {title} {Multiscale approach to
  equilibrating model polymer melts},}\ }\href@noop {} {\bibfield  {journal}
  {\bibinfo  {journal} {Phys. Rev. E}\ }\textbf {\bibinfo {volume} {94}},\
  \bibinfo {pages} {032502} (\bibinfo {year} {2016})}\BibitemShut {NoStop}%
\bibitem [{\citenamefont {Zhang}\ \emph {et~al.}(2014)\citenamefont {Zhang},
  \citenamefont {Moreira}, \citenamefont {Stuehn}, \citenamefont {Daoulas},\
  and\ \citenamefont {Kremer}}]{zhang2014equilibration}%
  \BibitemOpen
  \bibfield  {author} {\bibinfo {author} {\bibfnamefont {G.}~\bibnamefont
  {Zhang}}, \bibinfo {author} {\bibfnamefont {L.~A.}\ \bibnamefont {Moreira}},
  \bibinfo {author} {\bibfnamefont {T.}~\bibnamefont {Stuehn}}, \bibinfo
  {author} {\bibfnamefont {K.~C.}\ \bibnamefont {Daoulas}}, \ and\ \bibinfo
  {author} {\bibfnamefont {K.}~\bibnamefont {Kremer}},\ }\bibfield  {title}
  {\enquote {\bibinfo {title} {Equilibration of high molecular weight polymer
  melts: a hierarchical strategy},}\ }\href@noop {} {\bibfield  {journal}
  {\bibinfo  {journal} {ACS Macro. Lett.}\ }\textbf {\bibinfo {volume} {3}},\
  \bibinfo {pages} {198} (\bibinfo {year} {2014})}\BibitemShut {NoStop}%
\bibitem [{\citenamefont {Zhang}\ \emph {et~al.}(2015)\citenamefont {Zhang},
  \citenamefont {Stuehn}, \citenamefont {Daoulas},\ and\ \citenamefont
  {Kremer}}]{zhang2015communication}%
  \BibitemOpen
  \bibfield  {author} {\bibinfo {author} {\bibfnamefont {G.}~\bibnamefont
  {Zhang}}, \bibinfo {author} {\bibfnamefont {T.}~\bibnamefont {Stuehn}},
  \bibinfo {author} {\bibfnamefont {K.~C.}\ \bibnamefont {Daoulas}}, \ and\
  \bibinfo {author} {\bibfnamefont {K.}~\bibnamefont {Kremer}},\ }\bibfield
  {title} {\enquote {\bibinfo {title} {One size fits all: Equilibrating
  chemically different polymer liquids through universal long-wavelength
  description},}\ }\href@noop {} {\bibfield  {journal} {\bibinfo  {journal} {J.
  Chem. Phys.}\ }\textbf {\bibinfo {volume} {142}},\ \bibinfo {pages} {221102}
  (\bibinfo {year} {2015})}\BibitemShut {NoStop}%
\bibitem [{\citenamefont {Zhang}\ \emph {et~al.}(2019)\citenamefont {Zhang},
  \citenamefont {Chazirakis}, \citenamefont {Harmandaris}, \citenamefont
  {Stuehn}, \citenamefont {Daoulas},\ and\ \citenamefont
  {Kremer}}]{zhang2019hierarchical}%
  \BibitemOpen
  \bibfield  {author} {\bibinfo {author} {\bibfnamefont {G.}~\bibnamefont
  {Zhang}}, \bibinfo {author} {\bibfnamefont {A.}~\bibnamefont {Chazirakis}},
  \bibinfo {author} {\bibfnamefont {V.~A.}\ \bibnamefont {Harmandaris}},
  \bibinfo {author} {\bibfnamefont {T.}~\bibnamefont {Stuehn}}, \bibinfo
  {author} {\bibfnamefont {K.~C.}\ \bibnamefont {Daoulas}}, \ and\ \bibinfo
  {author} {\bibfnamefont {K.}~\bibnamefont {Kremer}},\ }\bibfield  {title}
  {\enquote {\bibinfo {title} {Hierarchical modelling of polystyrene melts:
  from soft blobs to atomistic resolution},}\ }\href@noop {} {\bibfield
  {journal} {\bibinfo  {journal} {Soft Matter}\ }\textbf {\bibinfo {volume}
  {15}},\ \bibinfo {pages} {289--302} (\bibinfo {year} {2019})}\BibitemShut
  {NoStop}%
\bibitem [{\citenamefont {Warner}\ and\ \citenamefont
  {Edwards}(1978)}]{WarnerEdwards}%
  \BibitemOpen
  \bibfield  {author} {\bibinfo {author} {\bibfnamefont {M.}~\bibnamefont
  {Warner}}\ and\ \bibinfo {author} {\bibfnamefont {S.}~\bibnamefont
  {Edwards}},\ }\bibfield  {title} {\enquote {\bibinfo {title} {Neutron
  scattering from strained polymer networks},}\ }\href@noop {} {\bibfield
  {journal} {\bibinfo  {journal} {J. Phys. A Math.}\ }\textbf {\bibinfo
  {volume} {11}},\ \bibinfo {pages} {1649} (\bibinfo {year}
  {1978})}\BibitemShut {NoStop}%
\bibitem [{\citenamefont {Read}\ and\ \citenamefont
  {McLeish}(1997)}]{read1997lozenge}%
  \BibitemOpen
  \bibfield  {author} {\bibinfo {author} {\bibfnamefont {D.}~\bibnamefont
  {Read}}\ and\ \bibinfo {author} {\bibfnamefont {T.}~\bibnamefont {McLeish}},\
  }\bibfield  {title} {\enquote {\bibinfo {title} {``lozenge''contour plots in
  scattering from polymer networks},}\ }\href@noop {} {\bibfield  {journal}
  {\bibinfo  {journal} {Phys. Rev. Lett.}\ }\textbf {\bibinfo {volume} {79}},\
  \bibinfo {pages} {87} (\bibinfo {year} {1997})}\BibitemShut {NoStop}%
\bibitem [{\citenamefont {Everaers}(1998)}]{everaers1998constrained}%
  \BibitemOpen
  \bibfield  {author} {\bibinfo {author} {\bibfnamefont {R.}~\bibnamefont
  {Everaers}},\ }\bibfield  {title} {\enquote {\bibinfo {title} {Constrained
  fluctuation theories of rubber elasticity: General results and an exactly
  solvable model},}\ }\href@noop {} {\bibfield  {journal} {\bibinfo  {journal}
  {Eur Phys. J. B.}\ }\textbf {\bibinfo {volume} {4}},\ \bibinfo {pages} {341}
  (\bibinfo {year} {1998})}\BibitemShut {NoStop}%
\bibitem [{\citenamefont {Mergell}\ and\ \citenamefont
  {Everaers}(2001)}]{mergell2001tube}%
  \BibitemOpen
  \bibfield  {author} {\bibinfo {author} {\bibfnamefont {B.}~\bibnamefont
  {Mergell}}\ and\ \bibinfo {author} {\bibfnamefont {R.}~\bibnamefont
  {Everaers}},\ }\bibfield  {title} {\enquote {\bibinfo {title} {Tube models
  for rubber- elastic systems},}\ }\href@noop {} {\bibfield  {journal}
  {\bibinfo  {journal} {Macromolecules}\ }\textbf {\bibinfo {volume} {34}},\
  \bibinfo {pages} {5675--5686} (\bibinfo {year} {2001})}\BibitemShut {NoStop}%
\bibitem [{\citenamefont {Faller}, \citenamefont {Kolb},\ and\ \citenamefont
  {M{\"u}ller-Plathe}(1999)}]{faller1999local}%
  \BibitemOpen
  \bibfield  {author} {\bibinfo {author} {\bibfnamefont {R.}~\bibnamefont
  {Faller}}, \bibinfo {author} {\bibfnamefont {A.}~\bibnamefont {Kolb}}, \ and\
  \bibinfo {author} {\bibfnamefont {F.}~\bibnamefont {M{\"u}ller-Plathe}},\
  }\bibfield  {title} {\enquote {\bibinfo {title} {Local chain ordering in
  amorphous polymer melts: Influence of chain stiffness},}\ }\href@noop {}
  {\bibfield  {journal} {\bibinfo  {journal} {Phys. Chem. Chem. Phys.}\
  }\textbf {\bibinfo {volume} {1}},\ \bibinfo {pages} {2071} (\bibinfo {year}
  {1999})}\BibitemShut {NoStop}%
\bibitem [{\citenamefont {Ngai}\ and\ \citenamefont
  {Plazek}(2007)}]{ngai2007temperature}%
  \BibitemOpen
  \bibfield  {author} {\bibinfo {author} {\bibfnamefont {K.~L.}\ \bibnamefont
  {Ngai}}\ and\ \bibinfo {author} {\bibfnamefont {D.~J.}\ \bibnamefont
  {Plazek}},\ }\bibfield  {title} {\enquote {\bibinfo {title} {Temperature
  dependences of the viscoelastic response of polymer systems},}\ }in\
  \href@noop {} {\emph {\bibinfo {booktitle} {Physical Properties of Polymers
  Handbook}}}\ (\bibinfo  {publisher} {Springer},\ \bibinfo {year} {2007})\ p.\
  \bibinfo {pages} {455}\BibitemShut {NoStop}%
\bibitem [{\citenamefont {Patlazhan}\ and\ \citenamefont
  {Remond}(2012)}]{patlazhan2012structural}%
  \BibitemOpen
  \bibfield  {author} {\bibinfo {author} {\bibfnamefont {S.}~\bibnamefont
  {Patlazhan}}\ and\ \bibinfo {author} {\bibfnamefont {Y.}~\bibnamefont
  {Remond}},\ }\bibfield  {title} {\enquote {\bibinfo {title} {Structural
  mechanics of semicrystalline polymers prior to the yield point: A review},}\
  }\href@noop {} {\bibfield  {journal} {\bibinfo  {journal} {J. Mater. Sci.}\
  }\textbf {\bibinfo {volume} {47}},\ \bibinfo {pages} {6749} (\bibinfo {year}
  {2012})}\BibitemShut {NoStop}%
\bibitem [{\citenamefont {Theodorou}\ and\ \citenamefont
  {Suter}(1986)}]{theodorou1986atomistic}%
  \BibitemOpen
  \bibfield  {author} {\bibinfo {author} {\bibfnamefont {D.~N.}\ \bibnamefont
  {Theodorou}}\ and\ \bibinfo {author} {\bibfnamefont {U.~W.}\ \bibnamefont
  {Suter}},\ }\bibfield  {title} {\enquote {\bibinfo {title} {Atomistic
  modeling of mechanical properties of polymeric glasses},}\ }\href@noop {}
  {\bibfield  {journal} {\bibinfo  {journal} {Macromolecules}\ }\textbf
  {\bibinfo {volume} {19}},\ \bibinfo {pages} {139} (\bibinfo {year}
  {1986})}\BibitemShut {NoStop}%
\bibitem [{\citenamefont {Kotelyanskii}, \citenamefont {Wagner},\ and\
  \citenamefont {Paulaitis}(1996)}]{kotelyanskii1996building}%
  \BibitemOpen
  \bibfield  {author} {\bibinfo {author} {\bibfnamefont {M.}~\bibnamefont
  {Kotelyanskii}}, \bibinfo {author} {\bibfnamefont {N.}~\bibnamefont
  {Wagner}}, \ and\ \bibinfo {author} {\bibfnamefont {M.~E.}\ \bibnamefont
  {Paulaitis}},\ }\bibfield  {title} {\enquote {\bibinfo {title} {Building
  large amorphous polymer structures: Atomistic simulation of glassy
  polystyrene},}\ }\href@noop {} {\bibfield  {journal} {\bibinfo  {journal}
  {Macromolecules}\ }\textbf {\bibinfo {volume} {29}},\ \bibinfo {pages} {8497}
  (\bibinfo {year} {1996})}\BibitemShut {NoStop}%
\bibitem [{\citenamefont {Doherty}\ \emph {et~al.}(1998)\citenamefont
  {Doherty}, \citenamefont {Holmes}, \citenamefont {Leung},\ and\ \citenamefont
  {Ross}}]{doherty1998polymerization}%
  \BibitemOpen
  \bibfield  {author} {\bibinfo {author} {\bibfnamefont {D.}~\bibnamefont
  {Doherty}}, \bibinfo {author} {\bibfnamefont {B.}~\bibnamefont {Holmes}},
  \bibinfo {author} {\bibfnamefont {P.}~\bibnamefont {Leung}}, \ and\ \bibinfo
  {author} {\bibfnamefont {R.}~\bibnamefont {Ross}},\ }\bibfield  {title}
  {\enquote {\bibinfo {title} {Polymerization molecular dynamics simulations.
  i. cross-linked atomistic models for poly (methacrylate) networks},}\
  }\href@noop {} {\bibfield  {journal} {\bibinfo  {journal} {Comput. Theor.
  Polym. S.}\ }\textbf {\bibinfo {volume} {8}},\ \bibinfo {pages} {169}
  (\bibinfo {year} {1998})}\BibitemShut {NoStop}%
\bibitem [{\citenamefont {Faller}\ \emph {et~al.}(2001)\citenamefont {Faller},
  \citenamefont {M{\"u}ller-Plathe}, \citenamefont {Doxastakis},\ and\
  \citenamefont {Theodorou}}]{faller2001local}%
  \BibitemOpen
  \bibfield  {author} {\bibinfo {author} {\bibfnamefont {R.}~\bibnamefont
  {Faller}}, \bibinfo {author} {\bibfnamefont {F.}~\bibnamefont
  {M{\"u}ller-Plathe}}, \bibinfo {author} {\bibfnamefont {M.}~\bibnamefont
  {Doxastakis}}, \ and\ \bibinfo {author} {\bibfnamefont {D.}~\bibnamefont
  {Theodorou}},\ }\bibfield  {title} {\enquote {\bibinfo {title} {Local
  structure and dynamics of trans-polyisoprene oligomers},}\ }\href@noop {}
  {\bibfield  {journal} {\bibinfo  {journal} {Macromolecules}\ }\textbf
  {\bibinfo {volume} {34}},\ \bibinfo {pages} {1436} (\bibinfo {year}
  {2001})}\BibitemShut {NoStop}%
\bibitem [{\citenamefont {Abrams}\ and\ \citenamefont
  {Kremer}(2003)}]{abrams2003combined}%
  \BibitemOpen
  \bibfield  {author} {\bibinfo {author} {\bibfnamefont {C.~F.}\ \bibnamefont
  {Abrams}}\ and\ \bibinfo {author} {\bibfnamefont {K.}~\bibnamefont
  {Kremer}},\ }\bibfield  {title} {\enquote {\bibinfo {title} {Combined
  coarse-grained and atomistic simulation of liquid bisphenol a-polycarbonate:
  Liquid packing and intramolecular structure},}\ }\href@noop {} {\bibfield
  {journal} {\bibinfo  {journal} {Macromolecules}\ }\textbf {\bibinfo {volume}
  {36}},\ \bibinfo {pages} {260} (\bibinfo {year} {2003})}\BibitemShut
  {NoStop}%
\bibitem [{\citenamefont {Milano}\ and\ \citenamefont
  {M{\"u}ller-Plathe}(2005)}]{milano2005mapping}%
  \BibitemOpen
  \bibfield  {author} {\bibinfo {author} {\bibfnamefont {G.}~\bibnamefont
  {Milano}}\ and\ \bibinfo {author} {\bibfnamefont {F.}~\bibnamefont
  {M{\"u}ller-Plathe}},\ }\bibfield  {title} {\enquote {\bibinfo {title}
  {Mapping atomistic simulations to mesoscopic models: A systematic
  coarse-graining procedure for vinyl polymer chains},}\ }\href@noop {}
  {\bibfield  {journal} {\bibinfo  {journal} {J. Phys. Chem. B.}\ }\textbf
  {\bibinfo {volume} {109}},\ \bibinfo {pages} {18609} (\bibinfo {year}
  {2005})}\BibitemShut {NoStop}%
\bibitem [{\citenamefont {Sun}\ and\ \citenamefont
  {Faller}(2005)}]{sun2005systematic}%
  \BibitemOpen
  \bibfield  {author} {\bibinfo {author} {\bibfnamefont {Q.}~\bibnamefont
  {Sun}}\ and\ \bibinfo {author} {\bibfnamefont {R.}~\bibnamefont {Faller}},\
  }\bibfield  {title} {\enquote {\bibinfo {title} {Systematic coarse-graining
  of atomistic models for simulation of polymeric systems},}\ }\href@noop {}
  {\bibfield  {journal} {\bibinfo  {journal} {Comput. Chem. Eng.}\ }\textbf
  {\bibinfo {volume} {29}},\ \bibinfo {pages} {2380} (\bibinfo {year}
  {2005})}\BibitemShut {NoStop}%
\bibitem [{\citenamefont {Tsolou}, \citenamefont {Mavrantzas},\ and\
  \citenamefont {Theodorou}(2005)}]{tsolou2005detailed}%
  \BibitemOpen
  \bibfield  {author} {\bibinfo {author} {\bibfnamefont {G.}~\bibnamefont
  {Tsolou}}, \bibinfo {author} {\bibfnamefont {V.~G.}\ \bibnamefont
  {Mavrantzas}}, \ and\ \bibinfo {author} {\bibfnamefont {D.~N.}\ \bibnamefont
  {Theodorou}},\ }\bibfield  {title} {\enquote {\bibinfo {title} {Detailed
  atomistic molecular dynamics simulation of cis-1, 4-poly (butadiene)},}\
  }\href@noop {} {\bibfield  {journal} {\bibinfo  {journal} {Macromolecules}\
  }\textbf {\bibinfo {volume} {38}},\ \bibinfo {pages} {1478} (\bibinfo {year}
  {2005})}\BibitemShut {NoStop}%
\bibitem [{\citenamefont {Tzoumanekas}\ and\ \citenamefont
  {Theodorou}(2006{\natexlab{a}})}]{tzoumanekas2006atomistic}%
  \BibitemOpen
  \bibfield  {author} {\bibinfo {author} {\bibfnamefont {C.}~\bibnamefont
  {Tzoumanekas}}\ and\ \bibinfo {author} {\bibfnamefont {D.~N.}\ \bibnamefont
  {Theodorou}},\ }\bibfield  {title} {\enquote {\bibinfo {title} {From
  atomistic simulations to slip-link models of entangled polymer melts:
  Hierarchical strategies for the prediction of rheological properties},}\
  }\href@noop {} {\bibfield  {journal} {\bibinfo  {journal} {Curr. Opin. Solid
  State Mater. Sci.}\ }\textbf {\bibinfo {volume} {10}},\ \bibinfo {pages} {61}
  (\bibinfo {year} {2006}{\natexlab{a}})}\BibitemShut {NoStop}%
\bibitem [{\citenamefont {Harmandaris}\ \emph {et~al.}(2006)\citenamefont
  {Harmandaris}, \citenamefont {Adhikari}, \citenamefont {van~der Vegt},\ and\
  \citenamefont {Kremer}}]{harmandaris2006hierarchical}%
  \BibitemOpen
  \bibfield  {author} {\bibinfo {author} {\bibfnamefont {V.}~\bibnamefont
  {Harmandaris}}, \bibinfo {author} {\bibfnamefont {N.}~\bibnamefont
  {Adhikari}}, \bibinfo {author} {\bibfnamefont {N.~F.}\ \bibnamefont {van~der
  Vegt}}, \ and\ \bibinfo {author} {\bibfnamefont {K.}~\bibnamefont {Kremer}},\
  }\bibfield  {title} {\enquote {\bibinfo {title} {Hierarchical modeling of
  polystyrene: From atomistic to coarse-grained simulations},}\ }\href@noop {}
  {\bibfield  {journal} {\bibinfo  {journal} {Macromolecules}\ }\textbf
  {\bibinfo {volume} {39}},\ \bibinfo {pages} {6708} (\bibinfo {year}
  {2006})}\BibitemShut {NoStop}%
\bibitem [{\citenamefont {Neyertz}\ and\ \citenamefont
  {Brown}(2008)}]{neyertz2008molecular}%
  \BibitemOpen
  \bibfield  {author} {\bibinfo {author} {\bibfnamefont {S.}~\bibnamefont
  {Neyertz}}\ and\ \bibinfo {author} {\bibfnamefont {D.}~\bibnamefont
  {Brown}},\ }\bibfield  {title} {\enquote {\bibinfo {title} {Molecular
  dynamics simulations of oxygen transport through a fully atomistic polyimide
  membrane},}\ }\href@noop {} {\bibfield  {journal} {\bibinfo  {journal}
  {Macromolecules}\ }\textbf {\bibinfo {volume} {41}},\ \bibinfo {pages} {2711}
  (\bibinfo {year} {2008})}\BibitemShut {NoStop}%
\bibitem [{\citenamefont {Padding}\ and\ \citenamefont
  {Briels}(2002)}]{padding2002time}%
  \BibitemOpen
  \bibfield  {author} {\bibinfo {author} {\bibfnamefont {J.}~\bibnamefont
  {Padding}}\ and\ \bibinfo {author} {\bibfnamefont {W.}~\bibnamefont
  {Briels}},\ }\bibfield  {title} {\enquote {\bibinfo {title} {Time and length
  scales of polymer melts studied by coarse-grained molecular dynamics
  simulations},}\ }\href@noop {} {\bibfield  {journal} {\bibinfo  {journal} {J.
  Chem. Phys.}\ }\textbf {\bibinfo {volume} {117}},\ \bibinfo {pages} {925}
  (\bibinfo {year} {2002})}\BibitemShut {NoStop}%
\bibitem [{\citenamefont {Liu}\ \emph {et~al.}(2013)\citenamefont {Liu},
  \citenamefont {Padding}, \citenamefont {den Otter},\ and\ \citenamefont
  {Briels}}]{liu2013coarse}%
  \BibitemOpen
  \bibfield  {author} {\bibinfo {author} {\bibfnamefont {L.}~\bibnamefont
  {Liu}}, \bibinfo {author} {\bibfnamefont {J.}~\bibnamefont {Padding}},
  \bibinfo {author} {\bibfnamefont {W.}~\bibnamefont {den Otter}}, \ and\
  \bibinfo {author} {\bibfnamefont {W.}~\bibnamefont {Briels}},\ }\bibfield
  {title} {\enquote {\bibinfo {title} {Coarse-grained simulations of moderately
  entangled star polyethylene melts},}\ }\href@noop {} {\bibfield  {journal}
  {\bibinfo  {journal} {J. Chem. Phys.}\ }\textbf {\bibinfo {volume} {138}},\
  \bibinfo {pages} {244912} (\bibinfo {year} {2013})}\BibitemShut {NoStop}%
\bibitem [{\citenamefont {Salerno}\ \emph {et~al.}(2016)\citenamefont
  {Salerno}, \citenamefont {Agrawal}, \citenamefont {Perahia},\ and\
  \citenamefont {Grest}}]{salerno2016resolving}%
  \BibitemOpen
  \bibfield  {author} {\bibinfo {author} {\bibfnamefont {K.~M.}\ \bibnamefont
  {Salerno}}, \bibinfo {author} {\bibfnamefont {A.}~\bibnamefont {Agrawal}},
  \bibinfo {author} {\bibfnamefont {D.}~\bibnamefont {Perahia}}, \ and\
  \bibinfo {author} {\bibfnamefont {G.~S.}\ \bibnamefont {Grest}},\ }\bibfield
  {title} {\enquote {\bibinfo {title} {Resolving dynamic properties of polymers
  through coarse-grained computational studies},}\ }\href@noop {} {\bibfield
  {journal} {\bibinfo  {journal} {Phys. Rev. Lett.}\ }\textbf {\bibinfo
  {volume} {116}},\ \bibinfo {pages} {058302} (\bibinfo {year}
  {2016})}\BibitemShut {NoStop}%
\bibitem [{\citenamefont {Faller}\ and\ \citenamefont
  {M{\"u}ller-Plathe}(2002)}]{faller2002modeling}%
  \BibitemOpen
  \bibfield  {author} {\bibinfo {author} {\bibfnamefont {R.}~\bibnamefont
  {Faller}}\ and\ \bibinfo {author} {\bibfnamefont {F.}~\bibnamefont
  {M{\"u}ller-Plathe}},\ }\bibfield  {title} {\enquote {\bibinfo {title}
  {Modeling of poly (isoprene) melts on different scales},}\ }\href@noop {}
  {\bibfield  {journal} {\bibinfo  {journal} {Polymer}\ }\textbf {\bibinfo
  {volume} {43}},\ \bibinfo {pages} {621} (\bibinfo {year} {2002})}\BibitemShut
  {NoStop}%
\bibitem [{\citenamefont {Faller}\ and\ \citenamefont
  {Reith}(2003)}]{faller2003properties}%
  \BibitemOpen
  \bibfield  {author} {\bibinfo {author} {\bibfnamefont {R.}~\bibnamefont
  {Faller}}\ and\ \bibinfo {author} {\bibfnamefont {D.}~\bibnamefont {Reith}},\
  }\bibfield  {title} {\enquote {\bibinfo {title} {Properties of poly
  (isoprene): Model building in the melt and in solution},}\ }\href@noop {}
  {\bibfield  {journal} {\bibinfo  {journal} {Macromolecules}\ }\textbf
  {\bibinfo {volume} {36}},\ \bibinfo {pages} {5406} (\bibinfo {year}
  {2003})}\BibitemShut {NoStop}%
\bibitem [{\citenamefont {Li}, \citenamefont {Kr{\"o}ger},\ and\ \citenamefont
  {Liu}(2011)}]{li2011primitive}%
  \BibitemOpen
  \bibfield  {author} {\bibinfo {author} {\bibfnamefont {Y.}~\bibnamefont
  {Li}}, \bibinfo {author} {\bibfnamefont {M.}~\bibnamefont {Kr{\"o}ger}}, \
  and\ \bibinfo {author} {\bibfnamefont {W.~K.}\ \bibnamefont {Liu}},\
  }\bibfield  {title} {\enquote {\bibinfo {title} {Primitive chain network
  study on uncrosslinked and crosslinked cis-polyisoprene polymers},}\
  }\href@noop {} {\bibfield  {journal} {\bibinfo  {journal} {Polymer}\ }\textbf
  {\bibinfo {volume} {52}},\ \bibinfo {pages} {5867} (\bibinfo {year}
  {2011})}\BibitemShut {NoStop}%
\bibitem [{\citenamefont {Maurel}\ \emph {et~al.}(2015)\citenamefont {Maurel},
  \citenamefont {Goujon}, \citenamefont {Schnell},\ and\ \citenamefont
  {Malfreyt}}]{maurel2015prediction}%
  \BibitemOpen
  \bibfield  {author} {\bibinfo {author} {\bibfnamefont {G.}~\bibnamefont
  {Maurel}}, \bibinfo {author} {\bibfnamefont {F.}~\bibnamefont {Goujon}},
  \bibinfo {author} {\bibfnamefont {B.}~\bibnamefont {Schnell}}, \ and\
  \bibinfo {author} {\bibfnamefont {P.}~\bibnamefont {Malfreyt}},\ }\bibfield
  {title} {\enquote {\bibinfo {title} {Prediction of structural and
  thermomechanical properties of polymers from multiscale simulations},}\
  }\href@noop {} {\bibfield  {journal} {\bibinfo  {journal} {RSC Adv.}\
  }\textbf {\bibinfo {volume} {5}},\ \bibinfo {pages} {14065} (\bibinfo {year}
  {2015})}\BibitemShut {NoStop}%
\bibitem [{\citenamefont {Chen}\ \emph {et~al.}(2007)\citenamefont {Chen},
  \citenamefont {Carbone}, \citenamefont {Cavalcanti}, \citenamefont {Milano},\
  and\ \citenamefont {M{\"u}ller-Plathe}}]{chen2007viscosity}%
  \BibitemOpen
  \bibfield  {author} {\bibinfo {author} {\bibfnamefont {X.}~\bibnamefont
  {Chen}}, \bibinfo {author} {\bibfnamefont {P.}~\bibnamefont {Carbone}},
  \bibinfo {author} {\bibfnamefont {W.~L.}\ \bibnamefont {Cavalcanti}},
  \bibinfo {author} {\bibfnamefont {G.}~\bibnamefont {Milano}}, \ and\ \bibinfo
  {author} {\bibfnamefont {F.}~\bibnamefont {M{\"u}ller-Plathe}},\ }\bibfield
  {title} {\enquote {\bibinfo {title} {Viscosity and structural alteration of a
  coarse-grained model of polystyrene under steady shear flow studied by
  reverse nonequilibrium molecular dynamics},}\ }\href@noop {} {\bibfield
  {journal} {\bibinfo  {journal} {Macromolecules}\ }\textbf {\bibinfo {volume}
  {40}},\ \bibinfo {pages} {8087} (\bibinfo {year} {2007})}\BibitemShut
  {NoStop}%
\bibitem [{\citenamefont {Fritz}\ \emph {et~al.}(2011)\citenamefont {Fritz},
  \citenamefont {Koschke}, \citenamefont {Harmandaris}, \citenamefont {van~der
  Vegt},\ and\ \citenamefont {Kremer}}]{fritz2011multiscale}%
  \BibitemOpen
  \bibfield  {author} {\bibinfo {author} {\bibfnamefont {D.}~\bibnamefont
  {Fritz}}, \bibinfo {author} {\bibfnamefont {K.}~\bibnamefont {Koschke}},
  \bibinfo {author} {\bibfnamefont {V.~A.}\ \bibnamefont {Harmandaris}},
  \bibinfo {author} {\bibfnamefont {N.~F.}\ \bibnamefont {van~der Vegt}}, \
  and\ \bibinfo {author} {\bibfnamefont {K.}~\bibnamefont {Kremer}},\
  }\bibfield  {title} {\enquote {\bibinfo {title} {Multiscale modeling of soft
  matter: Scaling of dynamics},}\ }\href@noop {} {\bibfield  {journal}
  {\bibinfo  {journal} {Phys. Chem. Chem. Phys.}\ }\textbf {\bibinfo {volume}
  {13}},\ \bibinfo {pages} {10412} (\bibinfo {year} {2011})}\BibitemShut
  {NoStop}%
\bibitem [{\citenamefont {Karimi-Varzaneh}, \citenamefont {Carbone},\ and\
  \citenamefont {M{\"u}ller-Plathe}(2008)}]{karimi2008fast}%
  \BibitemOpen
  \bibfield  {author} {\bibinfo {author} {\bibfnamefont {H.~A.}\ \bibnamefont
  {Karimi-Varzaneh}}, \bibinfo {author} {\bibfnamefont {P.}~\bibnamefont
  {Carbone}}, \ and\ \bibinfo {author} {\bibfnamefont {F.}~\bibnamefont
  {M{\"u}ller-Plathe}},\ }\bibfield  {title} {\enquote {\bibinfo {title} {Fast
  dynamics in coarse-grained polymer models: The effect of the hydrogen
  bonds},}\ }\href@noop {} {\bibfield  {journal} {\bibinfo  {journal} {J. Chem.
  Phys.}\ }\textbf {\bibinfo {volume} {129}},\ \bibinfo {pages} {154904}
  (\bibinfo {year} {2008})}\BibitemShut {NoStop}%
\bibitem [{\citenamefont {Eslami}, \citenamefont {Karimi-Varzaneh},\ and\
  \citenamefont {M{\"u}ller-Plathe}(2011)}]{eslami2011coarse}%
  \BibitemOpen
  \bibfield  {author} {\bibinfo {author} {\bibfnamefont {H.}~\bibnamefont
  {Eslami}}, \bibinfo {author} {\bibfnamefont {H.~A.}\ \bibnamefont
  {Karimi-Varzaneh}}, \ and\ \bibinfo {author} {\bibfnamefont {F.}~\bibnamefont
  {M{\"u}ller-Plathe}},\ }\bibfield  {title} {\enquote {\bibinfo {title}
  {Coarse-grained computer simulation of nanoconfined polyamide-6, 6},}\
  }\href@noop {} {\bibfield  {journal} {\bibinfo  {journal} {Macromolecules}\
  }\textbf {\bibinfo {volume} {44}},\ \bibinfo {pages} {3117} (\bibinfo {year}
  {2011})}\BibitemShut {NoStop}%
\bibitem [{\citenamefont {Chen}\ \emph {et~al.}(2008)\citenamefont {Chen},
  \citenamefont {Depa}, \citenamefont {Maranas},\ and\ \citenamefont
  {Sakai}}]{chen2008comparison}%
  \BibitemOpen
  \bibfield  {author} {\bibinfo {author} {\bibfnamefont {C.}~\bibnamefont
  {Chen}}, \bibinfo {author} {\bibfnamefont {P.}~\bibnamefont {Depa}}, \bibinfo
  {author} {\bibfnamefont {J.~K.}\ \bibnamefont {Maranas}}, \ and\ \bibinfo
  {author} {\bibfnamefont {V.~G.}\ \bibnamefont {Sakai}},\ }\bibfield  {title}
  {\enquote {\bibinfo {title} {Comparison of explicit atom, united atom, and
  coarse-grained simulations of poly (methyl methacrylate)},}\ }\href@noop {}
  {\bibfield  {journal} {\bibinfo  {journal} {J. Chem. Phys.}\ }\textbf
  {\bibinfo {volume} {128}},\ \bibinfo {pages} {124906} (\bibinfo {year}
  {2008})}\BibitemShut {NoStop}%
\bibitem [{\citenamefont {Tsch{\"o}p}\ \emph {et~al.}(1998)\citenamefont
  {Tsch{\"o}p}, \citenamefont {Kremer}, \citenamefont {Batoulis}, \citenamefont
  {B{\"u}rger},\ and\ \citenamefont {Hahn}}]{tschop1998simulation}%
  \BibitemOpen
  \bibfield  {author} {\bibinfo {author} {\bibfnamefont {W.}~\bibnamefont
  {Tsch{\"o}p}}, \bibinfo {author} {\bibfnamefont {K.}~\bibnamefont {Kremer}},
  \bibinfo {author} {\bibfnamefont {J.}~\bibnamefont {Batoulis}}, \bibinfo
  {author} {\bibfnamefont {T.}~\bibnamefont {B{\"u}rger}}, \ and\ \bibinfo
  {author} {\bibfnamefont {O.}~\bibnamefont {Hahn}},\ }\bibfield  {title}
  {\enquote {\bibinfo {title} {Simulation of polymer melts. i. coarse-graining
  procedure for polycarbonates},}\ }\href@noop {} {\bibfield  {journal}
  {\bibinfo  {journal} {Acta. Polym.}\ }\textbf {\bibinfo {volume} {49}},\
  \bibinfo {pages} {61} (\bibinfo {year} {1998})}\BibitemShut {NoStop}%
\bibitem [{\citenamefont {Hess}\ \emph {et~al.}(2006)\citenamefont {Hess},
  \citenamefont {Le{\'o}n}, \citenamefont {Van Der~Vegt},\ and\ \citenamefont
  {Kremer}}]{hess2006long}%
  \BibitemOpen
  \bibfield  {author} {\bibinfo {author} {\bibfnamefont {B.}~\bibnamefont
  {Hess}}, \bibinfo {author} {\bibfnamefont {S.}~\bibnamefont {Le{\'o}n}},
  \bibinfo {author} {\bibfnamefont {N.}~\bibnamefont {Van Der~Vegt}}, \ and\
  \bibinfo {author} {\bibfnamefont {K.}~\bibnamefont {Kremer}},\ }\bibfield
  {title} {\enquote {\bibinfo {title} {Long time atomistic polymer trajectories
  from coarse grained simulations: bisphenol-a polycarbonate},}\ }\href@noop {}
  {\bibfield  {journal} {\bibinfo  {journal} {Soft Matter}\ }\textbf {\bibinfo
  {volume} {2}},\ \bibinfo {pages} {409} (\bibinfo {year} {2006})}\BibitemShut
  {NoStop}%
\bibitem [{\citenamefont {Strauch}, \citenamefont {Yelash},\ and\ \citenamefont
  {Paul}(2009)}]{strauch2009coarse}%
  \BibitemOpen
  \bibfield  {author} {\bibinfo {author} {\bibfnamefont {T.}~\bibnamefont
  {Strauch}}, \bibinfo {author} {\bibfnamefont {L.}~\bibnamefont {Yelash}}, \
  and\ \bibinfo {author} {\bibfnamefont {W.}~\bibnamefont {Paul}},\ }\bibfield
  {title} {\enquote {\bibinfo {title} {A coarse-graining procedure for polymer
  melts applied to 1, 4-polybutadiene},}\ }\href@noop {} {\bibfield  {journal}
  {\bibinfo  {journal} {Phys. Chem. Chem. Phys.}\ }\textbf {\bibinfo {volume}
  {11}},\ \bibinfo {pages} {1942} (\bibinfo {year} {2009})}\BibitemShut
  {NoStop}%
\bibitem [{\citenamefont {Baschnagel}\ \emph {et~al.}(2000)\citenamefont
  {Baschnagel}, \citenamefont {Binder}, \citenamefont {Doruker}, \citenamefont
  {Gusev}, \citenamefont {Hahn}, \citenamefont {Kremer}, \citenamefont
  {Mattice}, \citenamefont {M{\"u}ller-Plathe}, \citenamefont {Murat},
  \citenamefont {Paul}, \citenamefont {Santos}, \citenamefont {Suter},\ and\
  \citenamefont {Tries}}]{baschnagel2000bridging}%
  \BibitemOpen
  \bibfield  {author} {\bibinfo {author} {\bibfnamefont {J.}~\bibnamefont
  {Baschnagel}}, \bibinfo {author} {\bibfnamefont {K.}~\bibnamefont {Binder}},
  \bibinfo {author} {\bibfnamefont {P.}~\bibnamefont {Doruker}}, \bibinfo
  {author} {\bibfnamefont {A.~A.}\ \bibnamefont {Gusev}}, \bibinfo {author}
  {\bibfnamefont {O.}~\bibnamefont {Hahn}}, \bibinfo {author} {\bibfnamefont
  {K.}~\bibnamefont {Kremer}}, \bibinfo {author} {\bibfnamefont {W.~L.}\
  \bibnamefont {Mattice}}, \bibinfo {author} {\bibfnamefont {F.}~\bibnamefont
  {M{\"u}ller-Plathe}}, \bibinfo {author} {\bibfnamefont {M.}~\bibnamefont
  {Murat}}, \bibinfo {author} {\bibfnamefont {W.}~\bibnamefont {Paul}},
  \bibinfo {author} {\bibfnamefont {S.}~\bibnamefont {Santos}}, \bibinfo
  {author} {\bibfnamefont {U.~W.}\ \bibnamefont {Suter}}, \ and\ \bibinfo
  {author} {\bibfnamefont {V.}~\bibnamefont {Tries}},\ }\bibfield  {title}
  {\enquote {\bibinfo {title} {Bridging the gap between atomistic and
  coarse-grained models of polymers: Status and perspectives},}\ }in\ \href
  {\doibase 10.1007/3-540-46778-5_2} {\emph {\bibinfo {booktitle}
  {Viscoelasticity, Atomistic Models, Statistical Chemistry}}}\ (\bibinfo
  {publisher} {Springer Berlin Heidelberg},\ \bibinfo {address} {Berlin,
  Heidelberg},\ \bibinfo {year} {2000})\ pp.\ \bibinfo {pages}
  {41--156}\BibitemShut {NoStop}%
\bibitem [{\citenamefont {Carbone}\ \emph {et~al.}(2008)\citenamefont
  {Carbone}, \citenamefont {Varzaneh}, \citenamefont {Chen},\ and\
  \citenamefont {M{\"u}ller-Plathe}}]{carbone2008transferability}%
  \BibitemOpen
  \bibfield  {author} {\bibinfo {author} {\bibfnamefont {P.}~\bibnamefont
  {Carbone}}, \bibinfo {author} {\bibfnamefont {H.~A.~K.}\ \bibnamefont
  {Varzaneh}}, \bibinfo {author} {\bibfnamefont {X.}~\bibnamefont {Chen}}, \
  and\ \bibinfo {author} {\bibfnamefont {F.}~\bibnamefont
  {M{\"u}ller-Plathe}},\ }\bibfield  {title} {\enquote {\bibinfo {title}
  {Transferability of coarse-grained force fields: The polymer case},}\
  }\href@noop {} {\bibfield  {journal} {\bibinfo  {journal} {J. Chem. Phys.}\
  }\textbf {\bibinfo {volume} {128}},\ \bibinfo {pages} {064904} (\bibinfo
  {year} {2008})}\BibitemShut {NoStop}%
\bibitem [{\citenamefont {Peter}\ and\ \citenamefont
  {Kremer}(2009)}]{MultiscalePeterKremerSOftMatter2009}%
  \BibitemOpen
  \bibfield  {author} {\bibinfo {author} {\bibfnamefont {C.}~\bibnamefont
  {Peter}}\ and\ \bibinfo {author} {\bibfnamefont {K.}~\bibnamefont {Kremer}},\
  }\bibfield  {title} {\enquote {\bibinfo {title} {Multiscale simulation of
  soft matter systems - from the atomistic to the coarse-grained level and
  back},}\ }\href@noop {} {\bibfield  {journal} {\bibinfo  {journal} {Soft
  Matter}\ }\textbf {\bibinfo {volume} {5}},\ \bibinfo {pages} {4357} (\bibinfo
  {year} {2009})}\BibitemShut {NoStop}%
\bibitem [{\citenamefont {Johnson}, \citenamefont {Head-Gordon},\ and\
  \citenamefont {Louis}(2007)}]{johnson2007representability}%
  \BibitemOpen
  \bibfield  {author} {\bibinfo {author} {\bibfnamefont {M.~E.}\ \bibnamefont
  {Johnson}}, \bibinfo {author} {\bibfnamefont {T.}~\bibnamefont
  {Head-Gordon}}, \ and\ \bibinfo {author} {\bibfnamefont {A.~A.}\ \bibnamefont
  {Louis}},\ }\bibfield  {title} {\enquote {\bibinfo {title} {Representability
  problems for coarse-grained water potentials},}\ }\href@noop {} {\bibfield
  {journal} {\bibinfo  {journal} {J. Chem. Phys.}\ }\textbf {\bibinfo {volume}
  {126}},\ \bibinfo {pages} {144509} (\bibinfo {year} {2007})}\BibitemShut
  {NoStop}%
\bibitem [{\citenamefont {de~Gennes}(1979{\natexlab{a}})}]{deGennes79}%
  \BibitemOpen
  \bibfield  {author} {\bibinfo {author} {\bibfnamefont {P.~G.}\ \bibnamefont
  {de~Gennes}},\ }\href@noop {} {\emph {\bibinfo {title} {Scaling Concepts in
  Polymer Physics}}}\ (\bibinfo  {publisher} {Cornell University Press},\
  \bibinfo {address} {Ithaca NY},\ \bibinfo {year} {1979})\BibitemShut
  {NoStop}%
\bibitem [{\citenamefont {Grest}\ and\ \citenamefont
  {Kremer}(1986)}]{grest1986molecular}%
  \BibitemOpen
  \bibfield  {author} {\bibinfo {author} {\bibfnamefont {G.~S.}\ \bibnamefont
  {Grest}}\ and\ \bibinfo {author} {\bibfnamefont {K.}~\bibnamefont {Kremer}},\
  }\bibfield  {title} {\enquote {\bibinfo {title} {Molecular dynamics
  simulation for polymers in the presence of a heat bath},}\ }\href@noop {}
  {\bibfield  {journal} {\bibinfo  {journal} {Phys. Rev. A}\ }\textbf {\bibinfo
  {volume} {33}},\ \bibinfo {pages} {3628} (\bibinfo {year}
  {1986})}\BibitemShut {NoStop}%
\bibitem [{\citenamefont {Faller}, \citenamefont {M{\"u}ller-Plathe},\ and\
  \citenamefont {Heuer}(2000)}]{faller2000local}%
  \BibitemOpen
  \bibfield  {author} {\bibinfo {author} {\bibfnamefont {R.}~\bibnamefont
  {Faller}}, \bibinfo {author} {\bibfnamefont {F.}~\bibnamefont
  {M{\"u}ller-Plathe}}, \ and\ \bibinfo {author} {\bibfnamefont
  {A.}~\bibnamefont {Heuer}},\ }\bibfield  {title} {\enquote {\bibinfo {title}
  {Local reorientation dynamics of semiflexible polymers in the melt},}\
  }\href@noop {} {\bibfield  {journal} {\bibinfo  {journal} {Macromolecules}\
  }\textbf {\bibinfo {volume} {33}},\ \bibinfo {pages} {6602} (\bibinfo {year}
  {2000})}\BibitemShut {NoStop}%
\bibitem [{\citenamefont {Faller}\ and\ \citenamefont
  {M{\"u}ller-Plathe}(2001)}]{faller2001chain}%
  \BibitemOpen
  \bibfield  {author} {\bibinfo {author} {\bibfnamefont {R.}~\bibnamefont
  {Faller}}\ and\ \bibinfo {author} {\bibfnamefont {F.}~\bibnamefont
  {M{\"u}ller-Plathe}},\ }\bibfield  {title} {\enquote {\bibinfo {title} {Chain
  stiffness intensifies the reptation characteristics of polymer dynamics in
  the melt},}\ }\href@noop {} {\bibfield  {journal} {\bibinfo  {journal} {Chem.
  Phys. Chem.}\ }\textbf {\bibinfo {volume} {2}},\ \bibinfo {pages} {180}
  (\bibinfo {year} {2001})}\BibitemShut {NoStop}%
\bibitem [{\citenamefont {Kremer}\ and\ \citenamefont
  {Grest}(1990{\natexlab{b}})}]{kremer1990dynamics}%
  \BibitemOpen
  \bibfield  {author} {\bibinfo {author} {\bibfnamefont {K.}~\bibnamefont
  {Kremer}}\ and\ \bibinfo {author} {\bibfnamefont {G.~S.}\ \bibnamefont
  {Grest}},\ }\bibfield  {title} {\enquote {\bibinfo {title} {Dynamics of
  entangled linear polymer melts: A molecular-dynamics simulation},}\
  }\href@noop {} {\bibfield  {journal} {\bibinfo  {journal} {J. Chem. Phys.}\
  }\textbf {\bibinfo {volume} {92}},\ \bibinfo {pages} {5057} (\bibinfo {year}
  {1990}{\natexlab{b}})}\BibitemShut {NoStop}%
\bibitem [{\citenamefont {Hou}\ \emph {et~al.}(2010)\citenamefont {Hou},
  \citenamefont {Svaneborg}, \citenamefont {Everaers},\ and\ \citenamefont
  {Grest}}]{hou2010stress}%
  \BibitemOpen
  \bibfield  {author} {\bibinfo {author} {\bibfnamefont {J.-X.}\ \bibnamefont
  {Hou}}, \bibinfo {author} {\bibfnamefont {C.}~\bibnamefont {Svaneborg}},
  \bibinfo {author} {\bibfnamefont {R.}~\bibnamefont {Everaers}}, \ and\
  \bibinfo {author} {\bibfnamefont {G.~S.}\ \bibnamefont {Grest}},\ }\bibfield
  {title} {\enquote {\bibinfo {title} {Stress relaxation in entangled polymer
  melts},}\ }\href@noop {} {\bibfield  {journal} {\bibinfo  {journal} {Phys.
  Rev. Lett.}\ }\textbf {\bibinfo {volume} {105}},\ \bibinfo {pages} {068301}
  (\bibinfo {year} {2010})}\BibitemShut {NoStop}%
\bibitem [{\citenamefont {Svaneborg}\ and\ \citenamefont
  {Everaers}(2020)}]{Svaneborg2020Characterization}%
  \BibitemOpen
  \bibfield  {author} {\bibinfo {author} {\bibfnamefont {C.}~\bibnamefont
  {Svaneborg}}\ and\ \bibinfo {author} {\bibfnamefont {R.}~\bibnamefont
  {Everaers}},\ }\bibfield  {title} {\enquote {\bibinfo {title} {Characteristic
  time and length scales in melts of kremer-grest bead-spring polymers with
  wormlike bending stiffness},}\ }\href {\doibase
  https://doi.org/10.1021/acs.macromol.9b02437} {\bibfield  {journal} {\bibinfo
   {journal} {Macromolecules}\ }\textbf {\bibinfo {volume} {53}},\ \bibinfo
  {pages} {1917} (\bibinfo {year} {2020})}\BibitemShut {NoStop}%
\bibitem [{\citenamefont {Grest}\ and\ \citenamefont
  {Kremer}(1990)}]{grest90a}%
  \BibitemOpen
  \bibfield  {author} {\bibinfo {author} {\bibfnamefont {G.~S.}\ \bibnamefont
  {Grest}}\ and\ \bibinfo {author} {\bibfnamefont {K.}~\bibnamefont {Kremer}},\
  }\bibfield  {title} {\enquote {\bibinfo {title} {Statistical properties of
  random cross-linked rubbers},}\ }\href@noop {} {\bibfield  {journal}
  {\bibinfo  {journal} {Macromolecules}\ }\textbf {\bibinfo {volume} {23}},\
  \bibinfo {pages} {4994} (\bibinfo {year} {1990})}\BibitemShut {NoStop}%
\bibitem [{\citenamefont {Duering}, \citenamefont {Kremer},\ and\ \citenamefont
  {Grest}(1994)}]{duering94}%
  \BibitemOpen
  \bibfield  {author} {\bibinfo {author} {\bibfnamefont {E.~R.}\ \bibnamefont
  {Duering}}, \bibinfo {author} {\bibfnamefont {K.}~\bibnamefont {Kremer}}, \
  and\ \bibinfo {author} {\bibfnamefont {G.~S.}\ \bibnamefont {Grest}},\
  }\bibfield  {title} {\enquote {\bibinfo {title} {Structure and relaxation of
  end-linked polymer networks},}\ }\href@noop {} {\bibfield  {journal}
  {\bibinfo  {journal} {J. Chem. Phys.}\ }\textbf {\bibinfo {volume} {101}},\
  \bibinfo {pages} {8169} (\bibinfo {year} {1994})}\BibitemShut {NoStop}%
\bibitem [{\citenamefont {Svaneborg}, \citenamefont {Grest},\ and\
  \citenamefont {Everaers}(2004)}]{SGE_prl_04}%
  \BibitemOpen
  \bibfield  {author} {\bibinfo {author} {\bibfnamefont {C.}~\bibnamefont
  {Svaneborg}}, \bibinfo {author} {\bibfnamefont {G.~S.}\ \bibnamefont
  {Grest}}, \ and\ \bibinfo {author} {\bibfnamefont {R.}~\bibnamefont
  {Everaers}},\ }\bibfield  {title} {\enquote {\bibinfo {title}
  {Strain-dependent localization, microscopic deformations, and macroscopic
  normal tensions in model polymer networks},}\ }\href@noop {} {\bibfield
  {journal} {\bibinfo  {journal} {Phys. Rev. Lett.}\ }\textbf {\bibinfo
  {volume} {93}},\ \bibinfo {pages} {257801} (\bibinfo {year}
  {2004})}\BibitemShut {NoStop}%
\bibitem [{\citenamefont {Svaneborg}\ \emph {et~al.}(2008)\citenamefont
  {Svaneborg}, \citenamefont {Everaers}, \citenamefont {Grest},\ and\
  \citenamefont {Curro}}]{svaneborg2008connectivity}%
  \BibitemOpen
  \bibfield  {author} {\bibinfo {author} {\bibfnamefont {C.}~\bibnamefont
  {Svaneborg}}, \bibinfo {author} {\bibfnamefont {R.}~\bibnamefont {Everaers}},
  \bibinfo {author} {\bibfnamefont {G.~S.}\ \bibnamefont {Grest}}, \ and\
  \bibinfo {author} {\bibfnamefont {J.~G.}\ \bibnamefont {Curro}},\ }\bibfield
  {title} {\enquote {\bibinfo {title} {Connectivity and entanglement stress
  contributions in strained polymer networks},}\ }\href@noop {} {\bibfield
  {journal} {\bibinfo  {journal} {Macromolecules}\ }\textbf {\bibinfo {volume}
  {41}},\ \bibinfo {pages} {4920} (\bibinfo {year} {2008})}\BibitemShut
  {NoStop}%
\bibitem [{\citenamefont {Gula}, \citenamefont {Karimi-Varzaneh},\ and\
  \citenamefont {Svaneborg}(2020)}]{Gula2020Entanglement}%
  \BibitemOpen
  \bibfield  {author} {\bibinfo {author} {\bibfnamefont {I.~A.}\ \bibnamefont
  {Gula}}, \bibinfo {author} {\bibfnamefont {H.~A.}\ \bibnamefont
  {Karimi-Varzaneh}}, \ and\ \bibinfo {author} {\bibfnamefont {C.}~\bibnamefont
  {Svaneborg}},\ }\bibfield  {title} {\enquote {\bibinfo {title} {Computational
  study of cross-link and entanglement contributions to the elastic properties
  of model pdms networks},}\ }\href {\doibase
  https://doi.org/10.1021/acs.macromol.0c00682} {\bibfield  {journal} {\bibinfo
   {journal} {Macromolecules}\ }\textbf {\bibinfo {volume} {53}},\ \bibinfo
  {pages} {6907} (\bibinfo {year} {2020})}\BibitemShut {NoStop}%
\bibitem [{\citenamefont {Everaers}\ and\ \citenamefont
  {Kremer}(1996)}]{everaers1996elastic}%
  \BibitemOpen
  \bibfield  {author} {\bibinfo {author} {\bibfnamefont {R.}~\bibnamefont
  {Everaers}}\ and\ \bibinfo {author} {\bibfnamefont {K.}~\bibnamefont
  {Kremer}},\ }\bibfield  {title} {\enquote {\bibinfo {title} {Elastic
  properties of polymer networks},}\ }in\ \href@noop {} {\emph {\bibinfo
  {booktitle} {J. Mol. Model.}}},\ Vol.~\bibinfo {volume} {2}\ (\bibinfo
  {organization} {Springer},\ \bibinfo {year} {1996})\ pp.\ \bibinfo {pages}
  {293--299}\BibitemShut {NoStop}%
\bibitem [{\citenamefont {Everaers}(1999)}]{everaers1999entanglement}%
  \BibitemOpen
  \bibfield  {author} {\bibinfo {author} {\bibfnamefont {R.}~\bibnamefont
  {Everaers}},\ }\bibfield  {title} {\enquote {\bibinfo {title} {Entanglement
  effects in defect-free model polymer networks},}\ }\href@noop {} {\bibfield
  {journal} {\bibinfo  {journal} {New J. Phys.}\ }\textbf {\bibinfo {volume}
  {1}},\ \bibinfo {pages} {12} (\bibinfo {year} {1999})}\BibitemShut {NoStop}%
\bibitem [{\citenamefont {Cao}\ and\ \citenamefont
  {Likhtman}(2015)}]{cao2015simulating}%
  \BibitemOpen
  \bibfield  {author} {\bibinfo {author} {\bibfnamefont {J.}~\bibnamefont
  {Cao}}\ and\ \bibinfo {author} {\bibfnamefont {A.~E.}\ \bibnamefont
  {Likhtman}},\ }\bibfield  {title} {\enquote {\bibinfo {title} {Simulating
  startup shear of entangled polymer melts},}\ }\href@noop {} {\bibfield
  {journal} {\bibinfo  {journal} {ACS Macro Lett.}\ }\textbf {\bibinfo {volume}
  {4}},\ \bibinfo {pages} {1376} (\bibinfo {year} {2015})}\BibitemShut
  {NoStop}%
\bibitem [{\citenamefont {O’Connor}, \citenamefont {Alvarez},\ and\
  \citenamefont {Robbins}(2018)}]{oconnor2018relating}%
  \BibitemOpen
  \bibfield  {author} {\bibinfo {author} {\bibfnamefont {T.~C.}\ \bibnamefont
  {O’Connor}}, \bibinfo {author} {\bibfnamefont {N.~J.}\ \bibnamefont
  {Alvarez}}, \ and\ \bibinfo {author} {\bibfnamefont {M.~O.}\ \bibnamefont
  {Robbins}},\ }\bibfield  {title} {\enquote {\bibinfo {title} {Relating chain
  conformations to extensional stress in entangled polymer melts},}\
  }\href@noop {} {\bibfield  {journal} {\bibinfo  {journal} {Phys. Rev. Lett.}\
  }\textbf {\bibinfo {volume} {121}},\ \bibinfo {pages} {047801} (\bibinfo
  {year} {2018})}\BibitemShut {NoStop}%
\bibitem [{\citenamefont {Xu}\ \emph {et~al.}(2018)\citenamefont {Xu},
  \citenamefont {Carrillo}, \citenamefont {Lam}, \citenamefont {Sumpter},\ and\
  \citenamefont {Wang}}]{xu2018molecular}%
  \BibitemOpen
  \bibfield  {author} {\bibinfo {author} {\bibfnamefont {W.-S.}\ \bibnamefont
  {Xu}}, \bibinfo {author} {\bibfnamefont {J.-M.~Y.}\ \bibnamefont {Carrillo}},
  \bibinfo {author} {\bibfnamefont {C.~N.}\ \bibnamefont {Lam}}, \bibinfo
  {author} {\bibfnamefont {B.~G.}\ \bibnamefont {Sumpter}}, \ and\ \bibinfo
  {author} {\bibfnamefont {Y.}~\bibnamefont {Wang}},\ }\bibfield  {title}
  {\enquote {\bibinfo {title} {Molecular dynamics investigation of the
  relaxation mechanism of entangled polymers after a large step deformation},}\
  }\href@noop {} {\bibfield  {journal} {\bibinfo  {journal} {ACS Macro Lett.}\
  }\textbf {\bibinfo {volume} {7}},\ \bibinfo {pages} {190} (\bibinfo {year}
  {2018})}\BibitemShut {NoStop}%
\bibitem [{\citenamefont {Hoy}\ and\ \citenamefont
  {Robbins}(2006)}]{hoy2006strain}%
  \BibitemOpen
  \bibfield  {author} {\bibinfo {author} {\bibfnamefont {R.~S.}\ \bibnamefont
  {Hoy}}\ and\ \bibinfo {author} {\bibfnamefont {M.~O.}\ \bibnamefont
  {Robbins}},\ }\bibfield  {title} {\enquote {\bibinfo {title} {Strain
  hardening of polymer glasses: Effect of entanglement density, temperature,
  and rate},}\ }\href@noop {} {\bibfield  {journal} {\bibinfo  {journal} {J.
  Polym. Sci. B Polym. Phys.}\ }\textbf {\bibinfo {volume} {44}},\ \bibinfo
  {pages} {3487} (\bibinfo {year} {2006})}\BibitemShut {NoStop}%
\bibitem [{\citenamefont {Hoy}(2011)}]{hoy2011understanding}%
  \BibitemOpen
  \bibfield  {author} {\bibinfo {author} {\bibfnamefont {R.~S.}\ \bibnamefont
  {Hoy}},\ }\bibfield  {title} {\enquote {\bibinfo {title} {Why is
  understanding glassy polymer mechanics so difficult?}}\ }\href@noop {}
  {\bibfield  {journal} {\bibinfo  {journal} {J. Polym. Sci. B Polym. Phys.}\
  }\textbf {\bibinfo {volume} {49}},\ \bibinfo {pages} {979--984} (\bibinfo
  {year} {2011})}\BibitemShut {NoStop}%
\bibitem [{\citenamefont {Sides}\ \emph {et~al.}(2004)\citenamefont {Sides},
  \citenamefont {Grest}, \citenamefont {Stevens},\ and\ \citenamefont
  {Plimpton}}]{sides2004effect}%
  \BibitemOpen
  \bibfield  {author} {\bibinfo {author} {\bibfnamefont {S.~W.}\ \bibnamefont
  {Sides}}, \bibinfo {author} {\bibfnamefont {G.~S.}\ \bibnamefont {Grest}},
  \bibinfo {author} {\bibfnamefont {M.~J.}\ \bibnamefont {Stevens}}, \ and\
  \bibinfo {author} {\bibfnamefont {S.~J.}\ \bibnamefont {Plimpton}},\
  }\bibfield  {title} {\enquote {\bibinfo {title} {Effect of end-tethered
  polymers on surface adhesion of glassy polymers},}\ }\href@noop {} {\bibfield
   {journal} {\bibinfo  {journal} {J. Polym. Sci., Part B: Polym. Phys.}\
  }\textbf {\bibinfo {volume} {42}},\ \bibinfo {pages} {199} (\bibinfo {year}
  {2004})}\BibitemShut {NoStop}%
\bibitem [{\citenamefont {Bobbili}\ and\ \citenamefont
  {Milner}(2020)}]{bobbili2020simulation}%
  \BibitemOpen
  \bibfield  {author} {\bibinfo {author} {\bibfnamefont {S.~V.}\ \bibnamefont
  {Bobbili}}\ and\ \bibinfo {author} {\bibfnamefont {S.~T.}\ \bibnamefont
  {Milner}},\ }\bibfield  {title} {\enquote {\bibinfo {title} {Simulation study
  of entanglement in semiflexible polymer melts and solutions},}\ }\href@noop
  {} {\bibfield  {journal} {\bibinfo  {journal} {Macromolecules}\ }\textbf
  {\bibinfo {volume} {53}},\ \bibinfo {pages} {3861--3872} (\bibinfo {year}
  {2020})}\BibitemShut {NoStop}%
\bibitem [{\citenamefont {Hahn}, \citenamefont {Site},\ and\ \citenamefont
  {Kremer}(2001)}]{hahn2001simulation}%
  \BibitemOpen
  \bibfield  {author} {\bibinfo {author} {\bibfnamefont {O.}~\bibnamefont
  {Hahn}}, \bibinfo {author} {\bibfnamefont {L.~D.}\ \bibnamefont {Site}}, \
  and\ \bibinfo {author} {\bibfnamefont {K.}~\bibnamefont {Kremer}},\
  }\bibfield  {title} {\enquote {\bibinfo {title} {Simulation of polymer melts:
  From spherical to ellipsoidal beads},}\ }\href@noop {} {\bibfield  {journal}
  {\bibinfo  {journal} {Macromol. Theory Simul.}\ }\textbf {\bibinfo {volume}
  {10}},\ \bibinfo {pages} {288} (\bibinfo {year} {2001})}\BibitemShut
  {NoStop}%
\bibitem [{\citenamefont {M{\"u}ller}(2013)}]{muller2013speeding}%
  \BibitemOpen
  \bibfield  {author} {\bibinfo {author} {\bibfnamefont {M.}~\bibnamefont
  {M{\"u}ller}},\ }\bibfield  {title} {\enquote {\bibinfo {title} {Speeding-up
  particle simulations of multicomponent polymer systems by coupling to
  continuum descriptions},}\ }in\ \href@noop {} {\emph {\bibinfo {booktitle}
  {Hybrid particle-continuum methods in computational materials physics}}},\
  Vol.~\bibinfo {volume} {46},\ \bibinfo {editor} {edited by\ \bibinfo {editor}
  {\bibfnamefont {M.~H.}\ \bibnamefont {M{\"u}ser}}, \bibinfo {editor}
  {\bibfnamefont {G.}~\bibnamefont {Sutmann}}, \ and\ \bibinfo {editor}
  {\bibfnamefont {R.~G.}\ \bibnamefont {Winkler}}}\ (\bibinfo  {publisher}
  {Forschungszentrum J{\"u}lich},\ \bibinfo {year} {2013})\ p.\ \bibinfo
  {pages} {127}\BibitemShut {NoStop}%
\bibitem [{\citenamefont {Wang}(2009)}]{wang2009studying}%
  \BibitemOpen
  \bibfield  {author} {\bibinfo {author} {\bibfnamefont {Q.}~\bibnamefont
  {Wang}},\ }\bibfield  {title} {\enquote {\bibinfo {title} {Studying soft
  matter with "soft" potentials: Fast lattice monte carlo simulations and
  corresponding lattice self-consistent field calculations},}\ }\href@noop {}
  {\bibfield  {journal} {\bibinfo  {journal} {Soft Matter}\ }\textbf {\bibinfo
  {volume} {5}},\ \bibinfo {pages} {4564} (\bibinfo {year} {2009})}\BibitemShut
  {NoStop}%
\bibitem [{\citenamefont {Everaers}\ \emph {et~al.}(2004)\citenamefont
  {Everaers}, \citenamefont {Sukumaran}, \citenamefont {Grest}, \citenamefont
  {Svaneborg}, \citenamefont {Sivasubramanian},\ and\ \citenamefont
  {Kremer}}]{PPA}%
  \BibitemOpen
  \bibfield  {author} {\bibinfo {author} {\bibfnamefont {R.}~\bibnamefont
  {Everaers}}, \bibinfo {author} {\bibfnamefont {S.~K.}\ \bibnamefont
  {Sukumaran}}, \bibinfo {author} {\bibfnamefont {G.~S.}\ \bibnamefont
  {Grest}}, \bibinfo {author} {\bibfnamefont {C.}~\bibnamefont {Svaneborg}},
  \bibinfo {author} {\bibfnamefont {A.}~\bibnamefont {Sivasubramanian}}, \ and\
  \bibinfo {author} {\bibfnamefont {K.}~\bibnamefont {Kremer}},\ }\bibfield
  {title} {\enquote {\bibinfo {title} {Rheology and microscopic topology of
  entangled polymeric liquids},}\ }\href@noop {} {\bibfield  {journal}
  {\bibinfo  {journal} {Science}\ }\textbf {\bibinfo {volume} {303}},\ \bibinfo
  {pages} {823} (\bibinfo {year} {2004})}\BibitemShut {NoStop}%
\bibitem [{\citenamefont {Theodorou}\ and\ \citenamefont
  {Suter}(1985)}]{theodorou1985detailed}%
  \BibitemOpen
  \bibfield  {author} {\bibinfo {author} {\bibfnamefont {D.~N.}\ \bibnamefont
  {Theodorou}}\ and\ \bibinfo {author} {\bibfnamefont {U.~W.}\ \bibnamefont
  {Suter}},\ }\bibfield  {title} {\enquote {\bibinfo {title} {Detailed
  molecular structure of a vinyl polymer glass},}\ }\href@noop {} {\bibfield
  {journal} {\bibinfo  {journal} {Macromolecules}\ }\textbf {\bibinfo {volume}
  {18}},\ \bibinfo {pages} {1467} (\bibinfo {year} {1985})}\BibitemShut
  {NoStop}%
\bibitem [{\citenamefont {Harmandaris}\ and\ \citenamefont
  {Kremer}(2009)}]{harmandaris2009dynamics}%
  \BibitemOpen
  \bibfield  {author} {\bibinfo {author} {\bibfnamefont {V.~A.}\ \bibnamefont
  {Harmandaris}}\ and\ \bibinfo {author} {\bibfnamefont {K.}~\bibnamefont
  {Kremer}},\ }\bibfield  {title} {\enquote {\bibinfo {title} {Dynamics of
  polystyrene melts through hierarchical multiscale simulations},}\ }\href@noop
  {} {\bibfield  {journal} {\bibinfo  {journal} {Macromolecules}\ }\textbf
  {\bibinfo {volume} {42}},\ \bibinfo {pages} {791} (\bibinfo {year}
  {2009})}\BibitemShut {NoStop}%
\bibitem [{\citenamefont {Weeks}, \citenamefont {Chandler},\ and\ \citenamefont
  {Andersen}(1971)}]{weeks1971role}%
  \BibitemOpen
  \bibfield  {author} {\bibinfo {author} {\bibfnamefont {J.~D.}\ \bibnamefont
  {Weeks}}, \bibinfo {author} {\bibfnamefont {D.}~\bibnamefont {Chandler}}, \
  and\ \bibinfo {author} {\bibfnamefont {H.~C.}\ \bibnamefont {Andersen}},\
  }\bibfield  {title} {\enquote {\bibinfo {title} {Role of repulsive forces in
  determining the equilibrium structure of simple liquids},}\ }\href@noop {}
  {\bibfield  {journal} {\bibinfo  {journal} {J. Chem. Phys.}\ }\textbf
  {\bibinfo {volume} {54}},\ \bibinfo {pages} {5237--5247} (\bibinfo {year}
  {1971})}\BibitemShut {NoStop}%
\bibitem [{Note1()}]{Note1}%
  \BibitemOpen
  \bibinfo {note} {We have retained the standard KG notation, where $\epsilon $
  denotes the energy scale of interactions and where the temperature is defined
  in units of $\epsilon /k_B$. In contrast, in our parameterisation of the KG
  model for commodity polymer melts~\protect \citep {Everaers2020Mapping} we
  have set the energy scale of all interactions to the thermal energy for the
  experimental target system, $\epsilon =k_B T_{exp}$, and thus rendered the
  model athermal. This has the advantage that the predicted chain structures
  become independent of temperature in agreement with experimental
  observations. For more details, we refer the reader to Ref.~\protect \citep
  {Everaers2020Mapping}.}\BibitemShut {Stop}%
\bibitem [{\citenamefont {Gr{\o}nbech-Jensen}\ and\ \citenamefont
  {Farago}(2013)}]{gronbech2013simple}%
  \BibitemOpen
  \bibfield  {author} {\bibinfo {author} {\bibfnamefont {N.}~\bibnamefont
  {Gr{\o}nbech-Jensen}}\ and\ \bibinfo {author} {\bibfnamefont
  {O.}~\bibnamefont {Farago}},\ }\bibfield  {title} {\enquote {\bibinfo {title}
  {A simple and effective verlet-type algorithm for simulating langevin
  dynamics},}\ }\href@noop {} {\bibfield  {journal} {\bibinfo  {journal} {Mol.
  Phys.}\ }\textbf {\bibinfo {volume} {111}},\ \bibinfo {pages} {983} (\bibinfo
  {year} {2013})}\BibitemShut {NoStop}%
\bibitem [{\citenamefont {Gr{\o}nbech-Jensen}, \citenamefont {Hayre},\ and\
  \citenamefont {Farago}(2014)}]{gronbech2014application}%
  \BibitemOpen
  \bibfield  {author} {\bibinfo {author} {\bibfnamefont {N.}~\bibnamefont
  {Gr{\o}nbech-Jensen}}, \bibinfo {author} {\bibfnamefont {N.~R.}\ \bibnamefont
  {Hayre}}, \ and\ \bibinfo {author} {\bibfnamefont {O.}~\bibnamefont
  {Farago}},\ }\bibfield  {title} {\enquote {\bibinfo {title} {Application of
  the g-jf discrete-time thermostat for fast and accurate molecular
  simulations},}\ }\href@noop {} {\bibfield  {journal} {\bibinfo  {journal}
  {Comp. Phys. Comm.}\ }\textbf {\bibinfo {volume} {185}},\ \bibinfo {pages}
  {524} (\bibinfo {year} {2014})}\BibitemShut {NoStop}%
\bibitem [{\citenamefont {Plimpton}(1995)}]{PlimptonLAMMPS}%
  \BibitemOpen
  \bibfield  {author} {\bibinfo {author} {\bibfnamefont {S.}~\bibnamefont
  {Plimpton}},\ }\bibfield  {title} {\enquote {\bibinfo {title} {Fast parallel
  algorithms for short-range molecular dynamics},}\ }\href@noop {} {\bibfield
  {journal} {\bibinfo  {journal} {J. Comp. Phys.}\ }\textbf {\bibinfo {volume}
  {117}},\ \bibinfo {pages} {1} (\bibinfo {year} {1995})}\BibitemShut {NoStop}%
\bibitem [{\citenamefont {Thompson}\ \emph {et~al.}(2022)\citenamefont
  {Thompson}, \citenamefont {Aktulga}, \citenamefont {Berger}, \citenamefont
  {Bolintineanu}, \citenamefont {Brown}, \citenamefont {Crozier}, \citenamefont
  {in~\'{}t Veld}, \citenamefont {Kohlmeyer}, \citenamefont {Moore},
  \citenamefont {Nguyen}, \citenamefont {Shan}, \citenamefont {Stevens},
  \citenamefont {Tranchida}, \citenamefont {Trott},\ and\ \citenamefont
  {Plimpton}}]{PlimptonLAMMPS2}%
  \BibitemOpen
  \bibfield  {author} {\bibinfo {author} {\bibfnamefont {A.~P.}\ \bibnamefont
  {Thompson}}, \bibinfo {author} {\bibfnamefont {H.~M.}\ \bibnamefont
  {Aktulga}}, \bibinfo {author} {\bibfnamefont {R.}~\bibnamefont {Berger}},
  \bibinfo {author} {\bibfnamefont {D.~S.}\ \bibnamefont {Bolintineanu}},
  \bibinfo {author} {\bibfnamefont {W.~M.}\ \bibnamefont {Brown}}, \bibinfo
  {author} {\bibfnamefont {P.~S.}\ \bibnamefont {Crozier}}, \bibinfo {author}
  {\bibfnamefont {P.~J.}\ \bibnamefont {in~\'{}t Veld}}, \bibinfo {author}
  {\bibfnamefont {A.}~\bibnamefont {Kohlmeyer}}, \bibinfo {author}
  {\bibfnamefont {S.~G.}\ \bibnamefont {Moore}}, \bibinfo {author}
  {\bibfnamefont {T.~D.}\ \bibnamefont {Nguyen}}, \bibinfo {author}
  {\bibfnamefont {R.}~\bibnamefont {Shan}}, \bibinfo {author} {\bibfnamefont
  {M.~J.}\ \bibnamefont {Stevens}}, \bibinfo {author} {\bibfnamefont
  {J.}~\bibnamefont {Tranchida}}, \bibinfo {author} {\bibfnamefont
  {C.}~\bibnamefont {Trott}}, \ and\ \bibinfo {author} {\bibfnamefont {S.~J.}\
  \bibnamefont {Plimpton}},\ }\bibfield  {title} {\enquote {\bibinfo {title}
  {{LAMMPS} a flexible simulation tool for particle based materials modeling at
  the atomic, meso, and continuum scales},}\ }\href {\doibase
  10.1016/j.cpc.2021.108171} {\bibfield  {journal} {\bibinfo  {journal} {Comp.
  Phys. Comm.}\ }\textbf {\bibinfo {volume} {271}},\ \bibinfo {pages} {108171}
  (\bibinfo {year} {2022})}\BibitemShut {NoStop}%
\bibitem [{\citenamefont {Kuhn}(1934)}]{Kuhn}%
  \BibitemOpen
  \bibfield  {author} {\bibinfo {author} {\bibfnamefont {W.}~\bibnamefont
  {Kuhn}},\ }\bibfield  {title} {\enquote {\bibinfo {title} {\"{U}ber die
  gestalt fadenf\"{o}rmiger molek\"{u}le in l\"{o}sung},}\ }\href@noop {}
  {\bibfield  {journal} {\bibinfo  {journal} {Kolloid.}\ }\textbf {\bibinfo
  {volume} {1}},\ \bibinfo {pages} {2} (\bibinfo {year} {1934})}\BibitemShut
  {NoStop}%
\bibitem [{\citenamefont {Lin}(1987)}]{lin87}%
  \BibitemOpen
  \bibfield  {author} {\bibinfo {author} {\bibfnamefont {Y.-H.}\ \bibnamefont
  {Lin}},\ }\bibfield  {title} {\enquote {\bibinfo {title} {Number of
  entanglement strands per cubed tube diameter, a fundamental aspect of
  topological universality in polymer viscoelasticity},}\ }\href@noop {}
  {\bibfield  {journal} {\bibinfo  {journal} {Macromolecules}\ }\textbf
  {\bibinfo {volume} {20}},\ \bibinfo {pages} {3080} (\bibinfo {year}
  {1987})}\BibitemShut {NoStop}%
\bibitem [{\citenamefont {Kavassalis}\ and\ \citenamefont
  {Noolandi}(1987)}]{kavassalis87}%
  \BibitemOpen
  \bibfield  {author} {\bibinfo {author} {\bibfnamefont {T.~A.}\ \bibnamefont
  {Kavassalis}}\ and\ \bibinfo {author} {\bibfnamefont {J.}~\bibnamefont
  {Noolandi}},\ }\bibfield  {title} {\enquote {\bibinfo {title} {New view of
  entanglements in dense polymer systems},}\ }\href@noop {} {\bibfield
  {journal} {\bibinfo  {journal} {Phys. Rev. Lett.}\ }\textbf {\bibinfo
  {volume} {59}},\ \bibinfo {pages} {2674} (\bibinfo {year}
  {1987})}\BibitemShut {NoStop}%
\bibitem [{\citenamefont {Fetters}, \citenamefont {Lohse},\ and\ \citenamefont
  {Colby}(2007)}]{fetters2007chain}%
  \BibitemOpen
  \bibfield  {author} {\bibinfo {author} {\bibfnamefont {L.~J.}\ \bibnamefont
  {Fetters}}, \bibinfo {author} {\bibfnamefont {D.~J.}\ \bibnamefont {Lohse}},
  \ and\ \bibinfo {author} {\bibfnamefont {R.~H.}\ \bibnamefont {Colby}},\
  }\bibfield  {title} {\enquote {\bibinfo {title} {Chain dimensions and
  entanglement spacings},}\ }in\ \href@noop {} {\emph {\bibinfo {booktitle}
  {Physical properties of polymers handbook}}},\ \bibinfo {editor} {edited by\
  \bibinfo {editor} {\bibfnamefont {J.}~\bibnamefont {Mark}}}\ (\bibinfo
  {publisher} {Springer},\ \bibinfo {year} {2007})\ p.\ \bibinfo {pages}
  {447}\BibitemShut {NoStop}%
\bibitem [{\citenamefont {Rosa}\ and\ \citenamefont
  {Everaers}(2014)}]{rosa2014ring}%
  \BibitemOpen
  \bibfield  {author} {\bibinfo {author} {\bibfnamefont {A.}~\bibnamefont
  {Rosa}}\ and\ \bibinfo {author} {\bibfnamefont {R.}~\bibnamefont
  {Everaers}},\ }\bibfield  {title} {\enquote {\bibinfo {title} {Ring polymers
  in the melt state: The physics of crumpling},}\ }\href@noop {} {\bibfield
  {journal} {\bibinfo  {journal} {Phys. Rev. Lett.}\ }\textbf {\bibinfo
  {volume} {112}},\ \bibinfo {pages} {118302} (\bibinfo {year}
  {2014})}\BibitemShut {NoStop}%
\bibitem [{\citenamefont {Everaers}(2012)}]{everaers2012topological}%
  \BibitemOpen
  \bibfield  {author} {\bibinfo {author} {\bibfnamefont {R.}~\bibnamefont
  {Everaers}},\ }\bibfield  {title} {\enquote {\bibinfo {title} {Topological
  versus rheological entanglement length in primitive-path analysis protocols,
  tube models, and slip-link models},}\ }\href@noop {} {\bibfield  {journal}
  {\bibinfo  {journal} {Phys. Rev. E.}\ }\textbf {\bibinfo {volume} {86}},\
  \bibinfo {pages} {022801} (\bibinfo {year} {2012})}\BibitemShut {NoStop}%
\bibitem [{\citenamefont {Uchida}, \citenamefont {Grest},\ and\ \citenamefont
  {Everaers}(2008)}]{Uchida_jcp_08}%
  \BibitemOpen
  \bibfield  {author} {\bibinfo {author} {\bibfnamefont {N.}~\bibnamefont
  {Uchida}}, \bibinfo {author} {\bibfnamefont {G.~S.}\ \bibnamefont {Grest}}, \
  and\ \bibinfo {author} {\bibfnamefont {R.}~\bibnamefont {Everaers}},\
  }\bibfield  {title} {\enquote {\bibinfo {title} {Viscoelasticity and
  primitive path analysis of entangled polymer liquids: From f-actin to
  polyethylene},}\ }\href@noop {} {\bibfield  {journal} {\bibinfo  {journal}
  {J. Chem. Phys.}\ }\textbf {\bibinfo {volume} {128}},\ \bibinfo {pages}
  {044902} (\bibinfo {year} {2008})}\BibitemShut {NoStop}%
\bibitem [{\citenamefont {Sukumaran}\ \emph {et~al.}(2005)\citenamefont
  {Sukumaran}, \citenamefont {Grest}, \citenamefont {Kremer},\ and\
  \citenamefont {Everaers}}]{sukumaran2005identifying}%
  \BibitemOpen
  \bibfield  {author} {\bibinfo {author} {\bibfnamefont {S.~K.}\ \bibnamefont
  {Sukumaran}}, \bibinfo {author} {\bibfnamefont {G.~S.}\ \bibnamefont
  {Grest}}, \bibinfo {author} {\bibfnamefont {K.}~\bibnamefont {Kremer}}, \
  and\ \bibinfo {author} {\bibfnamefont {R.}~\bibnamefont {Everaers}},\
  }\bibfield  {title} {\enquote {\bibinfo {title} {Identifying the primitive
  path mesh in entangled polymer liquids},}\ }\href@noop {} {\bibfield
  {journal} {\bibinfo  {journal} {J. Polym. Sci., Part B: Polym. Phys.}\
  }\textbf {\bibinfo {volume} {43}},\ \bibinfo {pages} {917} (\bibinfo {year}
  {2005})}\BibitemShut {NoStop}%
\bibitem [{\citenamefont {Kr{\"o}ger}(2005)}]{kroger2005shortest}%
  \BibitemOpen
  \bibfield  {author} {\bibinfo {author} {\bibfnamefont {M.}~\bibnamefont
  {Kr{\"o}ger}},\ }\bibfield  {title} {\enquote {\bibinfo {title} {Shortest
  multiple disconnected path for the analysis of entanglements in two-and
  three-dimensional polymeric systems},}\ }\href@noop {} {\bibfield  {journal}
  {\bibinfo  {journal} {Comput. Phys. Commun.}\ }\textbf {\bibinfo {volume}
  {168}},\ \bibinfo {pages} {209} (\bibinfo {year} {2005})}\BibitemShut
  {NoStop}%
\bibitem [{\citenamefont {Tzoumanekas}\ and\ \citenamefont
  {Theodorou}(2006{\natexlab{b}})}]{tzoumanekas2006topological}%
  \BibitemOpen
  \bibfield  {author} {\bibinfo {author} {\bibfnamefont {C.}~\bibnamefont
  {Tzoumanekas}}\ and\ \bibinfo {author} {\bibfnamefont {D.~N.}\ \bibnamefont
  {Theodorou}},\ }\bibfield  {title} {\enquote {\bibinfo {title} {Topological
  analysis of linear polymer melts: A statistical approach},}\ }\href@noop {}
  {\bibfield  {journal} {\bibinfo  {journal} {Macromolecules}\ }\textbf
  {\bibinfo {volume} {39}},\ \bibinfo {pages} {4592--4604} (\bibinfo {year}
  {2006}{\natexlab{b}})}\BibitemShut {NoStop}%
\bibitem [{\citenamefont {Kratky}\ and\ \citenamefont
  {Porod}(1949)}]{KratkyPorod}%
  \BibitemOpen
  \bibfield  {author} {\bibinfo {author} {\bibfnamefont {O.}~\bibnamefont
  {Kratky}}\ and\ \bibinfo {author} {\bibfnamefont {G.}~\bibnamefont {Porod}},\
  }\bibfield  {title} {\enquote {\bibinfo {title} {R\"{o}ntgenuntersuchung
  gel\"{o}ster fadenmolek\"{u}le},}\ }\href@noop {} {\bibfield  {journal}
  {\bibinfo  {journal} {Rec. Trav. Chim.}\ }\textbf {\bibinfo {volume} {68}},\
  \bibinfo {pages} {1106} (\bibinfo {year} {1949})}\BibitemShut {NoStop}%
\bibitem [{\citenamefont {Rouse}(1953)}]{rouse1953theory}%
  \BibitemOpen
  \bibfield  {author} {\bibinfo {author} {\bibfnamefont {P.~E.}\ \bibnamefont
  {Rouse}},\ }\bibfield  {title} {\enquote {\bibinfo {title} {A theory of the
  linear viscoelastic properties of dilute solutions of coiling polymers},}\
  }\href@noop {} {\bibfield  {journal} {\bibinfo  {journal} {J. Chem. Phys.}\
  }\textbf {\bibinfo {volume} {21}},\ \bibinfo {pages} {1272} (\bibinfo {year}
  {1953})}\BibitemShut {NoStop}%
\bibitem [{\citenamefont {Heinrich}, \citenamefont {Straube},\ and\
  \citenamefont {Helmis}(1988)}]{heinrich1988rubber}%
  \BibitemOpen
  \bibfield  {author} {\bibinfo {author} {\bibfnamefont {G.}~\bibnamefont
  {Heinrich}}, \bibinfo {author} {\bibfnamefont {E.}~\bibnamefont {Straube}}, \
  and\ \bibinfo {author} {\bibfnamefont {G.}~\bibnamefont {Helmis}},\
  }\bibfield  {title} {\enquote {\bibinfo {title} {Rubber elasticity of polymer
  networks: Theories},}\ }in\ \href@noop {} {\emph {\bibinfo {booktitle}
  {Polymer Physics}}}\ (\bibinfo  {publisher} {Springer Berlin Heidelberg},\
  \bibinfo {address} {Berlin, Heidelberg},\ \bibinfo {year} {1988})\ pp.\
  \bibinfo {pages} {33--87}\BibitemShut {NoStop}%
\bibitem [{\citenamefont {Rubinstein}\ and\ \citenamefont
  {Panyukov}(1997)}]{rubinstein1997nonaffine}%
  \BibitemOpen
  \bibfield  {author} {\bibinfo {author} {\bibfnamefont {M.}~\bibnamefont
  {Rubinstein}}\ and\ \bibinfo {author} {\bibfnamefont {S.}~\bibnamefont
  {Panyukov}},\ }\bibfield  {title} {\enquote {\bibinfo {title} {Nonaffine
  deformation and elasticity of polymer networks},}\ }\href@noop {} {\bibfield
  {journal} {\bibinfo  {journal} {Macromolecules}\ }\textbf {\bibinfo {volume}
  {30}},\ \bibinfo {pages} {8036} (\bibinfo {year} {1997})}\BibitemShut
  {NoStop}%
\bibitem [{\citenamefont {Edwards}(1967)}]{Edwards_procphyssoc_67}%
  \BibitemOpen
  \bibfield  {author} {\bibinfo {author} {\bibfnamefont {S.~F.}\ \bibnamefont
  {Edwards}},\ }\bibfield  {title} {\enquote {\bibinfo {title} {Statistical
  mechanics with topological constraints: I},}\ }\href@noop {} {\bibfield
  {journal} {\bibinfo  {journal} {Proc. Phys. Soc.}\ }\textbf {\bibinfo
  {volume} {91}},\ \bibinfo {pages} {513} (\bibinfo {year} {1967})}\BibitemShut
  {NoStop}%
\bibitem [{\citenamefont {Hoy}\ and\ \citenamefont
  {Robbins}(2005)}]{PhysRevE.72.061802}%
  \BibitemOpen
  \bibfield  {author} {\bibinfo {author} {\bibfnamefont {R.~S.}\ \bibnamefont
  {Hoy}}\ and\ \bibinfo {author} {\bibfnamefont {M.~O.}\ \bibnamefont
  {Robbins}},\ }\bibfield  {title} {\enquote {\bibinfo {title} {Effect of
  equilibration on primitive path analyses of entangled polymers},}\
  }\href@noop {} {\bibfield  {journal} {\bibinfo  {journal} {Phys. Rev. E.}\
  }\textbf {\bibinfo {volume} {72}},\ \bibinfo {pages} {061802} (\bibinfo
  {year} {2005})}\BibitemShut {NoStop}%
\bibitem [{\citenamefont {Schram}, \citenamefont {Rosa},\ and\ \citenamefont
  {Everaers}(2019)}]{schram2019local}%
  \BibitemOpen
  \bibfield  {author} {\bibinfo {author} {\bibfnamefont {R.~D.}\ \bibnamefont
  {Schram}}, \bibinfo {author} {\bibfnamefont {A.}~\bibnamefont {Rosa}}, \ and\
  \bibinfo {author} {\bibfnamefont {R.}~\bibnamefont {Everaers}},\ }\bibfield
  {title} {\enquote {\bibinfo {title} {Local loop opening in untangled ring
  polymer melts: A detailed “feynman test” of models for the large scale
  structure},}\ }\href@noop {} {\bibfield  {journal} {\bibinfo  {journal} {Soft
  Matter}\ }\textbf {\bibinfo {volume} {15}},\ \bibinfo {pages} {2418--2429}
  (\bibinfo {year} {2019})}\BibitemShut {NoStop}%
\bibitem [{\citenamefont {Binder}(1981)}]{binder1981finite}%
  \BibitemOpen
  \bibfield  {author} {\bibinfo {author} {\bibfnamefont {K.}~\bibnamefont
  {Binder}},\ }\bibfield  {title} {\enquote {\bibinfo {title} {Finite size
  scaling analysis of ising model block distribution functions},}\ }\href@noop
  {} {\bibfield  {journal} {\bibinfo  {journal} {Z. Physik B.}\ }\textbf
  {\bibinfo {volume} {43}},\ \bibinfo {pages} {119--140} (\bibinfo {year}
  {1981})}\BibitemShut {NoStop}%
\bibitem [{\citenamefont {Svaneborg}, \citenamefont {Grest},\ and\
  \citenamefont {Everaers}(2005)}]{SGE_epl_05}%
  \BibitemOpen
  \bibfield  {author} {\bibinfo {author} {\bibfnamefont {C.}~\bibnamefont
  {Svaneborg}}, \bibinfo {author} {\bibfnamefont {G.~S.}\ \bibnamefont
  {Grest}}, \ and\ \bibinfo {author} {\bibfnamefont {R.}~\bibnamefont
  {Everaers}},\ }\bibfield  {title} {\enquote {\bibinfo {title} {{Scattering
  from polymer networks under elongational strain}},}\ }\href@noop {}
  {\bibfield  {journal} {\bibinfo  {journal} {Europhys. Lett}\ }\textbf
  {\bibinfo {volume} {72}},\ \bibinfo {pages} {760} (\bibinfo {year}
  {2005})}\BibitemShut {NoStop}%
\bibitem [{\citenamefont {Tubiana}\ \emph {et~al.}(2021)\citenamefont
  {Tubiana}, \citenamefont {Kobayashi}, \citenamefont {Potestio}, \citenamefont
  {D{\"u}nweg}, \citenamefont {Kremer}, \citenamefont {Virnau},\ and\
  \citenamefont {Daoulas}}]{tubiana2021comparing}%
  \BibitemOpen
  \bibfield  {author} {\bibinfo {author} {\bibfnamefont {L.}~\bibnamefont
  {Tubiana}}, \bibinfo {author} {\bibfnamefont {H.}~\bibnamefont {Kobayashi}},
  \bibinfo {author} {\bibfnamefont {R.}~\bibnamefont {Potestio}}, \bibinfo
  {author} {\bibfnamefont {B.}~\bibnamefont {D{\"u}nweg}}, \bibinfo {author}
  {\bibfnamefont {K.}~\bibnamefont {Kremer}}, \bibinfo {author} {\bibfnamefont
  {P.}~\bibnamefont {Virnau}}, \ and\ \bibinfo {author} {\bibfnamefont
  {K.}~\bibnamefont {Daoulas}},\ }\bibfield  {title} {\enquote {\bibinfo
  {title} {Comparing equilibration schemes of high-molecular-weight polymer
  melts with topological indicators},}\ }\href@noop {} {\bibfield  {journal}
  {\bibinfo  {journal} {J. Phys. Condens.}\ }\textbf {\bibinfo {volume} {33}},\
  \bibinfo {pages} {204003} (\bibinfo {year} {2021})}\BibitemShut {NoStop}%
\bibitem [{\citenamefont {Rubinstein}\ and\ \citenamefont
  {Colby}(2003)}]{RubinsteinColby}%
  \BibitemOpen
  \bibfield  {author} {\bibinfo {author} {\bibfnamefont {M.}~\bibnamefont
  {Rubinstein}}\ and\ \bibinfo {author} {\bibfnamefont {R.~H.}\ \bibnamefont
  {Colby}},\ }\href@noop {} {\emph {\bibinfo {title} {Polymer physics}}}\
  (\bibinfo  {publisher} {Oxford University},\ \bibinfo {address} {New York},\
  \bibinfo {year} {2003})\BibitemShut {NoStop}%
\bibitem [{\citenamefont {de~Gennes}(1979{\natexlab{b}})}]{deGennes}%
  \BibitemOpen
  \bibfield  {author} {\bibinfo {author} {\bibfnamefont {P.-G.}\ \bibnamefont
  {de~Gennes}},\ }\href@noop {} {\emph {\bibinfo {title} {Scaling concepts in
  polymer physics}}}\ (\bibinfo  {publisher} {Cornell University Press},\
  \bibinfo {address} {New York},\ \bibinfo {year} {1979})\BibitemShut {NoStop}%
\bibitem [{\citenamefont {Dealy}, \citenamefont {Read},\ and\ \citenamefont
  {Larson}(2018)}]{dealy2018structure}%
  \BibitemOpen
  \bibfield  {author} {\bibinfo {author} {\bibfnamefont {J.~M.}\ \bibnamefont
  {Dealy}}, \bibinfo {author} {\bibfnamefont {D.~J.}\ \bibnamefont {Read}}, \
  and\ \bibinfo {author} {\bibfnamefont {R.~G.}\ \bibnamefont {Larson}},\
  }\href@noop {} {\emph {\bibinfo {title} {Structure and rheology of molten
  polymers: from structure to flow behavior and back again}}}\ (\bibinfo
  {publisher} {Carl Hanser Verlag GmbH Co KG},\ \bibinfo {year}
  {2018})\BibitemShut {NoStop}%
\bibitem [{\citenamefont {Wittmer}\ \emph {et~al.}(2004)\citenamefont
  {Wittmer}, \citenamefont {Meyer}, \citenamefont {Baschnagel}, \citenamefont
  {Johner}, \citenamefont {Obukhov}, \citenamefont {Mattioni}, \citenamefont
  {M{\"u}ller},\ and\ \citenamefont {Semenov}}]{Wittmer_Meyer_PRL04}%
  \BibitemOpen
  \bibfield  {author} {\bibinfo {author} {\bibfnamefont {J.~P.}\ \bibnamefont
  {Wittmer}}, \bibinfo {author} {\bibfnamefont {H.}~\bibnamefont {Meyer}},
  \bibinfo {author} {\bibfnamefont {J.}~\bibnamefont {Baschnagel}}, \bibinfo
  {author} {\bibfnamefont {A.}~\bibnamefont {Johner}}, \bibinfo {author}
  {\bibfnamefont {S.}~\bibnamefont {Obukhov}}, \bibinfo {author} {\bibfnamefont
  {L.}~\bibnamefont {Mattioni}}, \bibinfo {author} {\bibfnamefont
  {M.}~\bibnamefont {M{\"u}ller}}, \ and\ \bibinfo {author} {\bibfnamefont
  {A.~N.}\ \bibnamefont {Semenov}},\ }\bibfield  {title} {\enquote {\bibinfo
  {title} {Long range bond-bond correlations in dense polymer solutions},}\
  }\href@noop {} {\bibfield  {journal} {\bibinfo  {journal} {Phys. Rev. Lett.}\
  }\textbf {\bibinfo {volume} {93}},\ \bibinfo {pages} {147801} (\bibinfo
  {year} {2004})}\BibitemShut {NoStop}%
\bibitem [{\citenamefont {Wittmer}\ \emph
  {et~al.}(2007{\natexlab{a}})\citenamefont {Wittmer}, \citenamefont
  {Beckrich}, \citenamefont {Johner}, \citenamefont {Semenov}, \citenamefont
  {Obukhov}, \citenamefont {Meyer},\ and\ \citenamefont
  {Baschnagel}}]{wittmer2007polymer}%
  \BibitemOpen
  \bibfield  {author} {\bibinfo {author} {\bibfnamefont {J.~P.}\ \bibnamefont
  {Wittmer}}, \bibinfo {author} {\bibfnamefont {P.}~\bibnamefont {Beckrich}},
  \bibinfo {author} {\bibfnamefont {A.}~\bibnamefont {Johner}}, \bibinfo
  {author} {\bibfnamefont {A.~N.}\ \bibnamefont {Semenov}}, \bibinfo {author}
  {\bibfnamefont {S.~P.}\ \bibnamefont {Obukhov}}, \bibinfo {author}
  {\bibfnamefont {H.}~\bibnamefont {Meyer}}, \ and\ \bibinfo {author}
  {\bibfnamefont {J.}~\bibnamefont {Baschnagel}},\ }\bibfield  {title}
  {\enquote {\bibinfo {title} {Why polymer chains in a melt are not random
  walks},}\ }\href@noop {} {\bibfield  {journal} {\bibinfo  {journal}
  {Europhys. Lett.}\ }\textbf {\bibinfo {volume} {77}},\ \bibinfo {pages}
  {56003} (\bibinfo {year} {2007}{\natexlab{a}})}\BibitemShut {NoStop}%
\bibitem [{\citenamefont {Wittmer}\ \emph
  {et~al.}(2007{\natexlab{b}})\citenamefont {Wittmer}, \citenamefont
  {Beckrich}, \citenamefont {Meyer}, \citenamefont {Cavallo}, \citenamefont
  {Johner},\ and\ \citenamefont {Baschnagel}}]{wittmer2007intramolecular}%
  \BibitemOpen
  \bibfield  {author} {\bibinfo {author} {\bibfnamefont {J.~P.}\ \bibnamefont
  {Wittmer}}, \bibinfo {author} {\bibfnamefont {P.}~\bibnamefont {Beckrich}},
  \bibinfo {author} {\bibfnamefont {H.}~\bibnamefont {Meyer}}, \bibinfo
  {author} {\bibfnamefont {A.}~\bibnamefont {Cavallo}}, \bibinfo {author}
  {\bibfnamefont {A.}~\bibnamefont {Johner}}, \ and\ \bibinfo {author}
  {\bibfnamefont {J.}~\bibnamefont {Baschnagel}},\ }\bibfield  {title}
  {\enquote {\bibinfo {title} {Intramolecular long-range correlations in
  polymer melts: The segmental size distribution and its moments},}\
  }\href@noop {} {\bibfield  {journal} {\bibinfo  {journal} {Phys. Rev. E.}\
  }\textbf {\bibinfo {volume} {76}},\ \bibinfo {pages} {011803} (\bibinfo
  {year} {2007}{\natexlab{b}})}\BibitemShut {NoStop}%
\bibitem [{\citenamefont {Beckrich}\ \emph {et~al.}(2007)\citenamefont
  {Beckrich}, \citenamefont {Johner}, \citenamefont {Semenov}, \citenamefont
  {Obukhov}, \citenamefont {Benoit},\ and\ \citenamefont
  {Wittmer}}]{beckrich2007intramolecular}%
  \BibitemOpen
  \bibfield  {author} {\bibinfo {author} {\bibfnamefont {P.}~\bibnamefont
  {Beckrich}}, \bibinfo {author} {\bibfnamefont {A.}~\bibnamefont {Johner}},
  \bibinfo {author} {\bibfnamefont {A.~N.}\ \bibnamefont {Semenov}}, \bibinfo
  {author} {\bibfnamefont {S.~P.}\ \bibnamefont {Obukhov}}, \bibinfo {author}
  {\bibfnamefont {H.}~\bibnamefont {Benoit}}, \ and\ \bibinfo {author}
  {\bibfnamefont {J.~P.}\ \bibnamefont {Wittmer}},\ }\bibfield  {title}
  {\enquote {\bibinfo {title} {Intramolecular form factor in dense polymer
  systems: Systematic deviations from the debye formula},}\ }\href@noop {}
  {\bibfield  {journal} {\bibinfo  {journal} {Macromolecules}\ }\textbf
  {\bibinfo {volume} {40}},\ \bibinfo {pages} {3805} (\bibinfo {year}
  {2007})}\BibitemShut {NoStop}%
\bibitem [{\citenamefont {Semenov}(2010)}]{semenov2010bond}%
  \BibitemOpen
  \bibfield  {author} {\bibinfo {author} {\bibfnamefont {A.~N.}\ \bibnamefont
  {Semenov}},\ }\bibfield  {title} {\enquote {\bibinfo {title} {Bond-vector
  correlation functions in dense polymer systems},}\ }\href@noop {} {\bibfield
  {journal} {\bibinfo  {journal} {Macromolecules}\ }\textbf {\bibinfo {volume}
  {43}},\ \bibinfo {pages} {9139} (\bibinfo {year} {2010})}\BibitemShut
  {NoStop}%
\end{thebibliography}%

\end{document}